\documentclass{aa}  

\usepackage{graphicx}
\usepackage{txfonts}

\usepackage[export]{adjustbox}
\newcommand{\mum}{$\mathrm{\thinspace\mu m}$}
\usepackage{array}
\usepackage{multirow}
\usepackage{tabularx}
\usepackage{orcidlink}

\begin{document}

   \title{\emph{JWST} MIRI Flight Performance: Imaging}

    \author{Dan Dicken\inst{1}\orcidlink{0000-0003-0589-5969}
    \and
    Macarena García Marín\inst{2}\orcidlink{0000-0003-4801-0489}
    \and
    Irene Shivaei\inst{3,4}\orcidlink{0000-0003-4702-7561}
    \and
    Pierre Guillard\inst{5,6}\orcidlink{0000-0002-2421-1350} 
    \and
    Mattia Libralato\inst{7,8}\orcidlink{0000-0001-9673-7397}
    \and
    Alistair Glasse\inst{1}\orcidlink{0000-0002-2041-2462}
    \and
    Karl D.\ Gordon\inst{9,10}\orcidlink{0000-0001-5340-6774}
    \and
    Christophe Cossou\inst{11}\orcidlink{0000-0001-5350-4796}
    \and
    Patrick Kavanagh\inst{12}\orcidlink{0000-0001-6872-2358}
    \and
    Tea Temim\inst{13}\orcidlink{0000-0001-7380-3144}
    \and
    Nicolas Flagey\inst{9}\orcidlink{0000-0002-8763-1555}
    \and
    Pamela Klaassen\inst{1}\orcidlink{0000-0001-9443-0463}
    \and
    George H.\ Rieke\inst{3}\orcidlink{0000-0003-2303-6519}
    \and
    Gillian Wright\inst{1}\orcidlink{0000-0001-7416-7936}
    \and
    Stacey Alberts\inst{3}\orcidlink{0000-0002-8909-8782}
    \and
    Ruyman Azzollini\inst{4,12}\orcidlink{0000-0002-0438-0886}
    \and
    Javier Álvarez-Márquez\inst{4}\orcidlink{0000-0002-7093-1877}
    \and
    Patrice Bouchet\inst{11}\orcidlink{0000-0002-6018-3393}
    \and
    Stacey Bright\inst{9}\orcidlink{0000-0001-7951-7966}
    \and
    Misty Cracraft\inst{9}\orcidlink{0000-0002-7698-3002}
    \and
    Alain Coulais\inst{11,14}\orcidlink{0000-0001-6492-7719}
    \and
    Ors Hunor Detre\inst{15}\orcidlink{0000-0003-0585-4219}
    \and
    Mike Engesser\inst{9}\orcidlink{0000-0003-0209-674X}
    \and
    Ori D.\ Fox\inst{9}\orcidlink{0000-0003-2238-1572}
    \and
    Andras Gaspar\inst{3}\orcidlink{0000-0001-8612-3236}
    \and
    René Gastaud\inst{11}\orcidlink{0009-0007-5200-1362}
    \and
    Adrian M. Glauser\inst{16}\orcidlink{0000-0001-9250-1547}
    \and
    Dean C. Hines\inst{9}\orcidlink{0000-0003-4653-6161}
    \and
    Sarah Kendrew\inst{2}\orcidlink{0000-0002-7612-0469}
    \and
    Alvaro Labiano\inst{4,17}\orcidlink{0000-0002-0690-8824}
    \and
    Pierre-Oliver Lagage\inst{11}
    \and
    David Lee\inst{1}
    \and
    David R. Law\inst{9}\orcidlink{0000-0002-9402-186X}
    \and
    Jane E.\ Morrison\inst{3}\orcidlink{0000-0002-9288-9235}
    \and
    Alberto Noriega-Crespo\inst{9}\orcidlink{0000-0002-6296-8960}
    \and
    Olivia Jones\inst{1}\orcidlink{0000-0003-4870-5547}
    \and
    Polychronis Patapis\inst{16}\orcidlink{0000-0001-8718-3732}
    \and
    Silvia Scheithauer\inst{15}\orcidlink{0000-0003-4559-0721}
    \and
    G.~C.\ Sloan\inst{9,18}\orcidlink{0000-0003-4520-1044}
    \and
    Laszlo Tamas\inst{1}\orcidlink{0009-0001-8954-7138}
    }

    \institute{UK Astronomy Technology Centre, Royal Observatory Edinburgh, Blackford Hill, Edinburgh EH9 3HJ, UK
    \and
    European Space Agency, at Space Telescope Science Institute, 3700 San Martin Drive, Baltimore, MD 21218, USA
    \and
    Steward Observatory, University of Arizona, Tucson, AZ 85721, USA
    \and
    Centro de Astrobiolog\'{i}a (CAB), CSIC-INTA, Carretera de Ajalvir km 4, Torrej\'{o}n de Ardoz, 28850, Madrid, Spain
    \and
    Sorbonne Universit\'{e}, CNRS, UMR 7095, Institut d'Astrophysique de Paris, 98bis bd Arago, 75014 Paris, France
    \and
    Institut Universitaire de France, Minist{\`e}re de l'Enseignement Sup{\'e}rieur et de la Recherche, 1 rue Descartes, 75231 Paris Cedex 05, France
    \and
    AURA for the European Space Agency (ESA), Space Telescope Science Institute, 3700 San Martin Drive, Baltimore, MD 21218, USA
    \and
    INAF - Osservatorio Astronomico di Padova, Vicolo dell'Osservatorio 5, Padova, 35122, Italy
    \and
    Space Telescope Science Institute, 3700 San Martin Drive, Baltimore, MD 21218, USA
    \and
    Sterrenkundig Observatorium, Universiteit Gent, Gent, Belgium
    \and
    Université Paris-Saclay, CEA, IRFU, 91191, Gif-sur-Yvette, France
    \and
    School of Cosmic Physics, Dublin Institute for Advanced Studies, 31 Fitzwilliam Place, Dublin 2, Ireland
    \and
    Princeton University, 4 Ivy Ln, Princeton, NJ 08544, USA
    \and
    LERMA, Observatoire de Paris, Université PSL, Sorbonne Université, CNRS, Paris, France
    \and
    Max Planck Institute for Astronomy, K\"onigstuhl 17, 69117 Heidelberg, Germany
    \and
    Institute for Particle Physics and Astrophysics, ETH Zurich, Wolfgang-Paulistr. 27, 8093 Zurich, Switzerland
    \and
    Telespazio UK for the European Space Agency (ESA), ESAC,  Camino Bajo del Castillo s/n, 28692 Villanueva de la Cañada, Spain
    \and
    Department of Physics and Astronomy, University of North Carolina, Chapel Hill, NC 27599-3255, USA
     }

 
  \abstract
  {The Mid-Infrared Instrument (MIRI) aboard the \emph{James Webb Space Telescope (JWST)} provides the observatory with a huge advance in mid-infrared imaging and spectroscopy covering the wavelength range of 5 to 28\mum. This paper describes the performance and characteristics of the MIRI imager as understood during observatory commissioning activities, and through its first year of science operations. We discuss the measurements and results of the imager's point spread function, flux calibration, background, distortion and flat fields as well as results pertaining to best observing practices for MIRI imaging, and discuss known imaging artefacts that may be seen during or after data processing. Overall, we show that the MIRI imager has met or exceeded all its pre-flight requirements, and we expect it to make a significant contribution to mid-infrared science for the astronomy community for years to come. }


   \maketitle

\section{Introduction}
\label{sec:intro}
Forty years of mid-infrared (mid-IR, 5-30$\mathrm{\thinspace\mu m}$) imaging in space with observatories such as the \emph{Infrared Astronomical Satellite (IRAS)}, the \emph{Infrared Space Observatory (ISO)}, the \emph{Spitzer Space Telescope}, \emph{Akari}, and the \emph{Wide-Field Infrared Survey Explorer (WISE)} have unveiled a new view of the universe with a wealth of scientific reward. Mid-IR imaging not only enables us to see deep into the dust structures as well as the centre of our own Galaxy and Local Group, but also samples important dust continuum features such as Polycyclic Aromatic Hydrocarbons (PAH; 7.7 and 11.3$\mathrm{\thinspace\mu m}$) and silicates (10 and 18$\mathrm{\thinspace\mu m}$) that reveal the dominant energetic processes in  galaxies across cosmic time \citep{tielens2008,Li2020}. Moreover, deep wide-area surveys have shown that mid-IR photometric bands from \emph{Spitzer's} IRAC and MIPS instruments have proved to be powerful diagnostic tools for galaxy evolution in studying star formation processes \citep{elbaz2007, Noeske2007, leroy2008, Madau2014}, active galactic nuclei \citep{Richards2006,Daddi2007,Donley2012}, and even the very high redshift universe \citep{Stark2009, Smit2015}. The latest space-based IR observatory is \emph{JWST}, which has begun a golden age of IR astrophysics and it is the Mid-Infrared Instrument's (MIRI) imager that is providing imaging between 5 and 28 $\mathrm{\thinspace\mu m}$ with unparalleled resolution and sensitivity that will open new vistas for discovery in all areas of astronomy.

After launch, \emph{JWST} was commissioned in preparation for science operations in a phase of the mission that ran for 6 months up to the end of June 2022. An overview of the science performance of \emph{JWST} at the end of commissioning has been presented in \cite{rigby2022a} and specifically for MIRI in \cite{wright2023PASP}. After the launch and deployment of the observatory, the commissioning and preliminary calibration phase of the instruments took place in the latter half of the commissioning period. This is especially true for MIRI, which being the coldest instrument in the observatory, and the only instrument that is actively cooled, took the longest to cool down to its operating temperature. Therefore, most of MIRI commissioning observations were obtained in a relatively short time frame between late May and the end of June 2022. In order to conduct the MIRI imager commissioning efficiently, a detailed set of tests was outlined and prepared well in advance of launch, including algorithms for analysing the data. This work was referred to as Commissioning Analysis Projects (CAPs), and for the MIRI imager, contained analysis and characterisation of: basic functionality, the point spread function (PSF), plate scale and geometric distortion, glints and scattered light, flat fields, sky background, operations, image artefacts, performance checks, and initial photometric calibration. The results of the MIRI imager CAPs project are presented here. 

The MIRI imager, as well as the other MIRI modes and \emph{JWST} instruments, was put through extensive ground testing both at Goddard Space Flight Center and Houston Space Flight Center, where the latter involved a full end-to-end test with the flight optical telescope  assembly. However, although these demonstrated functionality and basic performance, the tests were limited for the MIRI imager because the test sources and backgrounds were optimised for the near-infrared instruments\footnote{The near-infrared instruments were prioritised in ground testing because they are the primary instruments for mirror alignment and phasing.} and not mid-IR  (5 to 28$\mathrm{\thinspace\mu m}$). Consequently, many aspects of the MIRI imager performance were not well characterised prior to launch, including the imager background, the flats and bright target limits as well as detector performance related to imaging. Therefore, the in-flight commissioning tests were important for demonstrating performance and preparing the initial calibrations to ready the instrument modes for science observations. 

We note that the aim of the in-flight commissioning tests was not to fully calibrate the MIRI imager, but to measure and check the in-flight performance compared with the instrument design specifications and the team's ground test measurements. The results presented here show the performance as understood from commissioning at the start of \emph{JWST} operations, and to the best of our knowledge from the first year of operations. The MIRI imager calibration and performance analysis will continue throughout the \emph{JWST} mission, and the reader should refer to \emph{JWST} documentation to get the most up to date information (\href{https://jwst-docs.stsci.edu/jwst-mid-infrared-instrument#JWSTMidInfraredInstrument-Imager}{\textit{JWST} Documentation - MIRI imaging}). 

This paper is organised as follows. We begin with an overview of the MIRI imager and summarising the key results of the imager commissioning program (Section~\ref{sec:overview} and~\ref{sec:key}). Next we discuss in detail the results of the MIRI imager commissioning including the point spread functions (PSFs) (Section~\ref{sec:psf}), flux calibration (Section~\ref{sec:calibration}), backgrounds (Section~\ref{sec:bkg}), distortion (Section~\ref{sec:distortion}) and flat fields (Section~\ref{sec:flats}). After which we describe some of the imaging artefacts that may be seen during and after data processing (Section~\ref{sec:artefacts}), and then discuss how the results of commissioning feed into the best observing practices for MIRI imaging (Section~\ref{sec:best}). Section~\ref{normalops} briefly summarizes experience with the MIRI imager during its first year of science operations.

\section{Imager Overview} \label{sec:overview}
\begin{figure*}
\includegraphics[width=0.8\textwidth, center]{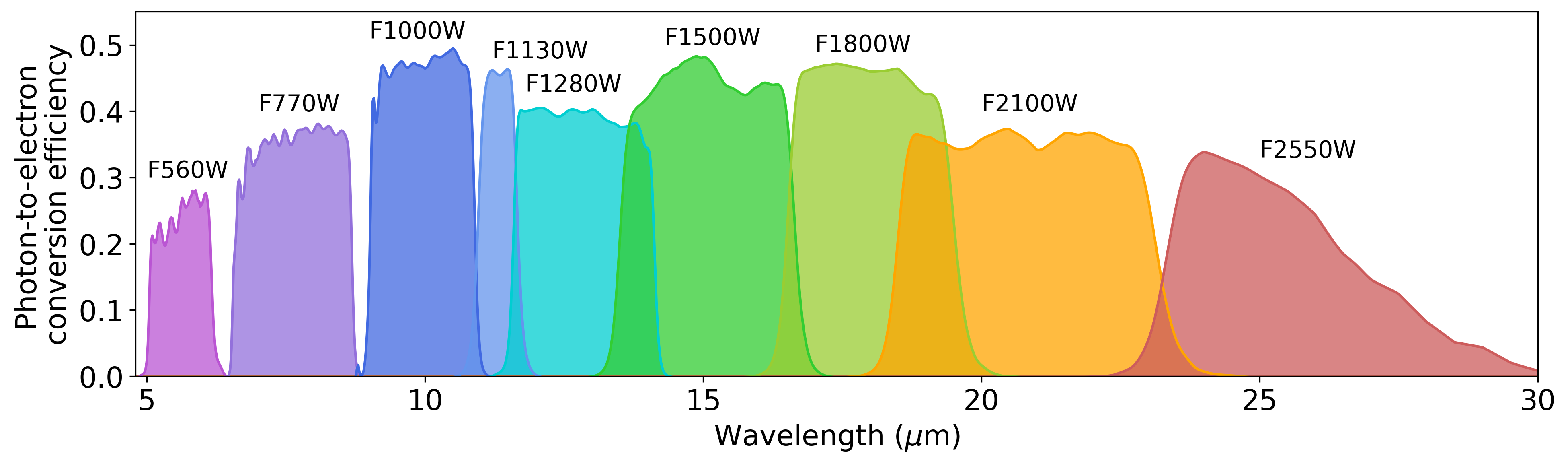}
\caption{MIRI imaging filter bandpasses.}
\label{fig:imager_bandpass}
\end{figure*}

\begin{figure*}
\includegraphics[width=0.8\textwidth, center]{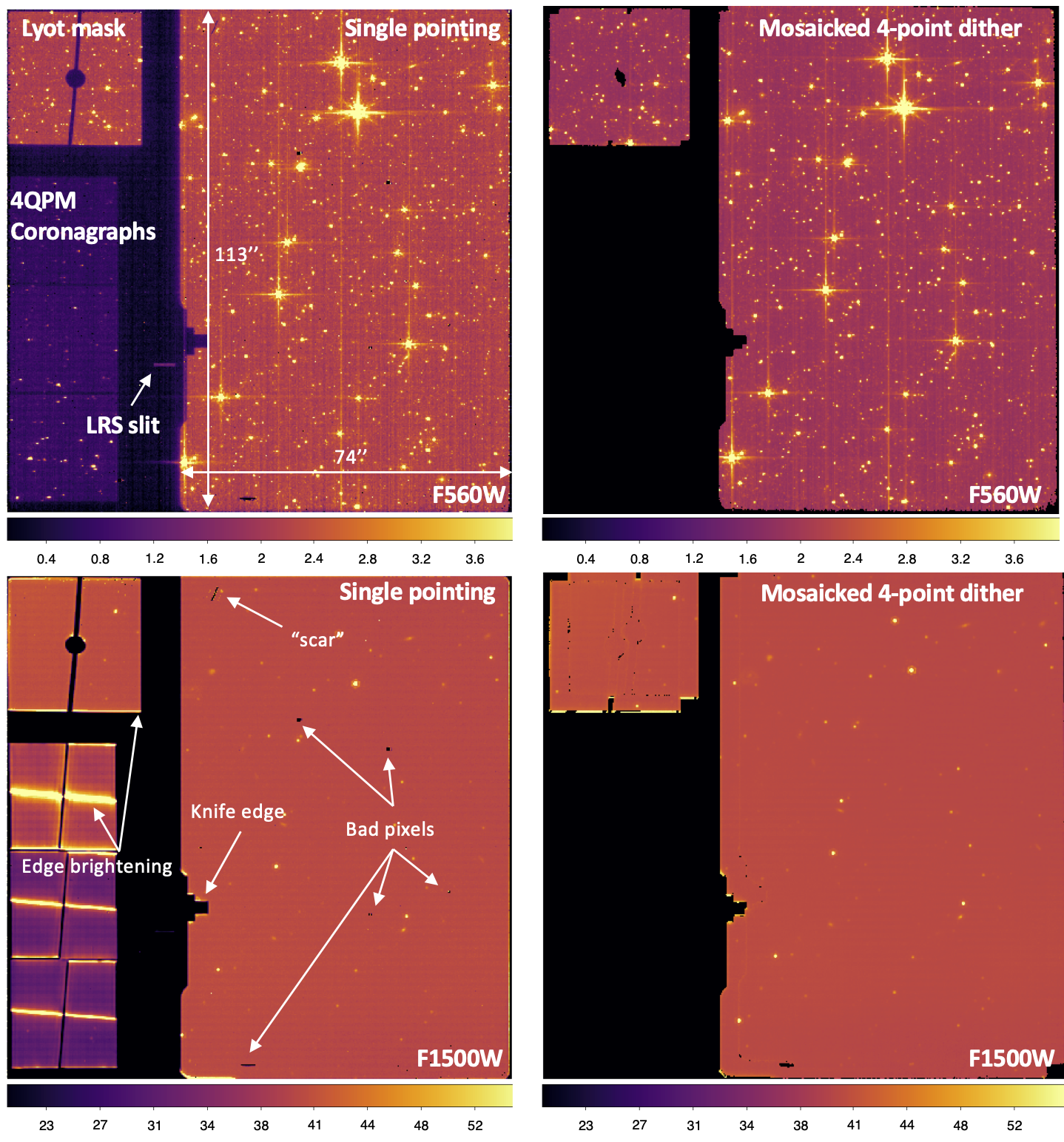}
\caption{Showing 5.6$\mathrm{\thinspace\mu m}$ (top) and 15.0$\mathrm{\thinspace\mu m}$ (bottom) imaging of the LMC from the commissioning program 1024 (observations 5 and 9). The images on the left show an example of a stage-2 \texttt{cal} FITS file, i.e., unrectified, flat-fielded single exposures. The images on the right show the resulting mosaic image of four dithers from the data set. The images are processed to level 3a in the pipeline. All units are MJy sr$^{-1}$. 
\label{fig:imager_examples}}
\end{figure*}

The MIRI imager (\citealt{bouchet2015}) consists of an all reflecting design with 9 primary filters (See Figure~\ref{fig:imager_bandpass}). The imager shares its detector focal plane system with MIRI's Low Resolution Spectrometer (LRS) and Coronagraphs, giving it a rectangular unobstructed field of view of $74 \times 113$ arcsec$^2$,  which is roughly a quarter of the size of the Near InfraRed Camera (NIRCam) imager field on \emph{JWST}, with a plate scale of $0.11$ arcsec pixel$^{-1}$. The imager full detector array is the entire 1024 x 1024 pixel square array shown in Figure~\ref{fig:imager_examples} that encompasses the imager field of view, the coronagraphs and LRS mask. The imager field of view covers 2/3 of the full array. The Lyot mask region is also included in imager science data products because the Lyot coronagraph has no additional optics.” The full imager array can be read out in FASTR1 or SLOWR1 mode with readout times per frame of $2.775$  and $23.890$ seconds respectively, where the slower readout mode can be used to reduce data volume \citep[see][]{morrison2023}. To prevent detector saturation on bright targets or backgrounds, MIRI imaging can also be obtained in subarray mode, which reads smaller sections of the detector at a faster rate by manipulating the detector clocking patterns \citep{ressler2015}. The subarrays, only available when using the FASTR1 readout pattern, are: BRIGHTSKY ($512 \times 512$ pixel$^2$, group time $0.865$ s), SUB256 ($256 \times 256$ pixel$^2$, group time $0.300$ s), SUB128 ($136 \times 128$ pixel$^2$, group time $0.119$ s) and SUB64 ($72 \times 64$ pixel$^2$, group time $0.085$ s). The basic parameters for the MIRI imager are provided in Table 1. The MIRI imaging filter bandpasses, in units of photon-to-electron conversion efficiency, are shown in Figure 1.

\begin{table}[t]
\begin{center}
\begin{tabular}{|c|c|c|c|}

\hline
Imaging Filters & $\lambda_0$ [$\mu$m] & R $=$ $\lambda/\Delta \lambda$ & FWHM\\
\hline
F560W  &  5.6 & 4.7 & 0.207 \\
F770W  &  7.7 & 3.5 & 0.269 \\
F1000W & 10.0 & 5.0 & 0.328 \\
F1130W & 11.3 & 16.1 & 0.375 \\
F1280W & 12.8 & 5.3 & 0.420 \\
F1500W & 15.0 & 5.0 & 0.488 \\ 
F1800W & 18.0 & 6.0 & 0.591 \\
F2100W & 21.0 & 4.2 & 0.674 \\
F2550W & 25.5 & 6.4 & 0.803 \\ 
\hline
\hline
Imaging Subarrays & Size (pixels) & Size (arcsecs) & Frame time\\
\hline
FULL       & $1024\times1032$  & $74"\times113"$  &  2.775s\\
BRIGHTSKY  & $512\times512$ & $56.3"\times56.3"$  & 0.865s\\
SUB256     & $256\times256$ & $28.2"\times28.2"$  & 0.300s\\
SUB128     & $128\times136$ & $14.1"\times14.1"$  & 0.119s\\
SUB64      & $64\times72$  & $7"\times7"$  & 0.085s\\
\hline
\end{tabular}
\caption{Top: MIRI imaging filters, central wavelengths, resolving powers and full width half maxima. Bottom: MIRI imaging subarrays, pixel and usable sizes, frame times. \label{table:basic_params}}
\end{center}
\end{table}

In Figure~\ref{fig:imager_examples}, we show examples of MIRI imaging from the commissioning program (Program Identification -- hereafter PID 1024 ; Observations 5 and 9) that were used for the calibration of the geometric distortion of the instrument (see Section~\ref{sec:distortion}). The images show a field in the Large Magellanic Cloud (LMC) taken both at 5.6  (top) and 15.0$\mathrm{\thinspace\mu m}$ (bottom). In the single pointed image, we clearly see the imager's rectangular field of view, and also the three 4-quadrant phase mask (4QPM) coronagraphs to the left of the image, the Lyot coronagraphic mask region to the top left, as well as the LRS slit that all share the same detector as the imager. The Lyot mask region is particularly important to MIRI imaging because the light arriving in the Lyot quadrant passes through the same optical elements as for the MIRI imager, with the exception of the Lyot stop. Thus, this small area can be used, calibrated and analysed as for the main imager region. The 4QPM coronagraphs include additional optical elements that require a calibration different from that of the imager. For these reasons,  mosaic images obtained from the official \textit{JWST} pipeline (the level 3 \texttt{i2d} FITS files) include both the imager and the Lyot regions, but have the 4QPM coronagraph regions masked (Figure~\ref{fig:imager_examples}, right column).

Also in Figure~\ref{fig:imager_examples} we can see a small rectangular protrusion into the imager field of view on the left hand edge (labelled in bottom-left panel). This is known as the ``knife edge,'' and was used to perform a ``knife edge'' optical test, whereby a point source could be moved across the edge of the detector and the signal measured as a function of position. This enables the PSF at the entrance focal plane (where the knife edge and coronagraphs are mounted in the focal plane structure) to be measured by fitting the signal distribution as a function of source position and comparing it to the optical model. 

We also identify regions of bad pixels in Figure~\ref{fig:imager_examples}, which appear as small dark patches in the single point images. Figure~\ref{fig:imager_examples}'s mosaicked images on the right show that these bad pixel areas are well mitigated with the 4-point dither pattern strategy of this observation. \textit{JWST} data products have a number of different flag identifiers for bad pixels including hot, open, dead and low quantum efficiency depending on the characteristics of the loss of performance(\href{https://jwst-pipeline.readthedocs.io/en/latest/jwst/references_general/references_general.html#data-quality-flags}{\textit{JWST Pipeline Documentation -- Data Quality Flags}}). In addition, the four columns and rows around the edge of the detector are also flagged as ``unreliable slope'' due to excess noise from these areas of the detector. There is also one column (number 385) that is masked in imaging because it does not give the correct output due to a manufacturing fault. This column lies to the left hand side of the imager field of view, and may be seen as a line of masked pixels in un-dithered data sets (pipeline stage 2 products). Also, to the top left of the field of view, the imager detector has a feature that looks like a diagonal stripe, centered at pixel (428,987) with a length of 16 pixels. This is sometimes referred to as the ``scar," and is due to a scratch on the surface of the detector. Although the pixels are responsive, this feature can scatter light in both directions perpendicular to the scratch (see \href{https://jwst-docs.stsci.edu/jwst-mid-infrared-instrument/miri-instrument-features-and-caveats#MIRIInstrumentFeaturesandCaveats-Imager%22scar%22}{\textit{JWST "scar"} Documentation} for more details). Excluding the edge pixels, the imager detector has less than 0.01\% of pixels flagged as bad in its first year or operation (see \href{https://jwst-docs.stsci.edu/jwst-mid-infrared-instrument/miri-features-and-caveats#MIRIFeaturesandCaveats-Detectorgeneralfeaturesandcaveats}{\textit{JWST imager Documentation} } for more details). 

Figure~\ref{fig:imager_examples} also illustrates the imager background change with wavelength (see Section~\ref{sec:bkg} and \citealt{glasse2015}) where the 5.6$\mathrm{\thinspace\mu m}$ imaging is MIRI's shortest wavelength and lowest background at around 2 MJy sr$^{-1}$. In longer wavelength imaging, shown in Figure~\ref{fig:imager_examples} at 15.5$\mathrm{\thinspace\mu m}$, edge brightening can be seen in the 4QPM coronagraphs as well as at the bottom of the Lyot mask region and around the knife edge. This feature, first seen in commissioning, is known as the ``glow-stick'' because of its appearance in the 4QPM coronagraphic masks \citep{boccaletti2022}. As can be seen in Figure~\ref{fig:imager_examples}, the feature gets brighter towards longer wavelengths but is not thought to have any adverse effects on MIRI imaging, although bright glow-stick features at long wavelength could cause row artefacts as discussed in Section~\ref{sec:row_column}.

Descriptions and references on the MIRI design, build and ground testing are in \cite{wright2015} and specifically for the imager in \cite{bouchet2015}. More information on the MIRI imager can be found in the \textit{JWST} documentation at the following link: \href{https://jwst-docs.stsci.edu/jwst-mid-infrared-instrument#JWSTMidInfraredInstrument-Imager}{\textit{JWST} Documentation - MIRI imaging}, which will include the most up-to-date information for observers during the mission.

\section{Key Imager Commissioning Results} \label{sec:key}

The primary aim of \textit{JWST}/MIRI commissioning was to verify that the instrument's on-orbit functionality and performance were consistent with the scientific capabilities estimated from ground testing. With over 3500 exposures taken in imaging mode, the commissioning analysis showed that MIRI imaging was equal to or better than expectations from pre-flight testing and, wherever possible, we mapped all the results to our requirements. The commissioning tests confirmed the correct functionality of the instrument mode including filters, slews, offsets and dither patterns as well as testing the templates of the Astronomer Proposal Tool (APT). No operational constraints or limitations on the instrument were necessary. 

In terms of image quality, the full width half maximum (FWHM) and encircled energies were measured to be within a few percent of expectations for all filters and field positions. Tests for stray light and optical ghosts showed no issues, with the exception of the glow-stick feature mentioned in the previous section, which is thought to have no impact on MIRI imaging. 

The team produced and tested calibration files including darks and flats and verified that the data pipeline successfully processed the data to scientific quality. With this the relative photometric response of imaging was shown to be determined better than 5\% at 80\% encircled energy including the response between different subarrays and filters. The low impact of persistent images and other artefacts was demonstrated as well as the recovery from the effects of cosmic rays including the effective mitigation of image artefacts using the process of annealing - raising the temperature of the detectors to release trapped charge. 

Lastly, MIRI's sensitivity was measured to be two orders of magnitude better than \textit{Spitzer} at 5.6~\mum\, and 1 order of magnitude better at 25.5~\mum\, making it the most sensitive astronomical imager to date in this wavelength region.

The most up-to-date and accurate estimate of the on-orbit sensitivity for imaging is captured in the Exposure Time Calculator (\href{https://jwst.etc.stsci.edu}{\textit{JWST} ETC - Home page}), and the reader is invited to refer to the latest release (version 3.0 at the time of writing) for definitive values. The estimated sensitivity prior to launch can be found in \citet{glasse2015}, where we note that the commissioning measurements \citep[implemented in the ETC 2.0 and summarised in][]{rigby2022a} show MIRI imaging sensitivity to be better by $10~\pm14 \%$ than pre-launch estimates.  

In summary, the commissioning observations were nearly all executed according to plan, and of sufficient quality to kick-start the on-orbit calibration program, which would be used as a baseline for the \textit{JWST} ETC and Cycle 2 proposal preparations. Calibrations will keep improving in time, as more data is acquired and the subtleties of the instrument behaviour are studied in depth. 

\section{Point Spread Functions} \label{sec:psf}
To verify the optical quality of the MIRI imager, and quantify detector artefacts, a commissioning program (PID~1028) was dedicated to the characterisation of the imager PSF. 
The objectives were three-fold: (1) measure the in-flight PSF properties with a high dynamic range for all MIRI bands; (2) evaluate and check the variations of the PSF properties across the field of view of the imager; and (3) reconstruct a ``super-resolved'' PSF at a spatial resolution higher than the diffraction limit at 5.6~\mum . The latter objective was, because the PSF is undersampled at this wavelength, to check the optical quality at 5.6~$\mu$m and perform a fine characterisation of the cross artefact. The cross artefact was known pre-launch and referred to as ``the cruciform", arising from the diffraction of infrared photons in the detector substrate \citep{gaspar2021}. In the following, we briefly describe the data analysis, and a general assessment of the optical quality made during commissioning. A separate paper discusses the PSF modelling done for photometry and precise astrometric calibration \citep{Libralato2023}. 

\subsection{Data acquisition and reduction}

All PID~1028 exposures were obtained in FASTR1 readout mode on 2022 May 24. Observations 1 to 5 were dedicated to the fine sampling of the PSF at F560W in 5 positions in the field of view.  We observed the MIRI standard star 2MASS J17430448+6655015 (A5V type) with the 16-points microscanning dither pattern, which is a non-standard dither pattern built to sample the surface of a pixel to enable the best PSF reconstruction possible.
The shifts between each dither position need to be random fractions of a pixel to mitigate the effects of intra-pixel gain variations, and the pointing accuracy was typically 2 mas. In Appendix~\ref{appendix_dithers}, we show the dither patterns and a comparison between the requested positions and the observed positions (Figure~\ref{fig:microscan_dithers}). The total exposure time per position was about 40 min (10 groups $\times$ 5 integrations per exposure, making a total of 80 exposures per position). This was designed to permit construction of a very high signal-to-noise ratio PSF with a dynamic range sufficient to evaluate the intensity in the wings of the PSF out to a large radius. 
The F560W data were processed up to stage-2 \texttt{cal} FITS products, and high-resolution PSFs were reconstructed from the cal.fits files using the deconvolution method described in \citet{Guillard2010}.

For all other filters, the target star was placed in the centre of the field of view (Observation 6), and all the data were reprocessed up to stage 3 (\texttt{i2d.fits}) with the version 1.8.3.dev17+gf4081ef2 of the \textit{JSWT} pipeline, and version 1019.pmap of the CRDS\footnote{Calibration Reference Data System, https://jwst-crds.stsci.edu/} context. 

\subsection{PSF cosmetics and metrics}

\begin{figure}
\includegraphics[width=0.5\textwidth, center]{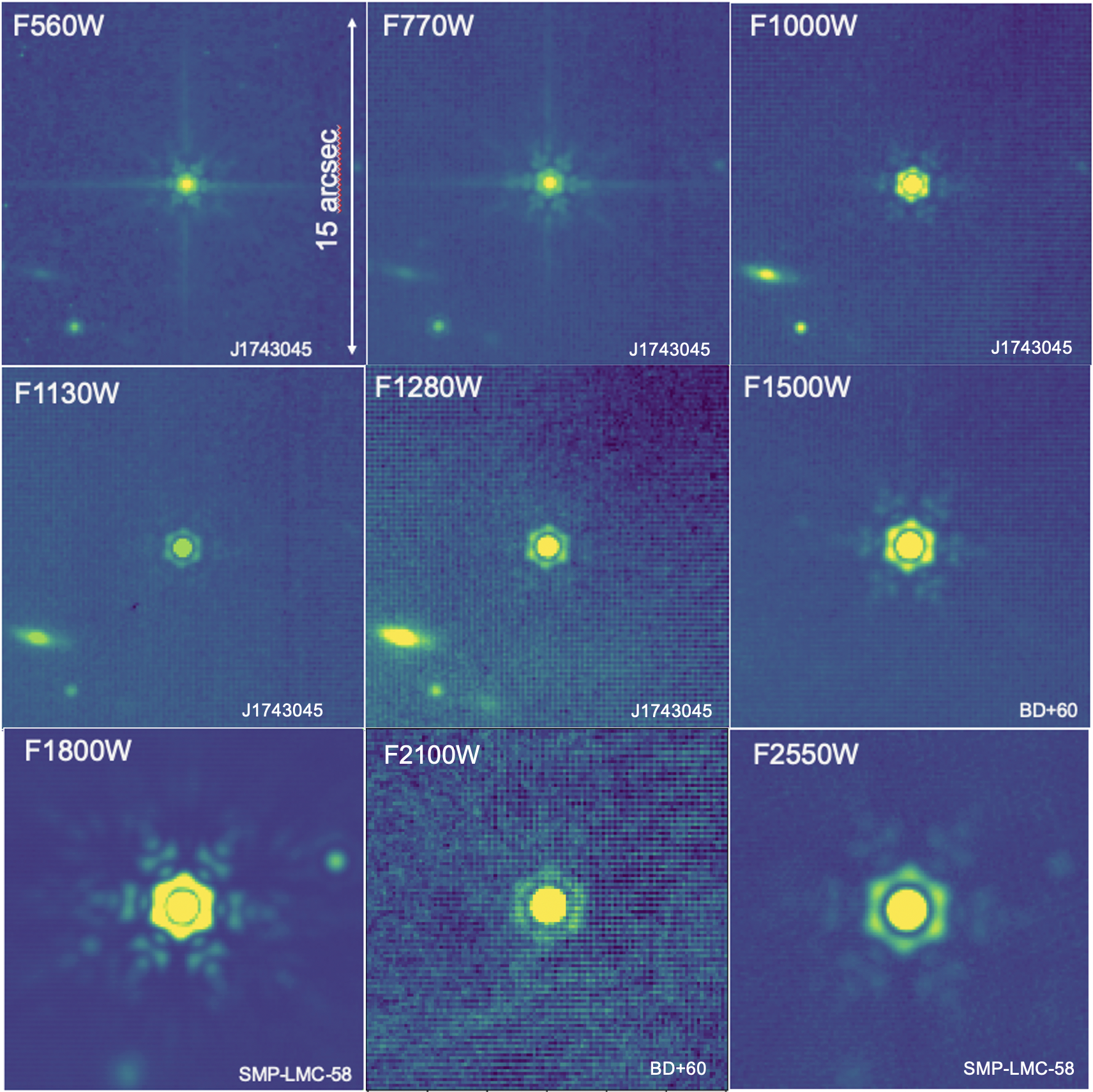}
\caption{PSF stamps for all MIRI filters from PID~1028. We show log-scale, flat-fielded, $15"\times 15"$ cropped images of the different sources used to compute PSF properties, indicated in the bottom right of each image: the 2 MIRI standard stars J1743045 and BD+60-1753, as well as the planetary nebula SMP-LMC-58, stronger at long wavelengths. Since this latter source has not been observed at 21 $\mu$m, we used BD+60-1753, which has a lower signal-to-noise. The cruciform artefact is clearly visible for the F560W and F770W filters. 
}
\label{fig:PSF_gallery}
\end{figure}

Cut-outs of PSF images taken during commissioning are shown in Figure~\ref{fig:PSF_gallery}. Those images were obtained from observations of two MIRI standard calibrators, J1743045 and BD+60-1753, as well as SMP-LMC-58 \citep[see][]{Jones2023}  at long wavelengths to optimise the signal-to-noise ratio. At wavelengths shorter than 10~\mum, the cruciform artefact is clearly visible, as it extends to radii larger than the PSF wings ($>100$ pixels, see section ~\ref{subsec:cruciform} for more details). 

\begin{table*}
\begin{center}
\begin{tabular}{|c|c|c|c|c|c|c|}
\hline
   & \multicolumn{2}{c}{Data (EE normalized to 5 arcsec)} &  \multicolumn{2}{c}{WebbPSF (EE normalized to 5 arcsec)} & \multicolumn{2}{c}{Data / WebbPSF} \\
\hline
   \multirow{2}*{Filter} & FWHM & EE within 1st &  FWHM & EE within 1st & FWHM & EE  \\
         & [arcsec] & dark Airy ring & [arcsec] & dark Airy ring & ratios & ratios \\
\hline
F560W & 0.207 & 55--63\% & 0.182 & 60.9\% & 1.14  & 0.9 -- 1.03 \\
F770W & 0.269 & 62\%  & 0.260 & 65.3\% & 1.03 & 0.95 \\
F1000W & 0.328 & 71\% & 0.321 & 70.9\% & 1.02 & 1 \\
F1130W & 0.375 & 73\% & 0.368 & 72.7\% & 1.02 & 1 \\
F1280W & 0.420 & 65\% & 0.412 & 72.8\% & 1.02 & 0.89 \\
F1500W & 0.488 & 77\% & 0.483 & 73.5\%  & 1.01 & 1.05 \\
F1800W & 0.591 & 74\% & 0.580 & 74.4\% & 1.02 & 0.99 \\ 
F2100W & 0.674 & 72\% & 0.665 & 75.1\% & 1.01 & 0.96 \\
F2550W & 0.803 & 68\% & 0.812 & 76.1\% & 0.98 & 0.89\\
\hline
\end{tabular}
\caption{MIRI imager PSF metrics, measured on commissioning data (2nd and 3rd columns), and WebbPSF models (4th and 5th columns). The last two columns tabulate FWHM and EEs (Encircled energies) ratios between data and models. FWHM are average values in x and y directions from fitting a bi-dimensional Airy function to the images.}
\label{table:PSF}
\end{center}
\end{table*}

\begin{figure}
\centering
\includegraphics[width=\linewidth]{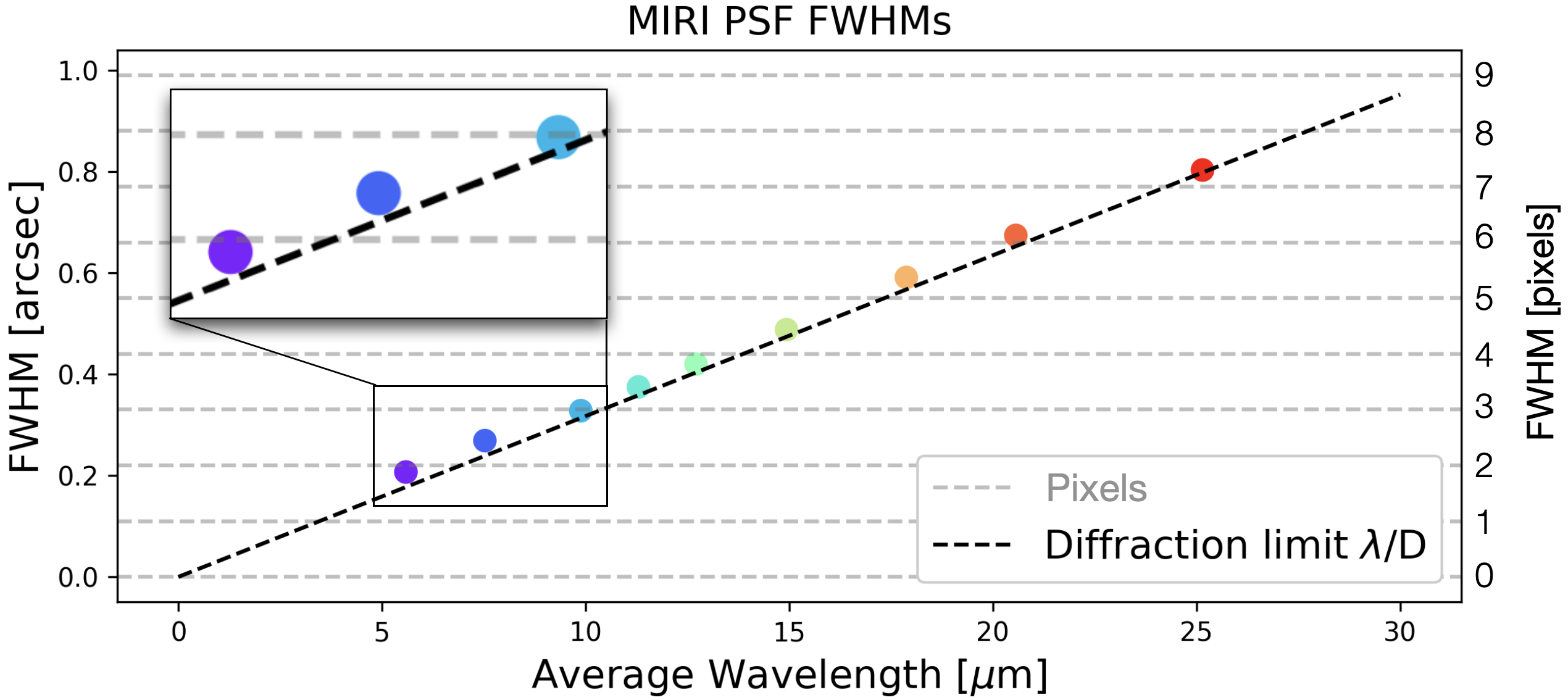}
\caption{FWHM in each filter, as measured in PSFs made with commissioning data. Each dot is the FWHM average in x and y directions from 2D Airy fits of the PSF images in each MIRI filter. The pixel size is 0.11", and light grey dashed lines materialise the pixels. The telescope diffraction limit is indicated by the black dashed line. The excess with respect to the diffraction limit at 5.6 and 7.7\mum\ is due to the broadening of the PSF induced by the diffraction and scattering in the detector substrate (see section ~\ref{subsec:cruciform}).}
\label{fig:MIRI_PSF_FWHM}
\end{figure}

Figure~\ref{fig:MIRI_PSF_FWHM} and Table~\ref{table:PSF} gather measurements of the PSF metrics and comparison with WebbPSF simulations, and show that the MIRI imaging performance is diffraction-limited at wavelengths longer than 10\mum\ (from F1000W and onward). The excess (respectively 14\% and 3\%) with respect to the diffraction limit at respectively 5.6 (F560W) and 7.7\mum\ (F770W) is due to the broadening of the PSF induced by the diffraction and scattering in the detector substrate \citep[see section ~\ref{subsec:cruciform} and][]{gaspar2021}.

\begin{figure*}
\includegraphics[width=\textwidth, center]{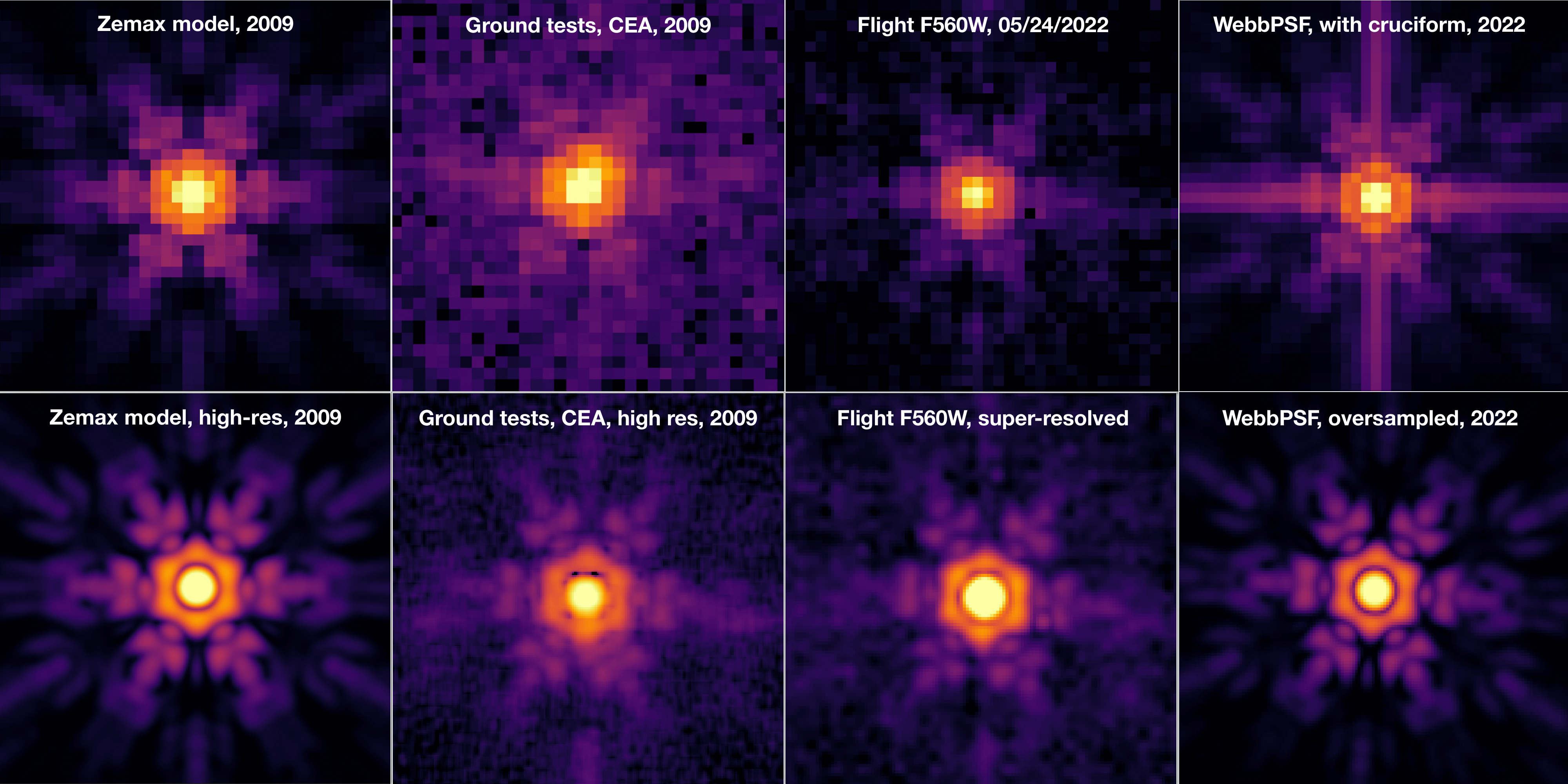}
\caption{Stamps of the F560W PSF core ($4"\times 4"$ boxes), showing the Zemax simulations (1st column), the ground-based PSF tests done at CEA with the imager ETM in 2008-2009 (2nd column),  the flight PSF commissioning data (3rd column), and the WebbPSF simulations computed during commissioning (last column). On the top right image, an approximate representation (exponential profile) of the cruciform was added on top of the WebbPSF model. All PSF images are shown on a log-scale, with same relative stretch, at the center of the imager field of view. The top row shows native resolution data with 0.11 arcsec pixel$^{-1}$ scale. The bottom row shows super-resolved PSFs. The Zemax and WebbPSF high-resolution simulations, as well as the reconstructed image of the super-resolved flight and ground-based PSFs, are shown with an oversampling factor of 4. The base of the cruciform (cross aligned in horizontal X and vertical Y directions) is visible both on ground-based and flight images.}
\label{fig:PSF_model_comp}
\end{figure*}

The under-sampling of the PSF at 5.6\mum\ makes it necessary to use a deconvolution technique, combined with fine-raster dithering, to reconstruct super-resolved PSFs, and we follow the methodology described in \citet{Guillard2010} to do so with commissioning data, lab data from ground-based test campaigns, and optical models (Zemax and WebbPSF simulations). Figure~\ref{fig:PSF_model_comp} shows a ``historical" gallery of F560W PSFs, which compares those different datasets and highlights the remarkable agreement between optical models and super-resolved PSFs reconstructed from the lab and flight data. The deconvolution of the sub-pixel, $4\times 4$-points dither data (microscanning pattern) performed during commissioning led to a gain in spatial resolution of a factor of $\approx 3$ (Figure~\ref{fig:PSF_model_comp}, 3rd column, bottom row). This allowed us to check the optical performance at 5.6\mum\ and perform a fine comparison between data and models.  

\begin{figure*}
\centering
\includegraphics[width=0.9\linewidth]{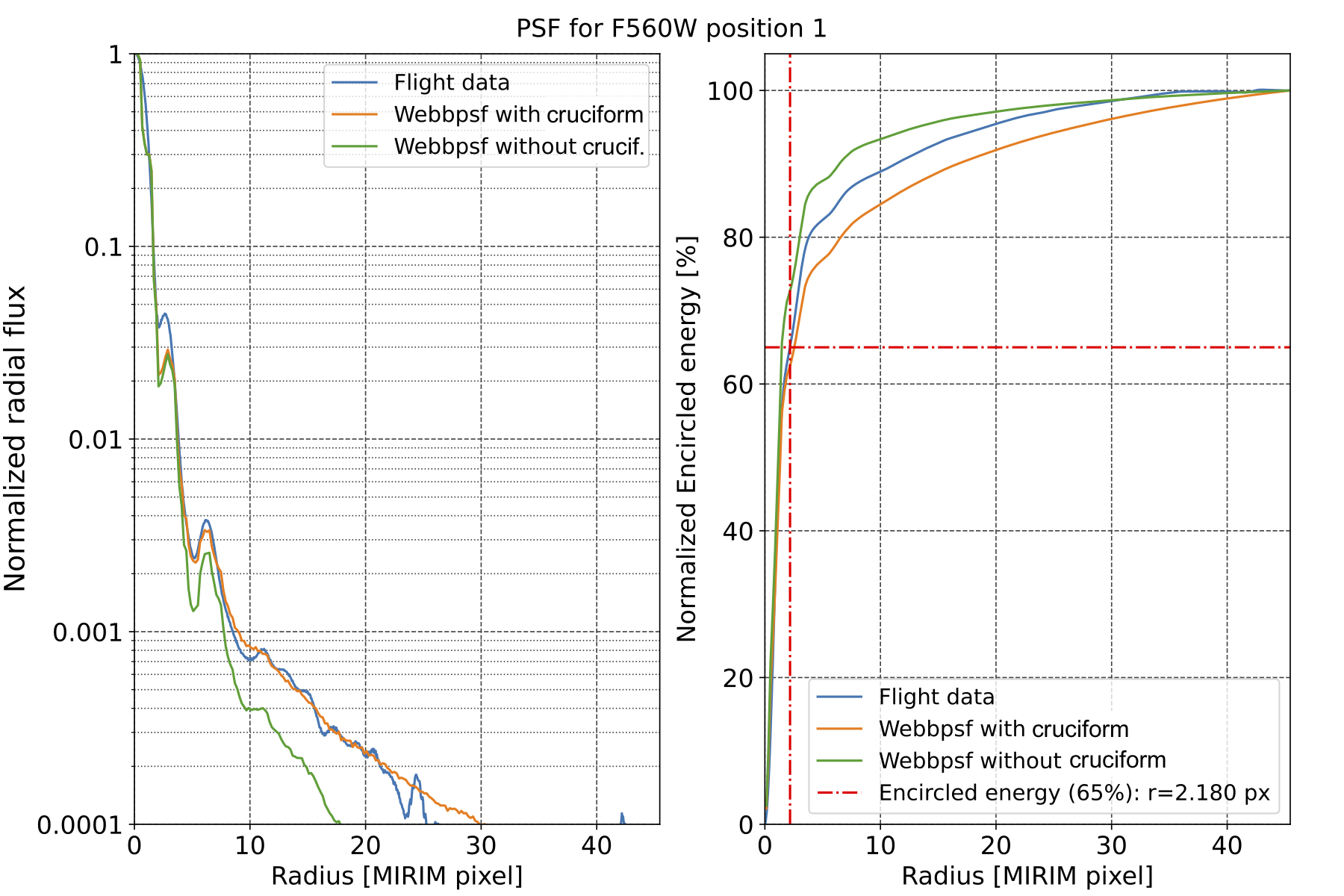}
\includegraphics[width=0.9\linewidth]{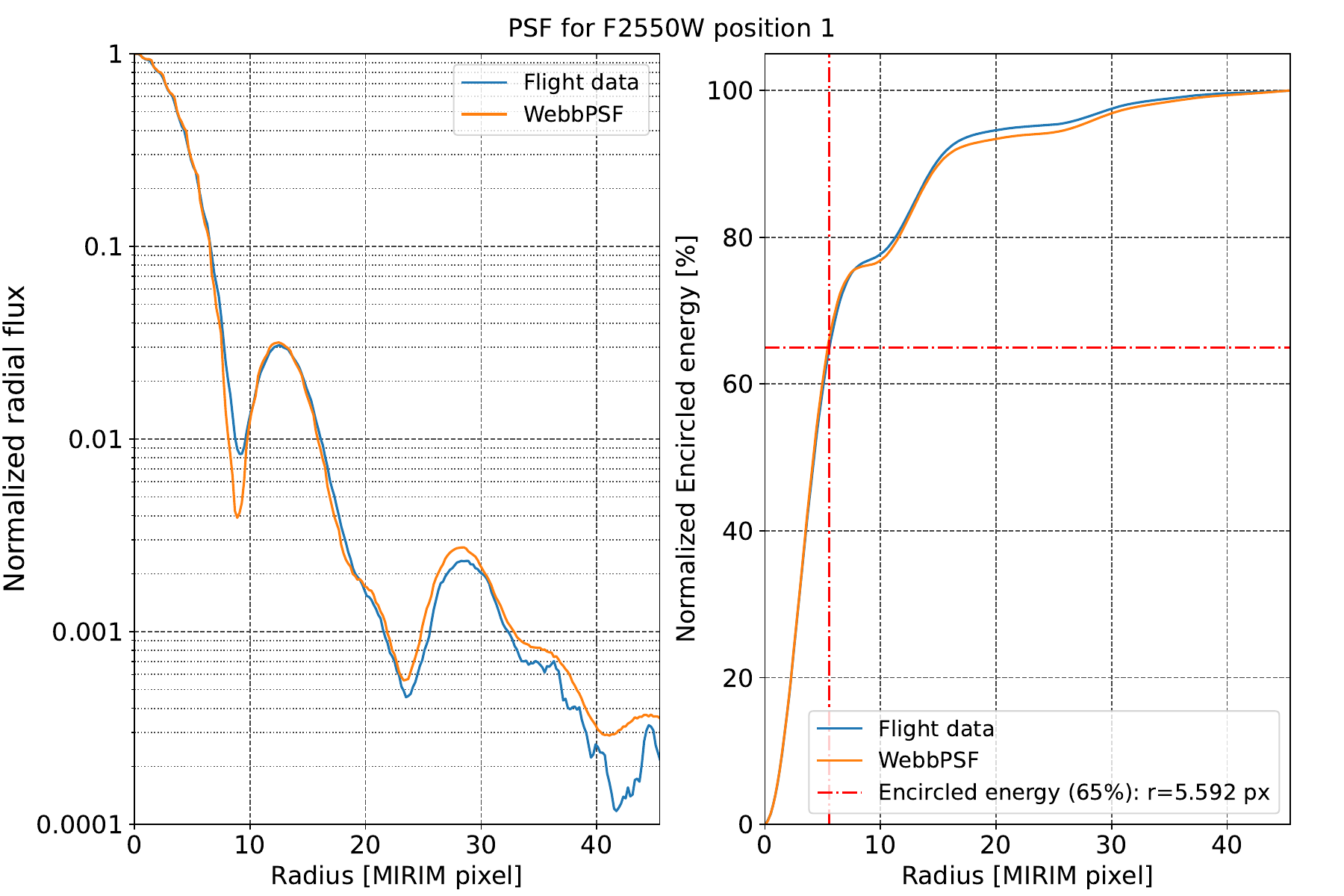}
\caption{PSF radial profiles at 5.6 and 25.5~$\mu$m: comparison between in-flight PSFs and simulated WebbPSFs. The left panels show radial profiles, normalised to the flux of the central pixel. The blue, green and orange curves show respectively: the flight data (see Figure~\ref{fig:PSF_gallery} for images), the WebbPSFs without the cruciform artefact, and the WebbPSFs with the cruciform artefact (see text for details). For the F560W filter, the profile at large radii (outside the secondary Airy ring at $r>6$~pixels) is dominated by the cruciform. At 25.5~\mum, there is no cruciform. The right panels show encircled energy (cumulative radial flux) profiles, normalized at a radius of 5~arcsecs ($\approx 45$ MIRIM pixels). The red dashed lines highlight the position of the radius containing 65\% of the total encircled energy within 5~arcsecs.}
\label{fig:PSF_F560W_F2500W}
\end{figure*}

In Figure ~\ref{fig:PSF_F560W_F2500W}, we show PSF radial profiles for the F560W and F2550W filters in the centre of the field of view of the MIRI imager, and we compare the data with WebbPSF simulations. The top panel displays the super-resolved F560W, which shows in particular that the cruciform artefact dominates the flux profile at radii longer than the secondary Airy ring (see  see Sect.~\ref{subsec:cruciform} for details). The cumulative radial flux profiles (Encircled Energy, EE, normalised at a radius of 5~arcsec or 45 pixels) are shown in the right panel of Figure ~\ref{fig:PSF_F560W_F2500W}. Note that WebbPSF models do not include all the detector effects that contribute to the PSF properties, like the cruciform for instance. At the time of commissioning, an approximate representation (exponential profile) of the cruciform spatial distribution was added to WebbPSF models (shown as orange profiles on Figure ~\ref{fig:PSF_F560W_F2500W}). The F560W flight data show that the actual PSF is more centrally concentrated than the WebbPSF simulations including the cruciform, because the exponential profile of the cross-artefact overestimates the cruciform flux in the central parts of the PSF. We also provide additional F560W PSF profiles for positions in the corners of the FoV in Appendix~\ref{appendix_PSF_profiles}. These profiles highlight an overall similar behaviour across the imager FoV, in agreement with the optical WebbPSF modelling, with the caveat that 
WebbPSFs in the two top corners of the FoV are extrapolated from other positions in the FoV. This introduces an artificial skewness in the modelled core profiles, which is not present in the fight data \citep[see Appendix~\ref{appendix_PSF_profiles} and][for more details]{Libralato2023}.

\subsection{The cross-shaped artefact, a.k.a the ``cruciform''}
\label{subsec:cruciform}

\begin{figure}
\includegraphics[width=0.5\textwidth, center]{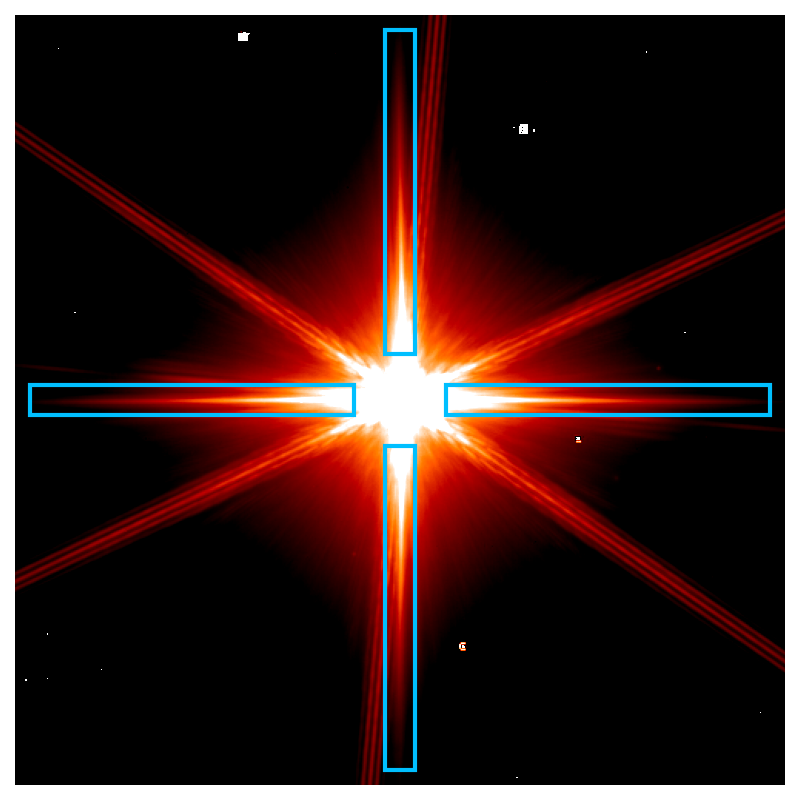}
\caption{Image of $\beta$ Doradus from the commissioning program PID 1023, where the blue rectangles highlight the cross-shaped artefact, a.k.a "the cruciform". This artefact is easily identified in the flight images for wavelengths shorter than 12\mum.}
\label{fig:betaDor_cross}
\end{figure}

Light that passes through a backside-illuminated detector wafer to the grid of frontside contacts that delineate the individual pixels and is reflected back into the detector by those contacts, and the gaps between contacts, will cause diffraction into wider angles than the incident ones. Some of the diffracted signal can be at an angle that causes total internal reflection within the detector, so the photons can be absorbed and detected far from their point of incidence onto the detector array. The resulting imaging artefact extends in directions dictated by the contact geometry, i.e. for the usual case of rectilinear contacts, the artefacts take the form of long lines emerging from the core image in orthogonal directions. This artefact was first identified and measured during ground-based test campaigns with an ETM (Engineering Test Model) detector (see Fig.~\ref{fig:PSF_model_comp}). \citet{gaspar2021} presented a physical and detailed model of the cruciform artefact. This behaviour is universal, but for most detector types is not observed because the absorption efficiency is sufficiently high to remove nearly all the incident photons before they reach the contacts. However, for Si:As impurity band conduction (IBC) detectors used between 5 and 10 $\mu$m, the reflected/diffracted signal is significant and yields a long-armed cross artefact i.e., the ``cruciform'', centred on the core image. This behaviour is illustrated in Figure~\ref{fig:betaDor_cross}, which shows that the cruciform artefact arising from observations of a very bright star, $\beta$ Dor, can extend up to the full dimensions of the imager field of view. The shape of the cruciform has also a spatial dependence on the FoV\footnote{The position of the cross-artefact shifts from the PSF centre and bends slightly when one moves around in the FoV. This is  thought to be caused by the non-normal incidence of the beam on the pixel grating lattices, which will be modelled in a future paper.}, which is currently not implemented in WebbPSF. This creates residuals when doing PSF-fitting with WebbPSF models \citep{Libralato2023}. In the case of MIRI imaging, this effect is not treated by the \emph{JWST} pipeline, and it is left to the user to model and remove the artefact if doing so is necessary for their particular science case. 

\section{Flux Calibration} \label{sec:calibration}
The imager flux calibration is provided as the reference file used by the pipeline in the PHOTOM step, that converts from units of DN~s$^{-1}$ to physical units of MJy~sr$^{-1}$. DN stands for Data Number and is the raw signal unit of the MIRI detectors. During commissioning, the flux calibration was based on two spectrophotometric calibrators (and one additional calibrator in subarray mode only) observed in two different epochs separated by about 12 days (PID~1027; see Table~\ref{tab:spectrophot_stars}). 
We also used the CALSPEC\footnote{https://www.stsci.edu/hst/instrumentation/reference-data-for-calibration-and-tools/astronomical-catalogs/calspec} stellar models that, in combination with the ground-based photon-conversion-efficiency, provided a theoretical measurement to compare to the in-flight aperture photometry (readers are referred to \citet{Gordon2022} for additional details). The data of these spectro-photometric calibrators were processed with the 1.5.3a0 version of the \textit{JWST} pipeline, version 11.16.6 of CRDS and context jwst\_0865.pmap. Some steps utilized dedicated commissioning reference files that at the time were not in CRDS. The calibration factors were derived as follows:

\begin{enumerate}
\item Process data down to the dither-combined images (i.e., pipeline stage 3 data products \texttt{i2d} FITS files). For this data reduction, the PHOTOM step was skipped, to preserve units of DN~s$^{-1}$.
\item Using source masking, generate and subtract a background model.
\item Identify the spectrophotometric calibrator.
\item Multiply by the average pixel area in steradian. 
\item Perform aperture photometry at different radius to produce the curve of growth, determine the 60\%, 70\% and 80\% of the encircled energy.
\item Subtract the local sky estimated as the median value of the pixels with an annulus centered on the calibrator.
\item Because the \textit{JWST} calibration pipeline assumes infinite aperture, perform the aperture correction using WebbPSF (we assume infinite aperture).
\item Define calibration factors as the ratio of the CALSPEC model-based photometric measurement per filter, and the measured aperture-corrected photometry.
\item Repeat process for each filter, and average results from different epochs.
\end{enumerate}

\begin{figure}
\includegraphics[width=0.5\textwidth, center]{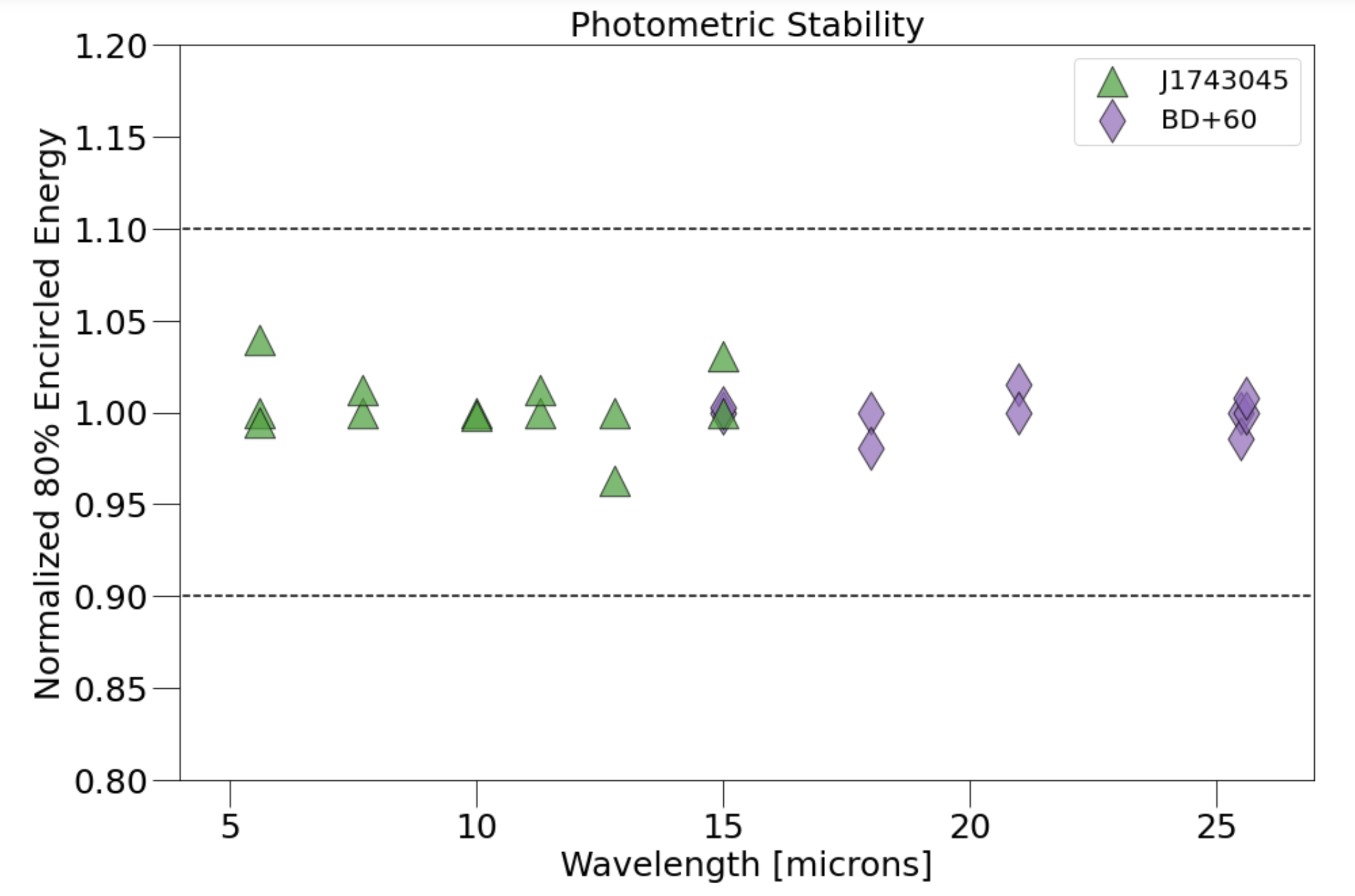}
\caption{MIRI imager photometric stability commissioning results, well within the 10\% pre-launch requirement (dashed line).}
\label{fig:phot_stability}
\end{figure}

The MIRI imager photometric stability was evaluated with a single repeat of the BD+60-1753 and J1743045 measurements in FULL array FASTR1 mode (as defined in Commissioning Activity Request MIRI-011). Data were taken a minimum of 9 and maximum of 12 days apart. As shown is Figure~\ref{fig:phot_stability}, all photometric measurements (80\% encircled energy) are consistent within 5\%, well in line with the requirements.

  
\begin{table*}
\centering
\resizebox{18cm}{!}{\begin{tabular}{|c|c|c|c|c|c|c|}
\hline
Calibrator & RA [h:m:s] & DEC [$^{\circ}$:$'$:$"$] & Spectral Type  & Filters & Subarray/Readout & Execution Date\\
\hline
2MASS J17430448+6655015  & 17:43:4.4886 & +66:55:1.62 & A5V & F560W, F770W, F1000W, F1130W, F1280W, F1500W & FULL/FASTR1 & 2022-05-28 19:56:0 \\
\hline
2MASS J17430448+6655015 & 17:43:4.4886 & +66:55:1.62   & A5V & F1500W & FULL/SLOWR1 & 2022-05-25 10:43:13 \\
\hline
BD +60 1753 & 17:24:52.2856 & +60:25:50.81 & A1V & F1500W, F1800W, F2100W, F2550W, F2550WR & FULL/FASTR1 & 2022-05-25 11:53:31\\
\hline
BD +60 1753 & 17:24:52.2856 & +60:25:50.81 & A1V & F560W, F770W, F1000W, F1500W & SUB256/FASTR1 & 2022-05-25 12:19:50\\
\hline
HD163466 & 17:52:25.3757 & +60:23:46.94 & A6V & F1000W, F1130W, F1280W, F1500W & SUB64/FASTR1 & 2022-05-25 12:57:45\\
\hline
2MASS J17430448+6655015  & 17:43:4.4886 & +66:55:1.62 & A5V & F560W, F770W, F1000W, F1130W, F1280W, F1500W & FULL/FASTR1 & 2022-06-07 12:52:46 \\
\hline
2MASS J17430448+6655015 & 17:43:4.4886 & +66:55:1.62   & A5V & F1500W & FULL/SLOWR1 & 2022-06-07 12:52:46 \\
\hline
BD +60 1753 & 17:24:52.2856 & +60:25:50.81 & A1V & F1500W, F1800W, F2100W, F2550W, F2550WR & FULL/FASTR1 & 2022-06-07 15:24:23\\
\hline
BD +60 1753 & 17:24:52.2856 & +60:25:50.81 & A1V & F1500W, F1800W, F2100W, F2550W, F2550WR & FULL/FASTR1 & 2022-06-13 22:55:21\\
\hline
\end{tabular}}
\caption{Spectrophotometric calibrators from commissioning program ``MIRI Imager Photometric Zero Points and Stability'', PID~1027.  All dates are UTC, and celestial coordinates are ICRS.
\label{tab:spectrophot_stars}}

\end{table*}

\section{Backgrounds} \label{sec:bkg}

The sensitivity of MIRI imaging is background limited at all bands. Hence, a proper background correction is needed for improving science performance of the imager. The background emission is dominated by astronomical origins (zodiacal light and Milky Way interstellar medium) at $\lambda \lesssim 12.5\,\mu$m and by the observatory stray light and thermal self-emission at longer wavelengths \citep[refer to][for a detailed analysis of \textit{JWST} background emission spectrum]{rigby2022b}. Here, we present the MIRI imager thermal background measurements that were performed during the commissioning period. We also provide recommendations for observations planning and data reduction to mitigate the effects of the background, particularly at longer wavelengths.  Finally, we describe our monitoring program and show how the background varies over time and as a function of telescope attitude in the two reddest MIRI imager filters.

\subsection{Observations and Methodology}
The commissioning program that was dedicated to assessing the MIRI thermal background level and variation is PID~1052. The program observed ``blank'' fields that were selected to not have any obvious sources in the WISE 3 bands combined images, in two different telescope attitude configurations, ``Hot'' and ``Cold'', to capture the extreme cases in terms of the amount of solar energy input onto the illuminated side of the sunshield. The Hot attitude is when the angle between the Sun-observatory line (sunline) and the telescope pointing direction is $90\pm5^{\circ}$, i.e., when the largest area of the sunshield is exposed to sunlight. The ``Cold'' attitude is when the angle is $130\pm 5^{\circ}$, when the telescope pointing is most away from the Sun. The goal was to assess the variation of the thermal background as seen by MIRI long-wavelength bands in these two bounding conditions.

The observations used MIRI filter bands F770W, F1280W, F1500W, F1800W, F2100W, and F2550W. These filters, covering all but the shortest wavelength band of MIRI, probe the regimes where different observatory components (tower, sunshield,  Optical Telescope Element) are expected to dominate. For example the sunshield emission, that represents one of the main contributors to the thermal background seen by MIRI, begins to contribute at 12\,$\mu$m. The F770W band was added to measure a wavelength region with no significant contributions from thermal emission. The F1800W filter observations were done in a $2 \times 2$ mosaic to evaluate the presence of spatial structure in the thermal emission at longer wavelengths. To measure the background level, the median of pixel values in a grid of 400 boxes ($26 \times 26$ pixel$^2$ each) is measured on each level 2 data (\texttt{cal} FITS files). The median and standard deviation of boxes are taken as background and its scatter, respectively.

\begin{table}
\centering
\resizebox{\columnwidth}{!}{\begin{tabular}{ |c|c|c| }
\hline
Filter & Average background [MJy/sr] & Variation between Cold and Hot attitudes (\%)  \\
\hline
F770W & 6.7   & 7.5\\
F1280W & 36.0  & 0.9\\
F1500W & 60.8  & 0.3\\
F1800W & 91.3  & 0.5\\
F2100W & 238.7 & 0.7\\
F2550W & 853.3 & 0.9
\\
\hline
\end{tabular}}
\caption{MIRI imager background measurements during commissioning}
\label{table:bkg}
\end{table}

\subsection{Background analysis results} 
\paragraph{Hot and Cold attitudes and temporal variation} 
The thermal infrared backgrounds in the F1280W, F1500W, F1800W, F2100W, and F2550W filters are shown in Table~\ref{table:bkg}. At long wavelengths ($>$12$\mu$m), where the thermal emission dominates, the background measurements in the two Hot and Cold bounding conditions are consistent within 1\%. This is in agreement with pre-launch predictions of the ETC.

A more significant background variation is observed in the F770W filter ($\sim 7\%$, Table~\ref{table:bkg}) during commissioning, with no correlation with the telescope orientation. This variation is also larger than that expected by the mirror temperature variation during the thermal stability test (moving from Hot to Cold attitude) by $\sim 0.2\,$K. In fact, this large variation in F770W is likely due to an astronomical origin. The F770W filter traces the emission from Polycyclic Aromatic Hydrocarbon (PAH) dust grains, which emit strongly at 7.7$\mu$m \citep{tielens2008}. The WISE 12$\mu$m images (which trace two strong PAH bands at 7.7 and 11.3\,$\mu$m) indicate that the pointing of the hot attitude observations was very close to a region with enhanced PAH emission. Therefore, the origin of the F770W background difference between the two attitudes is astrophysical. None of the other observed MIRI filters trace strong PAH bands to show the same effect, the only other filter that directly traces a PAH band is F1130W, which was not included in the Hot-Cold attitude test.

\begin{figure}
\includegraphics[width=.5\textwidth]{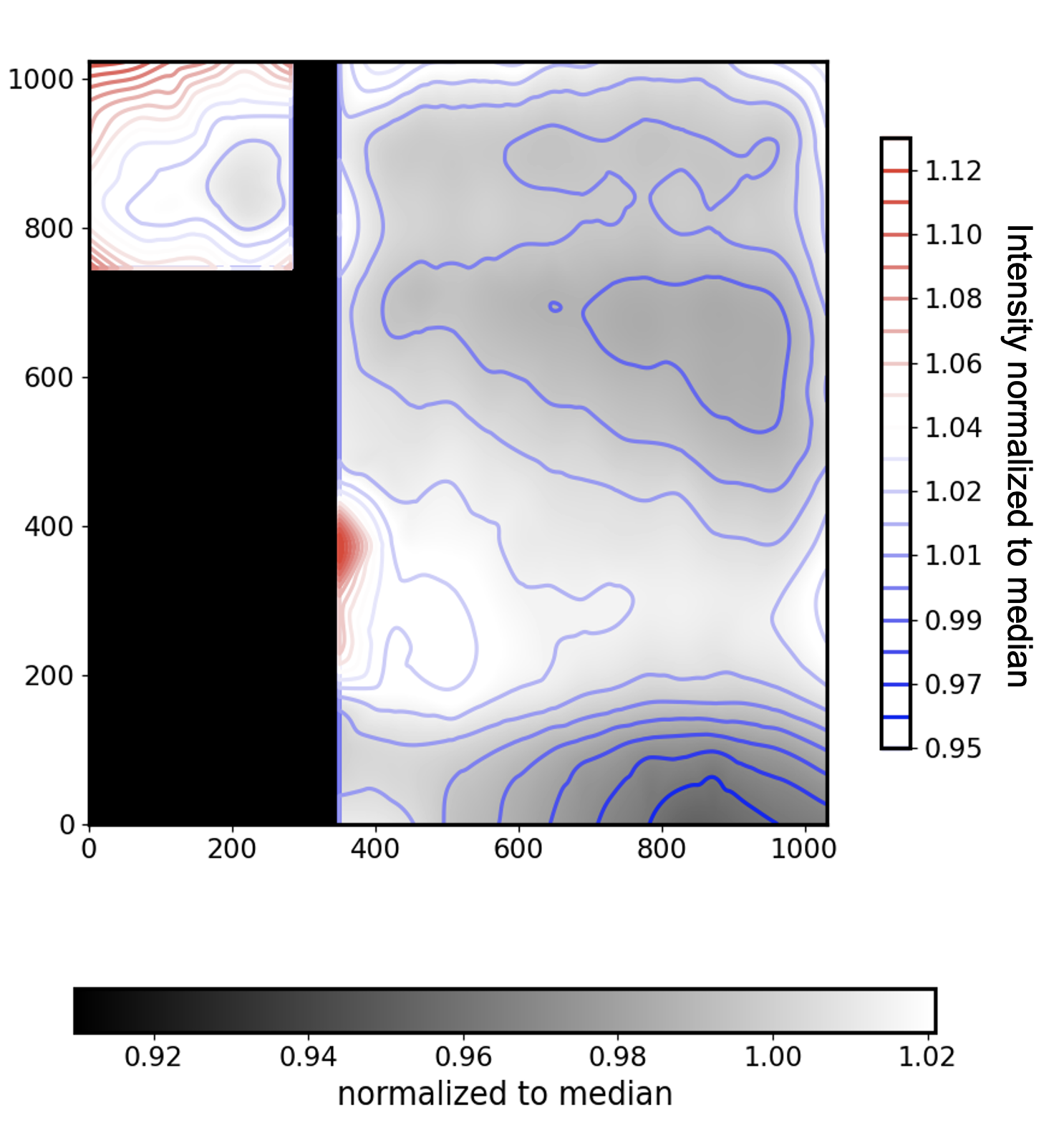}
\caption{MIRI F2100W background level 3 image from PID~1052 program. Sources are identified using multiple dithers and subtracted from the image. The image is normalized to the median value of the image. Colour contours show the spatial structure. Similar structure is observed in all MIRI bands.}
\label{fig:miri-f2100w-bkg}
\end{figure}

\paragraph{Background spatial structure variation} Figure~\ref{fig:miri-f2100w-bkg} shows the F2100W background image, where individual sources are subtracted using multiple dithers of the same pointing. A spatial structure that varies up to $\sim 10\%$ across the field is observed. This structure is persistent among all MIRI filters. To better understand the origin of this structure, we construct a master background by identifying bright sources (mostly point sources in these observations) through subtracting consecutive dithered images in each filter\footnote{MIRI dithers are significantly larger than the filters' FWHM, and hence, subtracting dithered images will cleanly subtract point sources.}, and replacing their value with interpolated sky value from other dithers. By normalizing the master backgrounds to their median value, and subtracting the Cold and Hot normalized backgrounds from each other in each filter, the residual image has a smooth structure with a symmetrical distribution centered at 0 and standard deviations of $<1\%$.

As the dominant observed spatial structure is constant with time and consistent among all MIRI filters, it is suspected to be due to imperfect flat field correction. Current flat field images (CRDS 11.16.15) reduce the variation to $< 5\%$ across the field. Background spatial variation needs to be revisited in the future with better flat field corrections as the flat field is currently the dominant source of variation.
In Section~\ref{sec:bkg-strategies}, we discuss strategies to correct this spatial structure.

\section{Distortion} \label{sec:distortion}

\begin{figure}
\includegraphics[width=240pt]{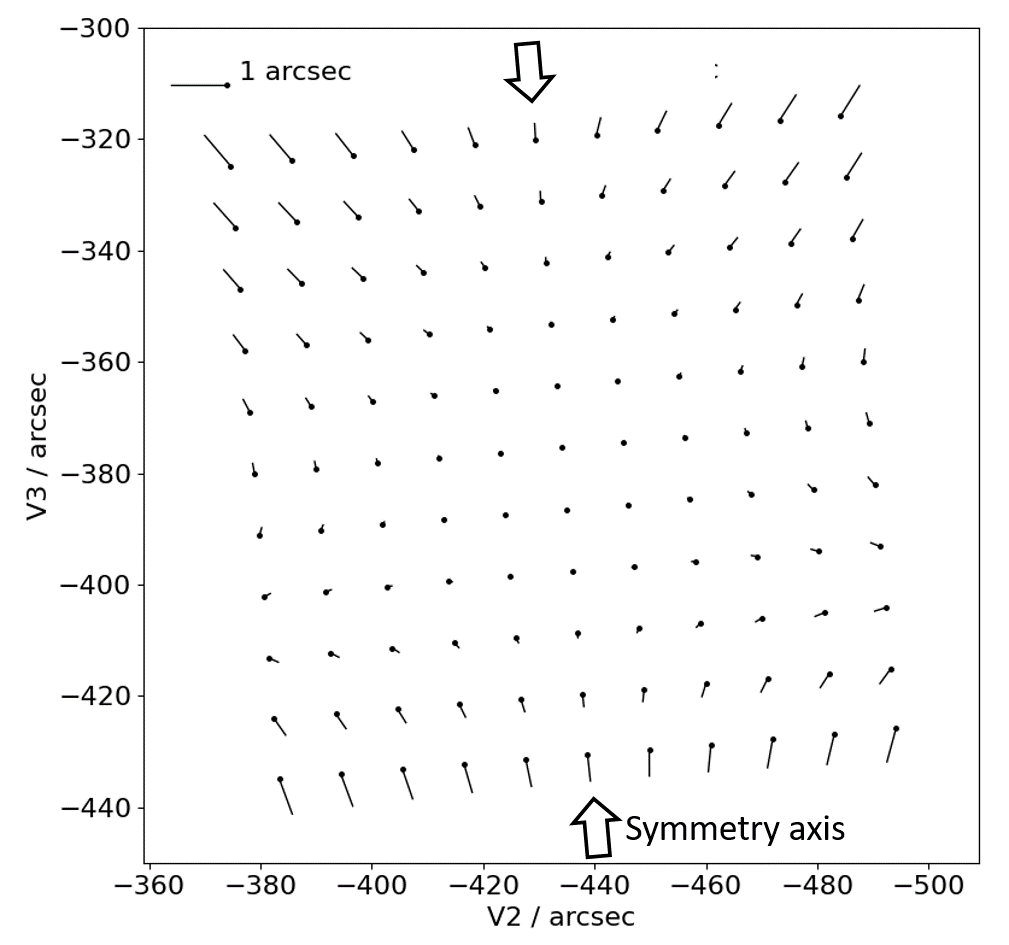}
\caption{Pre-flight predicted distortion map based on the MIRI optical design.  The points are projected from the an equispaced grid at the detector, with corners at pixels (10, 10) and (1013, 1013).  The vectors then plot the offset from the on sky positions obtained using the affine components of the distortion transforms, to the positions obtained using the full transforms.  
\label{fig:zemax_distortion}}
\end{figure}

Distortion across the imager field of view must be well characterised in order to support accurate astrometry and the photometric correction of non-uniformity in the on-sky projected area of detector pixels. The MIRI imager distortion map is shown in Figure~\ref{fig:zemax_distortion}, where image displacements of up to 1 arcsecond are seen at the corners of the field relative to the centre.  The observed `pin-cushion' like pattern is primarily a result of MIRI's off-axis all-reflective optical design, where the symmetry axis marked in the figure results from a symmetry plane in the optics which bisects the detector along a line parallel and close to the central column of the detector.  A smaller additional component arises from the observatory optics, with a displacement amplitude which varies with distance from the telescope optical axis.  

The distortion is geometrically well behaved, allowing us to represent the vector displacement at any point in the field using two sets of fourth order polynomial transforms, $\overrightarrow{X_{out}} = \overrightarrow{Y_{in}} A \overrightarrow{X_{in}}$ and $\overrightarrow{Y_{out}} = \overrightarrow{Y_{in}} B \overrightarrow{X_{in}}$.  $\overrightarrow{X}$ and $\overrightarrow{Y}$ are vectors of the form $\overrightarrow{X} = \overrightarrow{x^{i}}$, where $0 \le i < 5$ and $x$ is the focal plane coordinate.  $A$ and $B$ are 5x5 upper-diagonal matrices, where terms up to $x^{4}$ have been found to be sufficient to achieve a positional accuracy better than 5 milliarcseconds rms when mapping between pixel (row, column) and sky (V2, V3) focal plane coordinates.  In general, the $A$ and $B$ matrices are derived using a Singular Value Decomposition (SVD) method to find the optimum set of transforms which fit conjugate pairs of coordinates which have been measured accurately in the `sky' and `detector' focal planes.  For ground testing these coordinates were obtained from the Zemax optical model, with Figure~\ref{fig:zemax_distortion} showing the displacement of the projected regular grid of points compared to a purely affine transform, comprising offset, scaling and rotation only.

During commissioning and as one of its first on-sky observations (PID~1024), the MIRI imager was used to map a region of the \emph{JWST} Astrometric Calibration Field in the Large Magellanic Cloud. This field contains $>$200,000 isolated stars with $K$-band magnitudes ranging from $\sim$8 to $\sim$24 \citep{anderson2008} and positional accuracy of $\sim$1~mas \citep[see][and \href{https://jwst-docs.stsci.edu/jwst-data-calibration-considerations/jwst-data-absolute-astrometric-calibration}{\emph{JWST}-Data-Absolute-Astrometric-Calibration}]{sahlmann2017} and more recently, benefits from GAIA EDR3 astrometry \citep{gaia2021}. 

We used the positions, proper motions, extrapolated $K$-band fluxes, and expected observatory position angle to model and optimise our observations using the MIRI simulator, MIRISim \citep[][see Figure~\ref{fig:specsim}]{klaassen2021}.

\begin{figure}
\includegraphics[width=240pt]{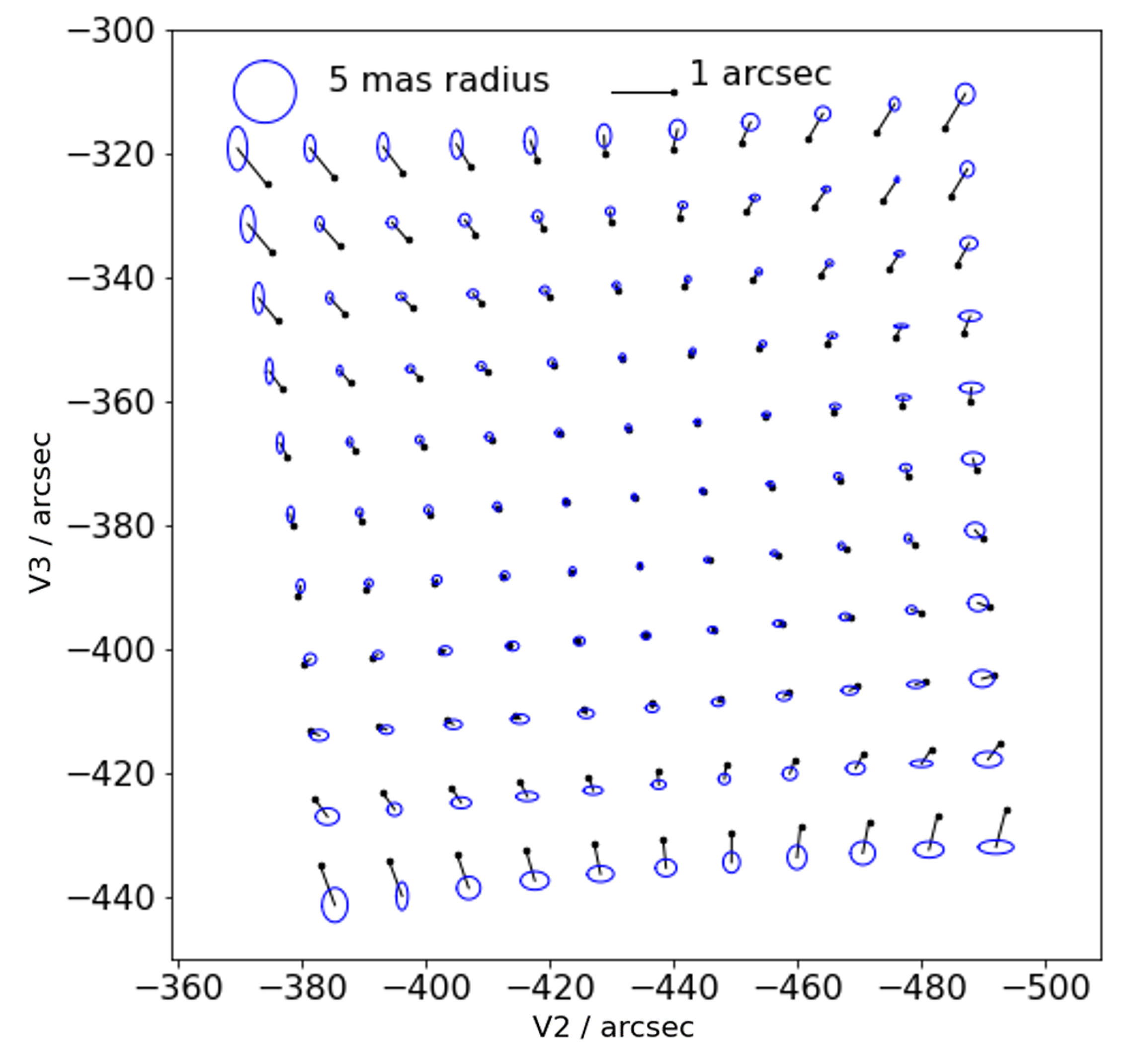}
\caption{Flight measured distortion map based on the MIRI F770W observations of the LMC astrometric field.  The 1 sigma root mean square position error is shown as field dependent ellipses.
\label{fig:dist_vector_field_flight}}
\end{figure}

Figure~\ref{fig:dist_vector_field_flight} then shows an example of the distortion vector field measured using two observations with the F770W filter, where the polynomial transform from detector to Gaia coordinates was calculated using the SVD method described above.  The agreement of the flight and optical model (Figure ~\ref{fig:zemax_distortion}) is seen to be excellent.

Approximately 2500 of the Gaia catalogued stars were identified with accurate centroids in MIRI's shortest wavelength filters, sufficient to provide a 1 sigma error well below the 5 mas target, as shown in Figure \ref{fig:dist_vector_field_flight}. At longer wavelengths the combination of increasing background limited shot noise and the Rayleigh-Jeans spectral shape of the targets causes this number to fall to just 20 or so identifications for the F2100W filter. This lack of long wavelength identifications is mitigated by the all-reflective optical design of the imager, such that the high order (i.e non-affine) elements of the distortion transforms in all bands is well described by a single set of transforms. Individual filters only differ by adding a boresight offset, as a result of the small variations (of the order of 1\arcsec) in the wedge angle of the refractive elements.

\begin{figure}
\includegraphics[width=240pt]{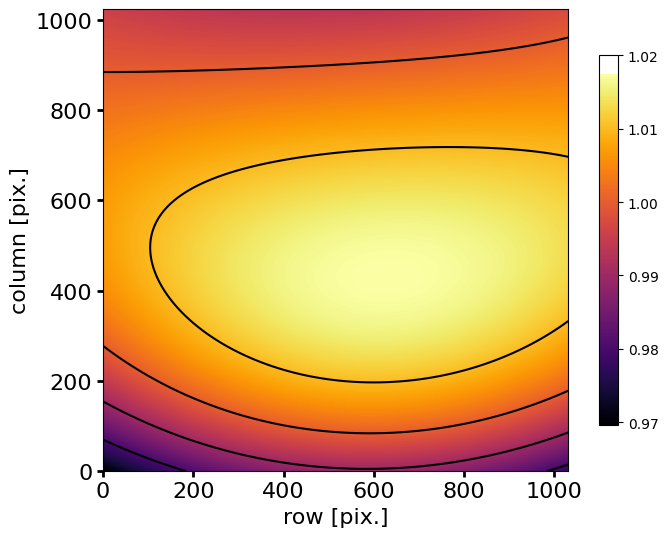}
\caption{The projected area on the sky for the MIRI imager, normalised to a pixel area of 0.0121 square arcseconds.
\label{fig:dist_area}}
\end{figure}

\begin{figure*}
\includegraphics[width=\linewidth]{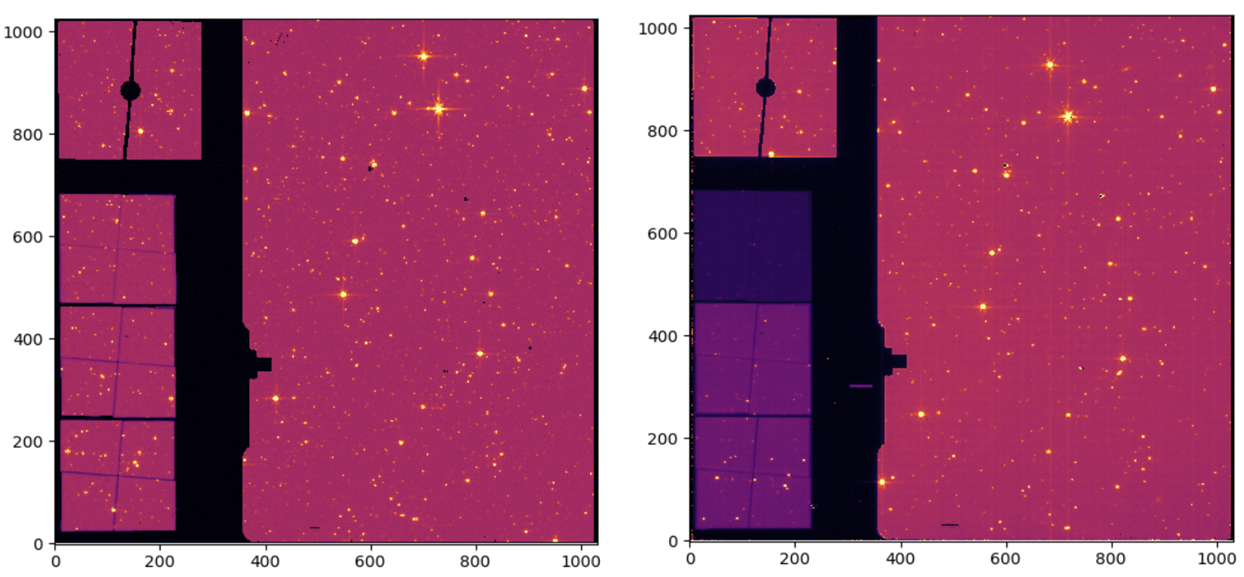}
\caption{– Left:  The MIRISim model prediction for the commissioning observation of the LMC astrometric field (using the F560W filter).   Right:  The commissioning observation (PID~1024).  The agreement between the images was an early encouraging indication that MIRI's performance was close to pre-flight predictions.  We note that the transmission of the coronagraphic phase masks at the left hand side of the field was set to unity in the MIRISim model.
\label{fig:specsim}}
\end{figure*}

The simulated and on-orbit measured observations are shown in Figure~\ref{fig:specsim} where the agreement in dynamic range and source count reflects the close match between MIRI's on-orbit performance and the predictions of pre-flight models.  This good correlation helped in the automated identification and cross-matching of stars in the images with their catalogue counterparts.

Non-parallelism between the faces of refractive elements in the imager optical train (the band-pass filters) cause the distortion map to be shifted on the detector, by tens up to approximately a hundred milli-arcseconds. The \emph{JWST} pipeline is therefore provided with an offset vector to account for the shift and support the accurate  co-alignment of images taken in multiple wavebands.

This impact of finite wedge angles in refractive elements is more severe for the operation of the 4QPM coronagraphs, where a  bright target must be positioned behind the coronagraphic null with an accuracy of 5 mas, after both the boresight offset due to the phase mask itself and that of the filter being used for target acquisition (TA) have been taken into account. In practice, the problem was solved on a case by case basis, with the effective pointing position measured for each TA filter, acquisition strategy and phase mask combination in turn.  A detailed description of this process may be found in \cite{boccaletti2022}. 

The on-sky projected area of a pixel across the entire detector is shown in Figure~\ref{fig:dist_area}, where it is calculated by evaluating the flight-measured on-sky position of all pixel corners and then calculating the geometric area of each quadrilateral.  The area varies between 0.970 to 1.017 times the baseline value of 0.0121 arcsec$^2$.

\section{Flat Fields} \label{sec:flats}
Pre-flight pixel flats were made from data taken with the MIRI Telescope Simulator during Flight Model testing at the Rutherford Appleton Laboratory and observations with the imager internal calibration source during instrument Cryo-Vacuum testing at NASA Goddard Space Flight Center. These tests did not include the telescope optical elements. From these data a considerable effort was made to create high quality pixel flats where features of the test set up and instrument meant that the data sets were not optimised for the task. Therefore, it was important in commissioning to verify that firstly the flats were applicable to flight data and to measure for the first time the low spatial frequency ``sky'' flat field, which is dominated by the telescope and observatory environment and thus could not be measured on the ground. 

The flat field commissioning activities for the imager were conducted in two programs which also overlap in their goals: one for the external flat fields (PID~1040) and another for the pixel flat (PID~1051). The input data for the commissioning flat analysis came from three sources: (1) an internal calibration lamp source i.e., a highly stable source to be used throughout the mission but with non-uniform illumination for the imager, (2) a low stellar density sky background, ideal for deriving the low frequency or ``sky" flat (regions of high zodiacal dust were chosen to improve the signal to noise), and (3) a high stellar density sky, which provides input to help correct the low frequency flat (a region of the LMC was used for the measurement). 

From ground testing we know that the MIRI imager pixel flats are wavelength-dependent and therefore a flat correction for each separate filter is needed. The main reason for that is the presence of at least two non-monochromatic features in the flat:
\begin{enumerate}
    \item There are large circular diffraction patterns that appear when the on-board source is being used, and those are due to diffraction at apertures in the imager calibration unit optics. They scale with wavelength, and multi-filter measurements are needed to correct them and extract the underlying pixel flat.
    \item During ground testing a set of circular spots that grew in wavelength were seen in the flats, and identified as contaminants on the detector surface.
\end{enumerate}

\begin{figure*}
\includegraphics[width=0.8\textwidth, center]{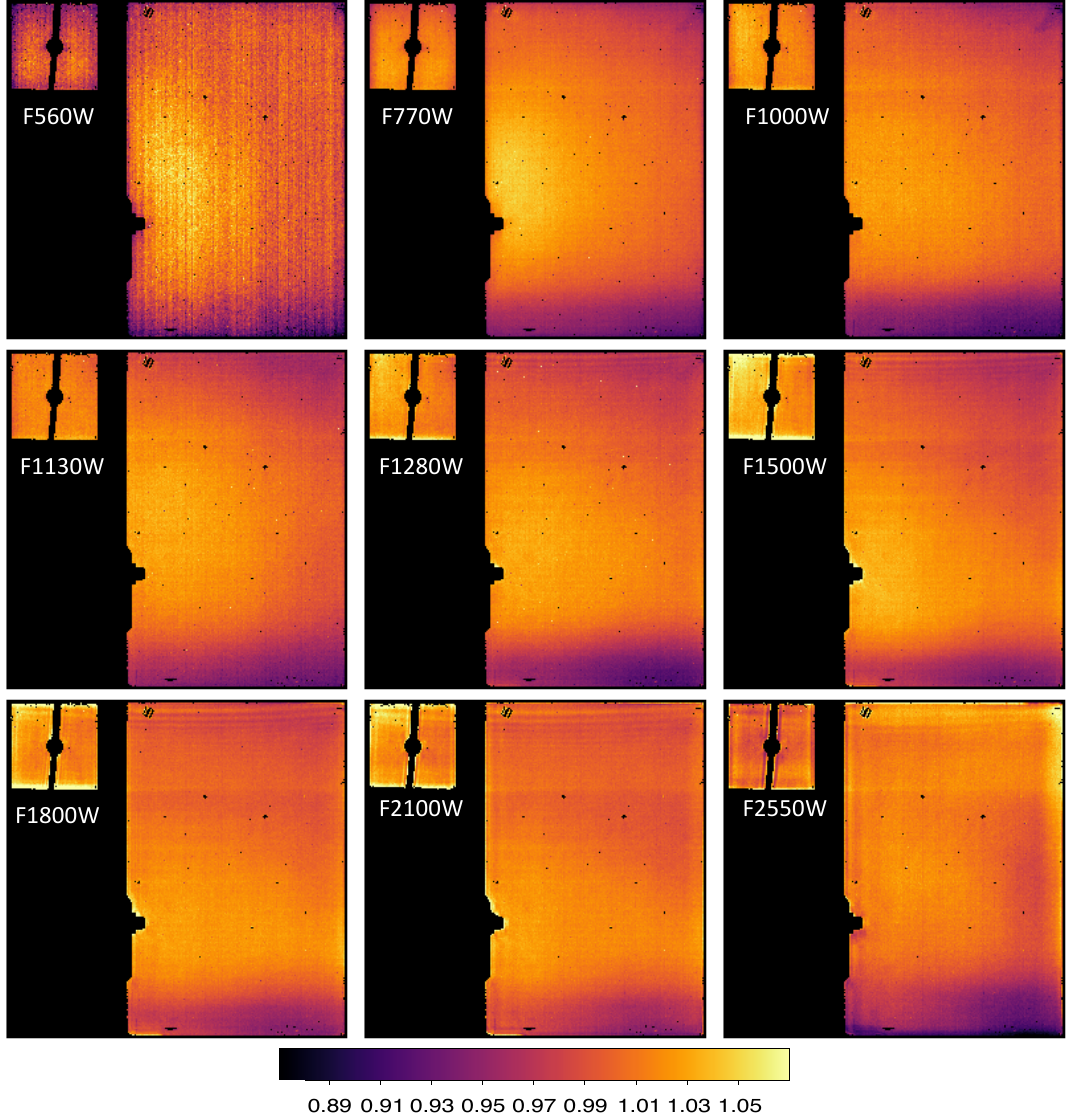}
\caption{Flat fields constructed using sigma-clipping of many observations aligned in array coordinates and filtering out sources that are in different array positions in different observations.}
\label{fig:flats}
\end{figure*}

The initial flat fields created in commissioning were constructed using sigma-clipping of many observations aligned in array coordinates, using extended emission that fills the imager's field of view, and filtering out sources that are in different array positions in different observations.
Observations taken from commissioning program PID~1040 of a large region in the Large Magellanic Cloud composed mostly of point sources were used initially for constructing the flat fields for all imaging filters.
After commissioning, the flat fields for F1000W, F1280W, F1500W, F1800W, and F2100W were updated by including additional observations.
These additional observations, taken in the first two months of science operations, were visually determined to be mostly composed of point sources. These flat fields are shown in Figure~\ref{fig:flats}.

The flat fields were constructed by properly combining multiple dithered on-sky observations and made independently between filters and using only full frame observations.
The Lyot region is included in the full frame flat fields as the observations are valid for regular imaging filters except for the regions eclipsed by the Lyot spot and support bar.
The eight pixels at the four edges of the detector were masked in the flat fields due to undesirable detector effects.
Subarray flat fields are created on-the-fly in the pipeline by cutting out the appropriate region from the full frame flat fields.

\section{Image artefacts} \label{sec:artefacts}

MIRI's Si:As blocked impurity band (BIB) detectors are state of the art for mid-infrared imaging. Nevertheless operating detectors at temperatures around 5 Kelvin is still challenging. In this section we describe some of the known non-ideal behaviours seen in MIRI's focal plane arrays as well of some of the corrections and mitigation available at the time or writing. Note that there is no effect or artefact that is known to significantly reduce MIRI's imaging performance. 

\subsection{Persistence}
\label{sec:persistence}
Persistent images are a common artefact of cooled infrared detectors - a memory effect where a residual of a previous image is seen in the subsequent exposure. Previous infrared space observatories such as Spitzer and WISE, which used similar detector technology to MIRI (Si:AS IBC detectors), recorded persistent structure in imaging \citep{hora2004, wright2010}. In these missions, the persistence structures were removed using a process known as annealing, where the detectors are temporarily heated by around 15 degrees to help release any trapped charge in the detection layers of the focal plane system. From ground testing, we knew that MIRI detectors would also, under certain conditions, exhibit persistence and a system of annealing was included in the design to help clear the detectors of persistence artefacts if necessary. Therefore, the investigation of persistence\footnote{Persistence as discussed in this paper is also commonly referred to as latents/latency but we, and the JWST documentation in general, avoid the word “latency” because it is also associated with properties that cannot be attributed to the persistence effect we describe. } was part of the commissioning activities for the imager where we aimed to confirm any memory effect was similar to the ground test expectation.

\begin{figure*}
\includegraphics[width=\linewidth]{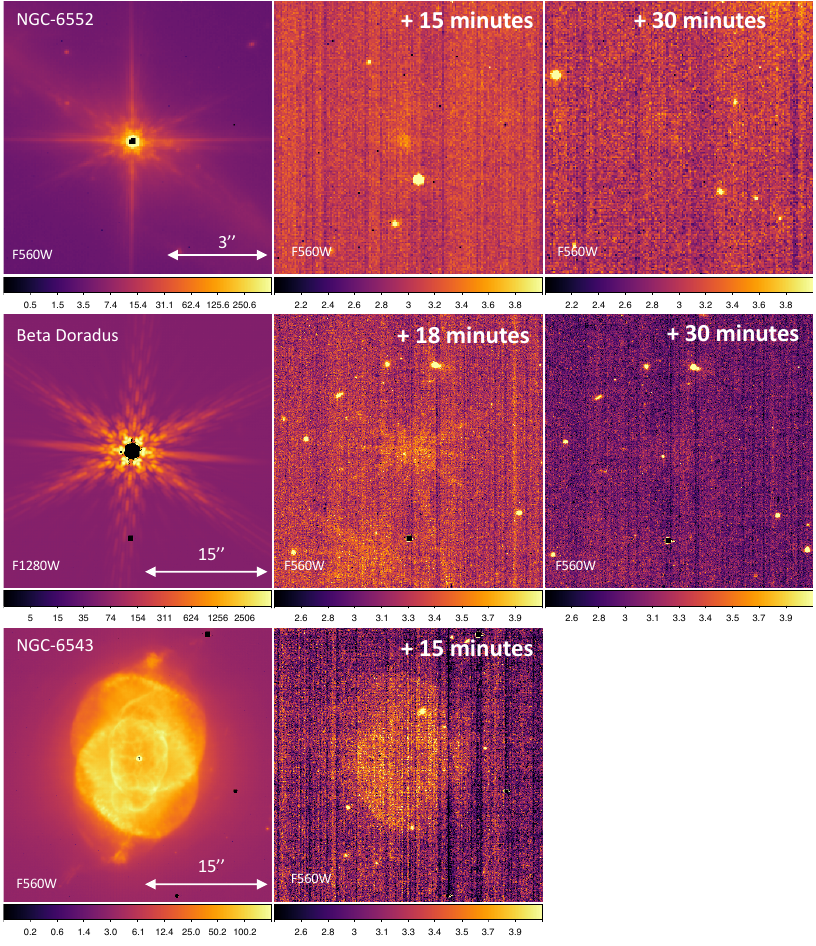}
\caption{Examples of persistence in flux/rate images (pipeline level 2a). Top row shows results from the persistence test in PID~1039. Each row, left to right, shows the same position on the detector. Left: image of the Seyfert galaxy (NGC~6552, $\beta$ Dor, or NGC~6543) that saturates the detector at its core. Centre: the next exposure taken 15 minutes after the first image - a faint persistence residual from the galaxy nucleus can just be seen in the image. Right: showing the same position 30 minutes after the first image was observed - the persistence residual is very faint. The images for $\beta$ Dor are from PID~1023. The bottom row shows persistence from the extended structure of the bright planetary nebula NGC~6543. Note the scale for the soak images on the left is log while that for the other images is linear and all units are DN  s$^{-1}$.
\label{fig:Persistence_examples}}
\end{figure*}

The principal commissioning activity to investigate persistence in MIRI detectors was designed to imprint an image of a bright target on the imager detector and analyse its decay (if any) for up to 40 minutes. Data from the commissioning activity can be found in PID~1039 where the target chosen for the imager test was the Seyfert galaxy NGC~6552: a compact, bright, mid-infrared source in the Northern continuous viewing zone for \emph{JWST}\footnote{A paper detailing the results of the imaging and MRS spectroscopic data for NGC~6552 has been published by \cite{alvarez-marquez2022}.}. The Seyfert galaxy was observed for nearly 10 minutes in the same position on the detector. The bright nucleus saturated the pixels at the galaxy centre in less than 3 groups ($<$ 10 seconds). With only 3 groups or less the pipeline could not fit a slope to the data and therefore no flux value is assigned to the nucleus pixels and is masked in the image (black pixels). The top row of Figure~\ref{fig:Persistence_examples} shows the primary result of the test where on the left is the galaxy image, or ``soak" image. The soak image is the one that contains the source of the persistence e.g. a bright target. The central and image to the right are subsequent images taken at the same detector position that the bright galaxy was observed, but a different sky position. These follow up images in Figure~\ref{fig:Persistence_examples} demonstrates that persistence from the galaxy nucleus is seen but at a very low level. The level of persistence seen 15 minutes after the galaxy was observed is $<$0.01\% of the flux of the target and appears to dissipate entirely by the next observation 30 minutes after the soak image ended. This provides us with initial evidence that persistence artefacts are likely to have a low impact on MIRI data, where such levels of persistence are easily averaged out in the process of mosaicking dithered imaging, which is the recommended imaging technique. 

Next, as part of the anneal verification and testing program (PID~1023), we observed the bright $K$mag $=$ 2 star $\beta$ Doradus in four MIRI imager filters (F560W, F770W, F1000W and F1280W). The tests were designed to provide an extreme test of detector saturation in order to test that the anneal recovery process works to remove any residual detector artefacts including persistence. $\beta$ Dor is an order of magnitude brighter than the Seyfert galaxy NGC~6552 at mid-infrared wavelengths where the core of the stars PSF saturated the detector in less than 1 frame ($<$2.7\,s) and was imaged for 250 seconds.  Before the anneal process was done we took several images to confirm any persistence artefacts from the bright star. The middle row of Figure~\ref{fig:Persistence_examples} shows the final image of $\beta$ Dor test at 12.8 $\mu$m as well as the subsequent 5.6\mum\ images taken of an off-source background close to the star approximately 18 and 30 minutes after the soak image. In the image, observed 18 minutes after the soak image, it is still possible to make out a persistent imprint of the 12.8 $\mu$m PSF of $\beta$ Dor. However, as for the Seyfert galaxy test, the contrast of the persistence image left by the star compared to the background is very low, only a few tenths of a DN s$^{-1}$ above the nominal background. Likewise, the background imaged 30 minutes after the last observation of $\beta$ Dor shows very little persistence from the extreme hard saturated observations. We re-iterate here that these images show persistence from a single pointing and the average image of the stack of dithered images in the pipeline mosaicking step will further reduce the impact on the final image product. Choosing large dither patterns will also help reduce the impact of persistence for compact sources to avoid imaging a bright object on area already affected by persistence. The $\beta$ Dor test confirmed the low impact of persistence for extremely bright targets as well as for longer wavelength data up to 12.8\mum. In addition this test provided evidence that is was not necessary to have to anneal after bright observation to remove persistence artefacts as used in previous missions with similar detectors e.g. WISE. 

Lastly, in the bottom row of Figure~\ref{fig:Persistence_examples} we show an example of persistence from an extended object, the Cat's Eye nebula NGC~6543. The mid-infrared bright nebula was observed for 10 minutes in the same detector position. This example shows that the planetary nebula is the brightest at its core and most of its extended structure is not saturated or only partially saturated in the soak image. However, the persistence image, taken approximately 15 minutes after, traces the extended structure of the target. This confirms what was seen in ground tests: that persistence is dependent on the soak time as well as the flux from the source. An image does not have to be saturated to cause persistence. It also demonstrates that persistence from extended objects can be more easily seen in imaging because the spatially extended low contrast artefacts can clearly be distinguished from the background unlike point source persistence. 

\begin{figure*}
\includegraphics[width=\linewidth]{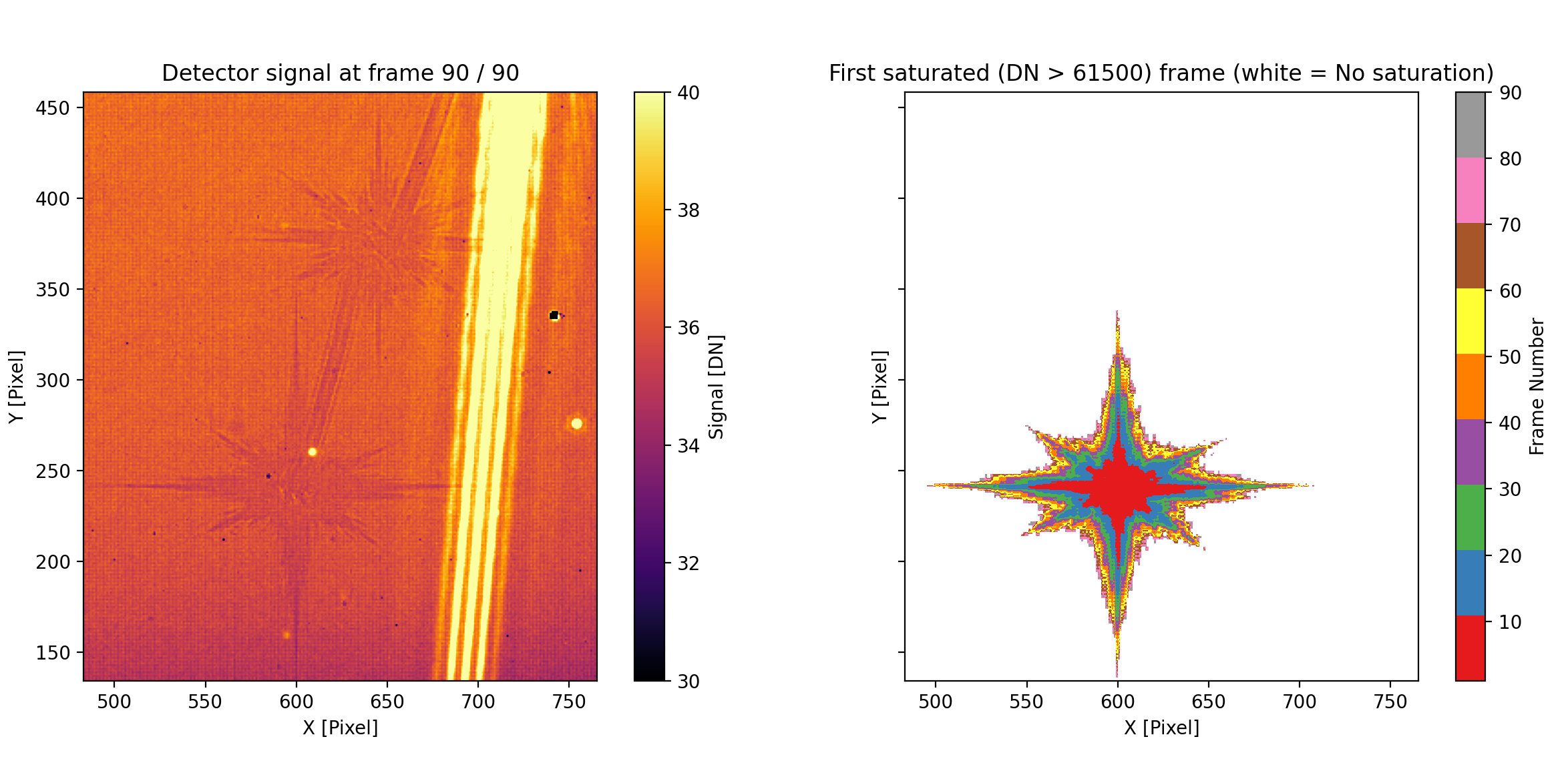}
\caption{Left: image of $\beta$ Dor at 10~$\mu$m at the position where the bright star was observed in the previous two observations at 5.6 and 7.7~$\mu$m. The persistence images of the star in the two positions can be seen in low contrast as well as that of the slew between positions. Right: identification of the pixels that are marked as ``hard'' saturated in the first position observation at 5.6\mum. We can see the persistence image matches well the image structure of the hard saturated pixels from the image. 
\label{fig:persistence_saturation}}
\end{figure*}

In Figure~\ref{fig:persistence_saturation} we show observations of $\beta$ Dor the F1000W image from the test - the 3rd positional offset of the star. The image is a cut out of the part of the detector image where the previous two offset positions of the bright star were made using the F560W and F770W filters (See Figure \ref{fig:beta_doradus} for reference). In this image we can see a persistence artefact from the bright star at these two positions and persistence from the slew between them - again at low contrast compared to the background. Also in Figure~\ref{fig:persistence_saturation} in the right panel we show which of the pixels in the first position image at F560W reached  saturation - defined here as more than 61500 counts. The observation was made with 1 integration with 90 groups and the plot is colour coded by the group number in which the pixel saturated. For example a pixel that reached 61500 counts in less than 10 groups is marked in red. If the pixel reached 61500 counts in groups 10 to 20 of the integration it is marked in blue and so forth. Figure~\ref{fig:persistence_saturation} then clearly shows that the central core of the star saturated first followed by pixels further out from the centre. Figure~\ref{fig:persistence_saturation} also demonstrates that the persistence signature left by the bright star, seen in the left panel of Figure~\ref{fig:persistence_saturation}, corresponds well to the pixels of the detector that reached  saturation before the end of the ramp, as marked in the right panel of Figure~\ref{fig:persistence_saturation}. We note here that this is not just to the areas of highest flux. From ground testing we expect short integration, high flux, saturating imaging to produce relatively strong persistence that decays quickly whereas bright unsaturated imaging, with long soak times, to have relatively weaker persistence but longer decay times.

In summary, the persistence artefacts from bright sources seen seen in MIRI during the commissioning period are very low contrast and dissipate within approximately 30 minutes of the observation of origin. This means that strong persistence between different scientific programs that would affect data quality for science appears unlikely given the long observatory overheads, especially slew times. Although, persistence within a science program may be seen on timescales of 15 minutes or less. However, there are many factors that contribute to whether or not a persistence artefact is made or seen. Notably the depth of the subsequent image is important as seen in the next section, where deep imaging can reveal very low contrast persistence that would otherwise not be seen. We note that the pipeline has persistence tools for the near-IR detectors but these are not applicable to MIRI data at the time of writing. Lastly, we know that increasing the number of integrations can mitigate persistence or, more specifically, reducing the counts per pixel per integration. Using this strategy must be balanced with keeping enough groups per integration for calibration, typically more than 10 when possible. 

\subsubsection{Slew persistence} \label{sec:slew_persistence}
As shown in Figure~\ref{fig:Persistence_examples} and \ref{fig:persistence_saturation} persistence artefacts can also be seen when a bright target slews across the imager field of view. This has the potential to be the most common persistence artefact in imaging due to the short lead time ($<$ 5 minutes) between the slew and the start of observations. It is also worth noting that slew persistence can be seen in a different filter than the one used in the previous observation. Because the movement of the MIRI filter wheel is the last activity before a MIRI imaging observation, an object saturating the detector in the filter used for the previous image, and thus during the subsequent slew, may well appear fainter and non-saturating when the next filter is chosen and the object imaged.

\begin{figure}
\centering\includegraphics[width=1\linewidth]{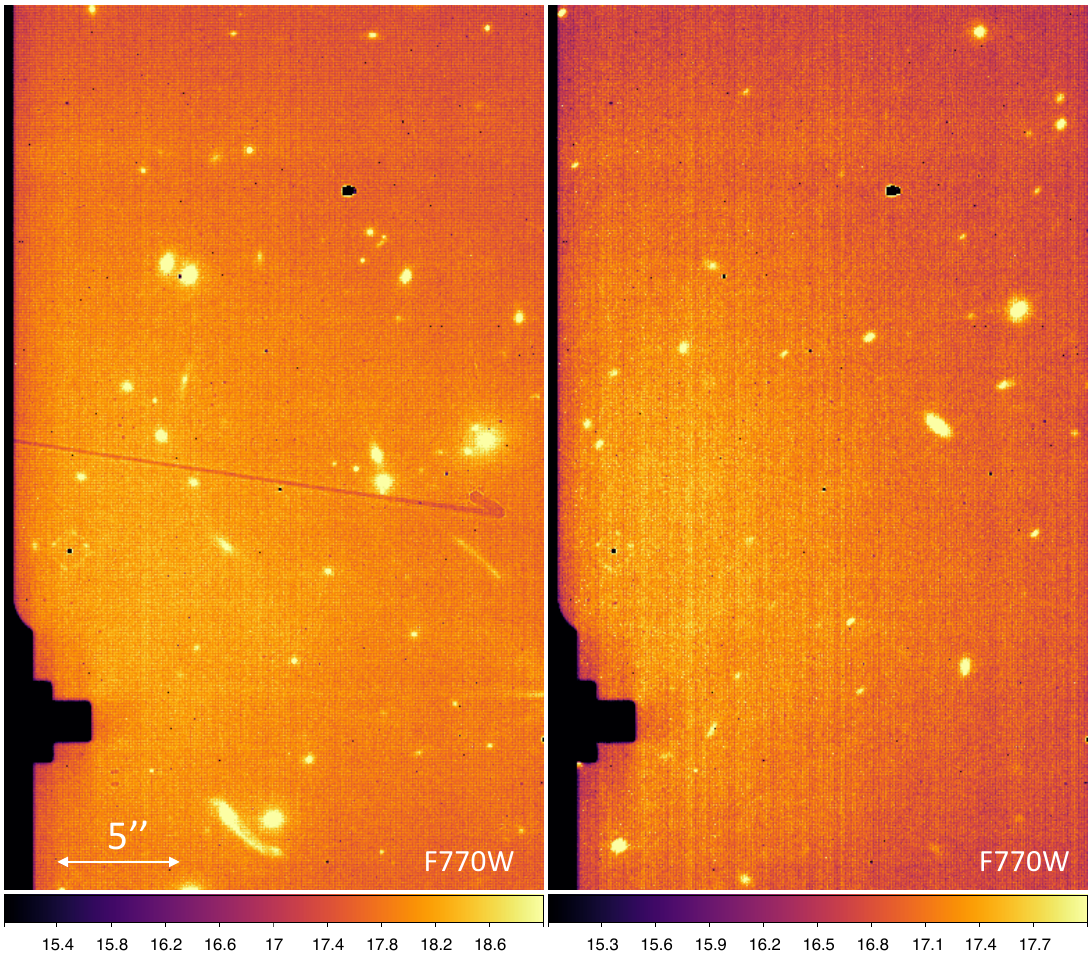}
\caption{Left: level 2a flux image from the ERO program of the lensed cluster observations of SMACS0723 (PID~2736). Slew persistence that resembles a check mark can be seen in the image taken 5 hours after the bright star that was the origin of the slew persistence. Right: image taken 12 hours later in program PID~1124. The slew persistence can just be made out in this image.  \label{fig:SMACS0723}}
\end{figure}

A prominent example of slew persistence appeared throughout the whole seven hour data set of the Early Release Observation (ERO) program targeting the lensed cluster SMACS0723. Figure~\ref{fig:SMACS0723} shows the first exposure of the data set where an elongated ``check mark" shaped persistence streak can be seen in the image (PID~2736). The origin of this slew persistence appears to be a $K$= 4 star (HD~92209) imaged with the F2300C filter as the last observation in a commissioning test (PID~1045, Observation 75) taken approximately 4 hours before the observations of SMACS0723. The ``check mark" shape comes from the initial slew followed by a guide star correction to put the star in the centre of imager field of view. Note we still see this slew persistence at the end of the SMAC0723 observations, some 11 hours after the star of origin was observed. This confirms that long-lived persistence can exist in MIRI imaging which is also consistent with ground detector tests made at NASA's Jet Propulsion Laboratory that saw persistence lasting at least 10 hours. However, these long-lived persistence are at a very low contrast, where the difference between the persistence artefact and the background is less than 0.5 DN s$^{-1}$. Why the slew persistence in this case was particularly long-lived is not known at the time of writing.

Another possible source of persistence in imaging is from the movement of the filter wheel at the start of an observation. The filter wheel is moved for an observation after the target is positioned in the field of view of the imager. As the filter wheel turns, a target may leave a persistence signature when passing through a filter in which it is very bright. In particular the F-Lens filter\footnote{The MIRI F-Lens was used during ground testing to form a pupil image from which optical alignment can be verified. This test was repeated in flight to determine whether the pupil alignment had changed and to check for vignetting. The in-flight measurement showed MIRI remained aligned and met requirements reference in \citet{kubalak2016}.}, which has a wide wavelength throughput, can leave an imprint on the detector. An example can be seen in Section \ref{sec:bright_targets}, Figure~\ref{fig:beta_doradus}. In the 5.6~$\mu$m image there is an imprint behind the star that resembles the \emph{JWST} mirrors and this is in fact the persistence image of the F-Lens filter passing across the imager.

\subsubsection{Persistence from the background}\label{sec:artefacts-bkg}

\begin{figure}
\centering\includegraphics[width=.9\linewidth]{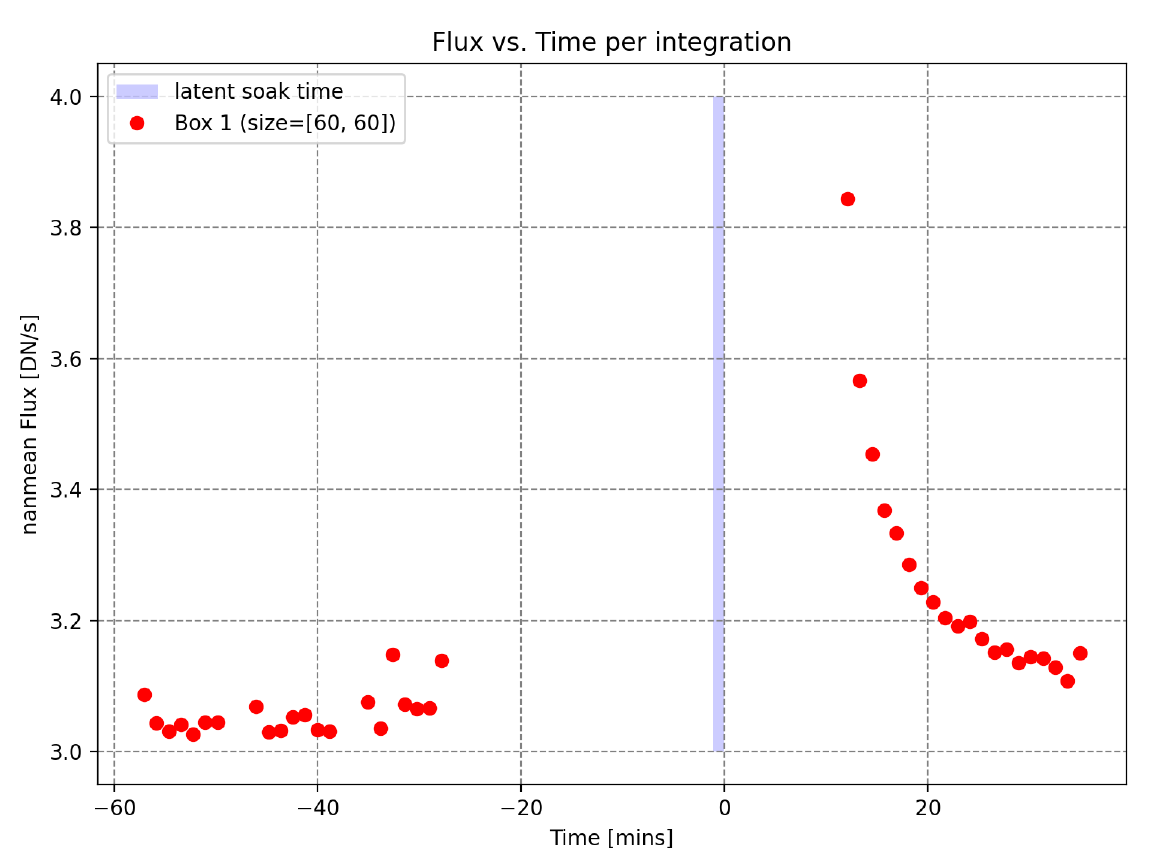}
\caption{Flux (DN/s) per integration of the background in 5.6 $\mu$m imaging observations taken before and after an observation at 25.5 $\mu$m. The time and duration of the 25.5 $\mu$m imaging observation is represented by the blue shaded region. The data displayed was processed to level 2a. \label{fig:25um_persistence}}
\end{figure}

As discussed in Section \ref{sec:bkg} the thermal background of \emph{JWST} at MIRI's longest wavelengths has a high flux \citep{glasse2015} and this can also be a source of persistence. In particular, if an observation at short wavelength follows directly after an observation at long wavelength a persistence from the higher background image may be seen at short wavelength. During the commissioning of MIRI we tested the impact of persistence from high background observations in program PID~1039. Figure~\ref{fig:25um_persistence} shows the results where we plot the background flux at 5.6\mum\ from the short wavelength integrations taken before and after a 300 second observation at 25.5\mum. The plot shows that the background flux clearly increased in the 5.6\mum\ images taken after the observation at 25.5\mum\ and then decreases back towards the nominal background level measured before the long wavelength observation. In fact, we see that the background does not quite recover to the nominal 5.6\mum\ observation level even 30 minutes after the long wavelength observation. Looking back at Figure \ref{fig:Persistence_examples}, in the centre row of the figure we can see the change in background at 5.6\mum\ 18 minutes and 30 minutes after the first image due to the persistence from the background of the 12.8\mum\ observation of $\beta$ Dor taken before. It is important to note that this flux change in the background is small - around 0.5 DN/s over 30 minutes, but nonetheless this can be significant for imaging and therefore dithered and mosaicked imaging may require background matching in the pipeline. This change in the background is unlikely to affect most science programs. However, as a general rule, we recommend observers to transition from short to long wavelength filters to help avoid any persistence contamination of this kind. Finally, we note that if the MIRI imager has been left in it nominal idle mode in a long wavelength filter for hours or even days this can leave a very long duration (several hours) persistence signature in the background. But again the change in the background flux measured is very small but it can have a notable effect on the quality of very low background calibration data i.e. in the acquisition of MIRI darks.

\subsubsection{Persistence from cosmic rays}
\label{sec:persistence_cosmic}
\begin{figure}
\centering\includegraphics[width=1\linewidth]{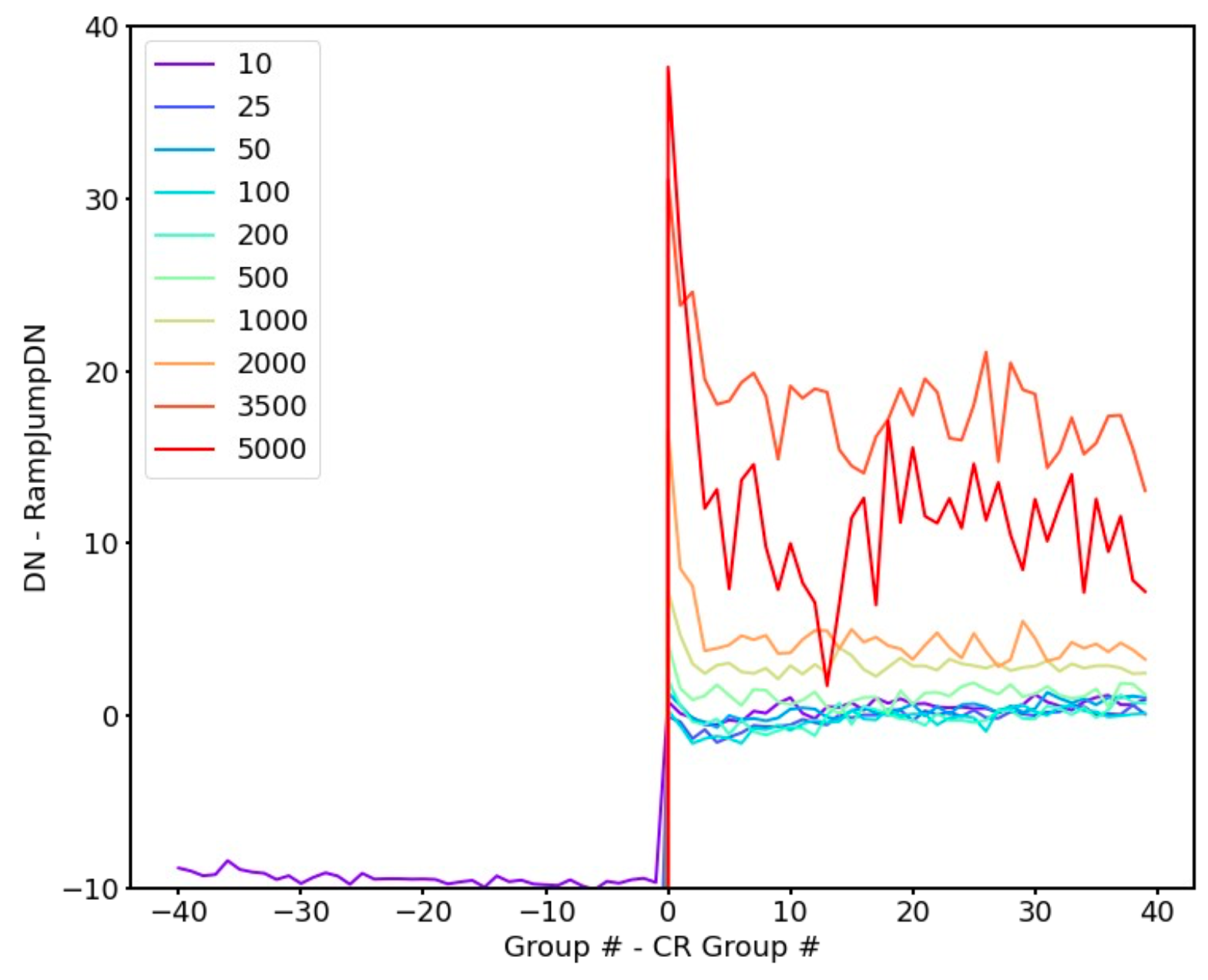}
\caption{plot of ramp group vs signal value showing ramps that have had cosmic ray hits of increasing intensity. The signal values are normalised to the signal value at the time of the cosmic ray hit. For higher intensity cosmic ray hits a decay in the signal after the initial jump is seen - which is persistence. \label{fig:cosmic_persistence}}
\end{figure}

\begin{figure}
\centering\includegraphics[width=1\linewidth]{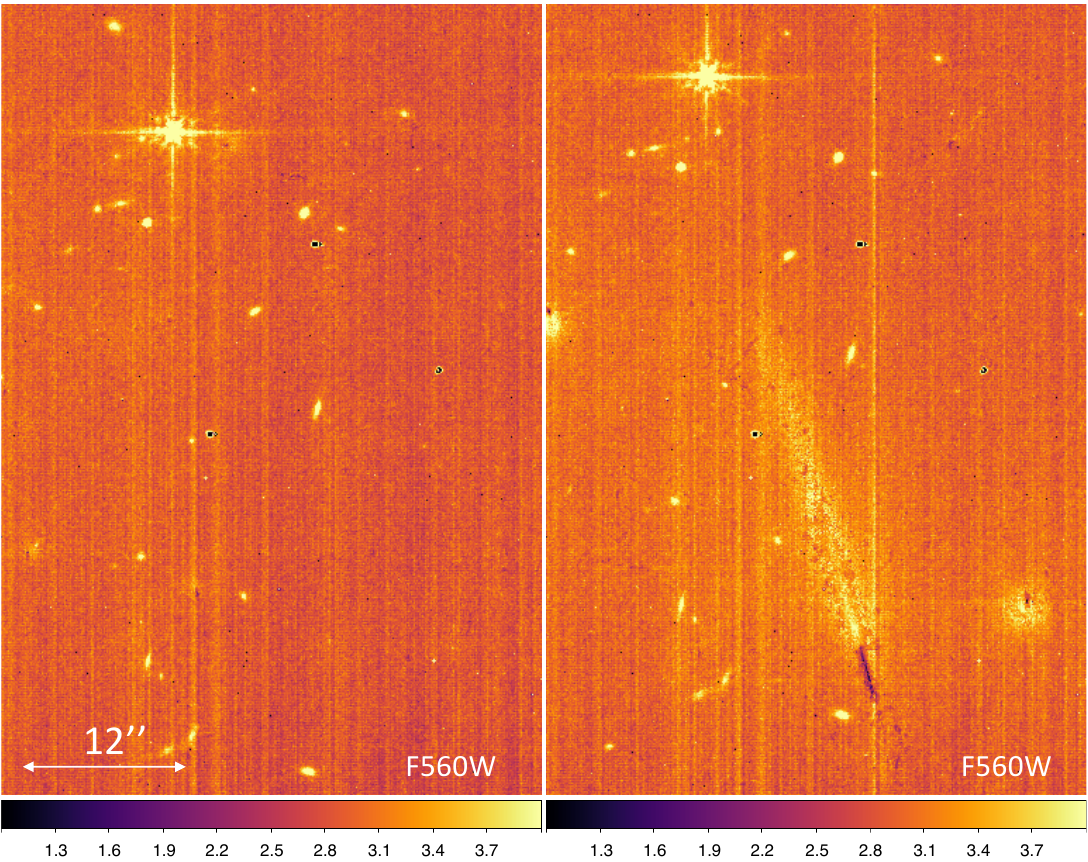}
\caption{Images of a calibration star from commissioning activity PID~1023 - Observations 13 (left) and 14 (right), part of a dithered sequence. The right image shows the appearance of cosmic shower artefacts in the detector image - an elongated structure in the centre right of the image as well as several more spherical ``showers" that are not part of the sky image towards the upper left and lower right of the image. The high power cosmic ray also pulls-up the value of the entire row in which it hit the detector. Images are taken at 5.6~$\mu$m and processed to level 2a in the pipeline with flux units of DN s$^{-1}$. The target star at the top of the image can be used as a reference for the position on the sky.  \label{fig:showers}}
\end{figure}

The sample up-the-ramp method, where the flux is derived by measuring the slope of the ramp created by successive non-destructive reads of the detector \citep{morrison2023}, is used by all \emph{JWST} instruments. When a cosmic ray hits a pixel, adding unwanted signal to the data, the slope of the two ramps either side of the hit can be measured separately and the flux derived. Using this technique cosmic rays are very effectively removed in MIRI data with the \emph{JWST} pipeline's jump algorithm in the processing of data from level 1b to level 2a. However, in commissioning the team noticed that some cosmic ray artefacts were not removed. An example of such an artefact can be seen in Figure~\ref{fig:showers} which we refer to as cosmic ray ``showers"\footnote{ \emph{JWST}'s near-infrared detectors have a similar artefact known as ``snow balls" but the underlying mechanism for showers in MIRI data is different than that of snow balls in the near-infrared detectors.}. After further analysis we discovered that showers were the residuals of high energy cosmic ray hits where the energy arriving in the detector is spread over more than one pixel as well as secondary hits. The residual signal from these powerful cosmic rays can then leave persistence artefacts which we see as a shower in the flux image.  Showers in MIRI detectors have a variety of different forms from spherical to very elongated and can spread out over hundreds of pixels. The residual showers will have the largest impact on short wavelength data with low backgrounds e.g., deep imaging programs, where the background is dominated by cosmic showers and cannot be easily averaged out in the dithered mosaicked image. Figure~\ref{fig:cosmic_persistence} shows examples of the persistence from cosmic rays as the signal from the cosmic ray is seen to decay after the jump, where the amplitude of the effect increases with the cosmic ray power. The cosmic ray example in Figure~\ref{fig:showers} also pulls up the values of the entire row of pixels that it hits and this row pull-up can also leave a persistence signature. 

The persistence from cosmic showers can be either positive or negative compared to the background. This is dependent on how fast the persistence is decaying relative to the background flux in the image. For example, negative persistence from a powerful cosmic ray can occur due to the persistence decaying quickly over the duration of an integration thus pulling down the slope measurement of the nominal flux of the background. Persistence from a cosmic ray with a slower decay over the duration of an integration will be seen as positive in flux compared to the background as the persistence flux adds to the nominal flux of the background. 

It is challenging to automatically flag the persistence from these cosmic rays because the pipeline looks for changes in flux between consecutive groups while the signal added from the ``showers'' is usually small compared to the flux measured for the background. A number of improvements to the jump algorithm have already been implemented in the \emph{JWST} pipeline. Further improvements are ongoing and the reader should refer to the pipeline documentation for more information about cosmic shower rejection. As discussed in Section \ref{sec:deep_obs}, increasing the number of integrations appears to be a good strategy for reducing the impact of cosmic showers in deep imaging due to the increased redundancy allowed by taking the mean of the data sets in each dithered position. 

\subsection{Row and column artefacts}
\label{sec:row_column}
Si:As BIB detectors are prone to rows and columns artefacts that come from the target that are in high contrast to the background. Such artefacts were also seen in the Spitzer/IRAC detectors, especially channels 3 and 4, as well as in the MIPS silicon IBC arrays, described as a ``jailbar” effect. The effect in MIRI's detectors was extensively studied during several ground test campaigns with flight spare focal plane arrays and flight clone electronics at NASA's Jet Propulsion Laboratory. The results of those tests are presented in \cite{dicken2022}. The key result from those tests is that, although the row and column artefacts are seen in the flux images, they actually result from a change to the signal measured in the ramp in level 1 data. The signal change results in a distortion or bend in the flux ramp that is related to how quickly the bright contrasting source saturates the pixels in the affected rows and columns. We also found that the row and column effects are often seen together but have different properties and therefore have different origins or mechanisms that produce them. One key difference is that the column artefact is seen only in the columns that contain the bright source but the row artefact can be seen in the rows above (in the read direction) the bright pixels causing the artifact. Finally, because these effects are seen in a number of different array types, covering two different multiplexers and detector types (Si:As, In:Sb) with MIRI and Spitzer, this confirms that the artefacts originate in the multiplexer rather than in the detector layers.

\begin{figure}
\centering\includegraphics[width=1\linewidth]{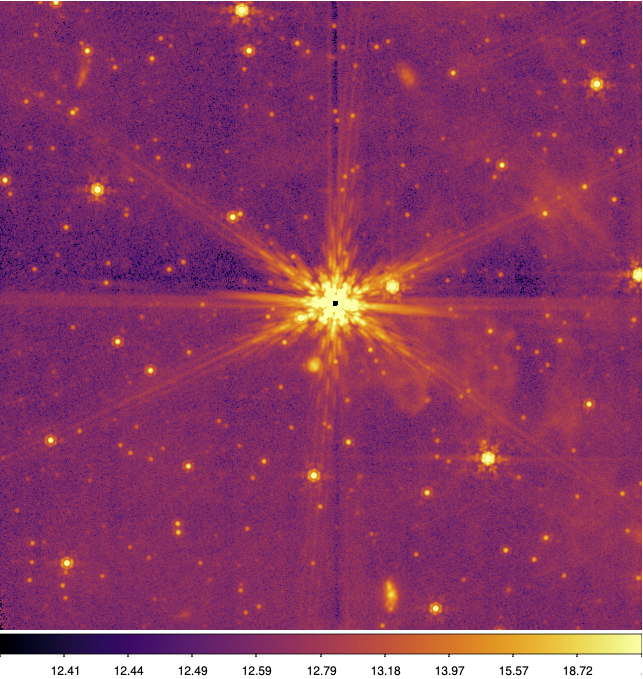}
\caption{F1000W image of the spectral calibration nebula SMP-LMC-58 from the commissioning program PID~1090. The image scale has been cut to highlight the structure in the background. The central core of the bright planetary nebula can be seen to cause column and row structure in the imager where the column structure creates a vertical stripe that is lower than the mean background and the row structure creates a stripe that is higher than the nominal background.\label{fig:PN_PID1090}}
\end{figure}

Figure~\ref{fig:PN_PID1090} shows a commissioning F1000W image of a bright spectral calibration, the planetary nebula SMP-LMC-58. The bright central core of the nebula causes row and column structure in the image which is seen as a column with lower flux than the background above and below the nebula while rows with higher flux than the nominal background either side of the nebula. The image is shown in detector coordinates so the row and column artefacts follows the XY directions of the pixel grid in the detector plane, whereas the PSF can be differentiated because it is an angle of a few degrees. This image is the mosaiked combined image of a 4-point dither observation, showing that dithering does not mitigate row and column artefacts. The artefact is tied to the source and not the detector, so the process of dithering, moving the object to different positions in the detector cannot mitigate the row and column artefacts. And hence is reinforced in the stacking and averaging of the mosaicked image process.

Commissioning analysis shows that we see row and column artefacts at all wavelengths of the MIRI imager - a result we were not able to verify in pre-launch ground testing due to the low backgrounds of the test environments. Additionally, in flight, the column artefact appears to manifest differently above and below the source of origin whereas in pre-flight testing the artefact was identical above and below the source of origin. This can be seen in Figure~\ref{fig:PN_PID1090} where the column artefact above the nebula is more enhanced in the flux image than below. This is currently not well understood. We also see less incidence of row artefacts than we were expecting from ground testing. This is likely because the row artefacts are seen at lower contrast than the column artefact effect where they can be spread over many rows in the read direction up the detector beyond the source of origin. 

At the time of writing there is no pipeline correction for row and column artefacts. The MIRI team have trialled several column and row filtering tools, similar to those available for Spitzer data, with good initial results. For any future development of a pipeline correction for row and column artefacts see the MIRI \emph{JWST} documentation pages. 

\subsection{Stripes at low background}
When showing low background imaging at high contrast, such as the 5.6\mum\ imaging in Figure~\ref{fig:Persistence_examples} and \ref{fig:showers}, striping in the background can be seen in the level 2 processed images. This striping is an artefact left over from the reset anomaly correction which is part of the dark correction in the pipeline \citep[see][]{morrison2023}. The dark correction works well to remove the reset structure but in commissioning it became clear that the dark and reset anomaly change in time. Therefore, a dark correction data product made from data at a different epoch than the data being processed may not completely subtract the anomaly as it should. This issue is mitigated as the background increases with wavelength for imaging because the dark subtraction becomes less dominant as photon noise increases with higher background emission. The reset anomaly is also more dominant in the first 25 groups of data, therefore longer ramps tend to see reduced striping at low backgrounds. At the time of writing, corrections for this issue are still being worked on and the reader should refer to \emph{JWST} documentation for future updates. 

\section{Recommended Best Practices} \label{sec:best}
\subsection{Dithering}
\label{sec:dithering}
Dithering in MIRI imaging is a recommended technique that provides a variety of benefits. It improves the PSF sampling, which for MIRI is mostly relevant at wavelengths shorter than 7~$\mu$m, where the PSF is slightly undersampled. Dithering also facilitates the subtraction of the background for point source imaging where much the image is ``source" free, as well as mitigating the impact of bad pixels (see Figure~\ref{fig:imager_examples}), and minimising detector effects. A minimum of four dither positions or more is recommended to allow redundancy when creating the mosaic to mitigate against the effects of bad pixels, detector artefacts and cosmic ray residuals. 

As shown on Section~\ref{sec:persistence_cosmic}, cosmic ray showers are prevalent in MIRI data and can have a significant impact on mosaicked imaging. When a high energy cosmic ray causes a residual the size of tens of pixels (known as a cosmic ray shower in MIRI) it can appear as an artefact in the mosaic when the dither pattern step size is smaller than the cosmic ray shower residual. To mitigate this in MIRI imaging, the MIRI team now recommends dither patterns with long steps in between positions, which will facilitate their removal during the dithered image combination step as this reduces the chance of overlap of artefacts between dither positions. In particular, the Cycling pattern with size "LARGE" has been very effective at alleviating the residuals from showers as well as removing sources from the background through image stacking. 

\begin{figure*}
\includegraphics[width=\linewidth]{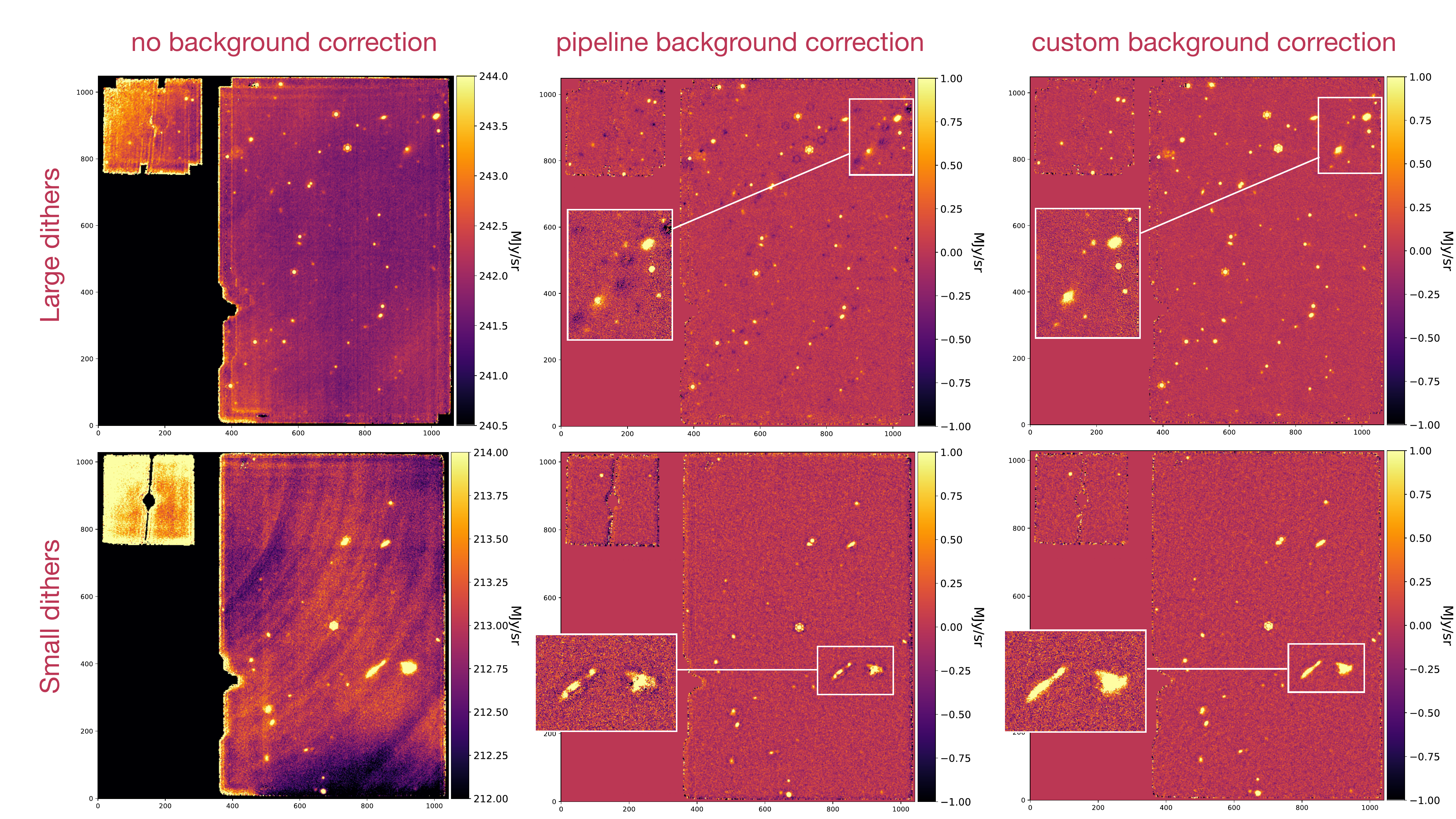}
\caption{Demonstration of the effect of dither pattern size on the background subtraction in MIRI F2100W observations (top: PID~1024, bottom: PID~1028). First column shows the combined level 3 (\texttt{i2d}) image without background correction. Second and third columns show the background-subtracted combined level 3 images using the \emph{JWST} pipeline (middle) and manually (right). In both cases the master background is constructed using individual science dithers (i.e., no dedicated background observations). The pipeline uses all the dithers and make a sigma-clipped mean-stack, while in the manual method we exclude dithers that are close to each other and interpolate the source values with background from other dithers (see Section~\ref{sec:bkg-strategies} for details). 
The manual (custom) background subtraction returns better results than that of the pipeline. However, both strategies fail as a good solution for observations with dithers smaller than $\sim 1.5\times$ FWHM. Background images constructed from such dithers have source residuals and over-subtract the flux of any source that is resolved.
The inset panels are zoom-in examples for visual demonstration. The bottom left image shows arcs in the background across time image. These are referred to as "tree-rings" and can be seen in short wavelength - low background  - data. They are a detector artefact and associated with the reset anomaly which is removed in pipeline dark subtraction \citep{morrison2023}. 
\label{fig:bkgsub}}
\end{figure*}

\subsection{Background subtraction strategies} \label{sec:bkg-strategies}
All MIRI observations require background subtraction, particularly at longer wavelengths where the telescope thermal emission is significant.
As discussed in Section~\ref{sec:bkg} (Figure~\ref{fig:miri-f2100w-bkg}), spatial variation can be seen across the background field in observations of all MIRI filters, which hinders accurate data analysis of faint sources and structures. These features are often the residual of imperfect flat field correction and/or dark/reset anomaly corrections \citep{morrison2023} and can be removed in the background subtraction step. However, as more calibration data is taken in cycles 1 and 2, we expect these features to be mostly removed prior to the background correction step. We note that as the dominant spatial structure seen in MIRI data is due to flat field (i.e., a multiplicative effect), correcting it via background {\em subtraction} results in incorrect fluxes. However, this effect was measured to be $<5\%$ across the field of view and is often dominated by other flux measurement uncertainties.

Observations of extended sources that fill most of the MIRI imager field of view (e.g., nearby galaxies) will benefit from dedicated background observations, while various dithers of fields with point-like sources (e.g., extra-galactic fields) can be used as background images themselves. Here, we discuss two strategies to construct and subtract a background image from individual dithers: 1) using the background subtraction step in level 2 of the \emph{JWST} pipeline, 2) manually outside of the pipeline performed on level 2 products. We explain the two methods and their differences below.

\paragraph{ \emph{JWST} pipeline}
Background observations (either dedicated backgrounds or science images of sparse fields) can be input into the \emph{JWST} pipeline in background subtraction step as part of the \texttt{calwebb\_image2} (stage 2) process.
If more than one background observation is given to the pipeline, they will be combined into a sigma-clipped mean before being subtracted from each of the science data images. An example notebook on how to set up the background subtraction step is provided on GitHub here: \href{https://github.com/STScI-MIRI/Imaging_ExampleNB/blob/main/Pipeline_demo_subtract_imager_background.ipynb}{Background Subtraction Demo}. Background-subtracted \texttt{cal} files will then be fed to the stage 3 of the pipeline to make the final combined image.

We demonstrate the result of this step using F2100W observations of two commissioning programs PID~1024 and 1028 (see Figure~\ref{fig:bkgsub}). 
PID~1024 used four dithers. Two dithers in each pair are very close to each other ($\sim 0.2$\arcsec) but the two pairs have a gap of $\sim 5$\arcsec. Using these four point dithered images as the background shows an improvement over the combined image without background subtraction is significant. However, because of the closeness of the dithers in each pair, the sources are not completely removed in the clipped mean background. Hence, we see residual, negative, images in the pipeline image result. Pipeline background subtraction works best if all the dithers are sufficiently large (more than a few FWHM of the PSF). If there are individual \texttt{cal} files not sufficiently far from each other, we recommend to not use them as ``background images''.

\paragraph{Manual background construction}
Users can also construct a master background using either multiple dedicated background observations or multiple dithers of point-source-like science observations. The general methodology is to identify and remove sources from individual dithers and then, combine the source-free images to produce a single master background. Below, we explain one example for this general methodology that has been successful with MIRI data.

Instead of a using sigma-clipped mean to combine the background images, as is done in the pipeline, we use difference imaging for each two dithers to identify sources in the first image, and replace the image values of the source with those from the second image (which should be only background and without source emission). This can be done for each pair with sufficiently large dithers (a few times the MIRI filter's FWHM). Then, the median of the source-subtracted single dithers will be the master background. 

The background level might change from one observation to the other, particularly in longer-wavelength observations if individual dither are taken far from each other in time as the telescope thermal emission changes. To have a zero-level background subtracted image, we also renormalize the master background-subtracted individual dithers to have a zero median over the field. 

This method also works well only if the dithers are well separated from each other (multiples of MIRI FWHMs). Examples are shown in the right panel of Figure~\ref{fig:bkgsub}. MIRI imaging standard 4-dither pattern is ideal for this method. Smaller dither patterns that are designed for parallel modes (e.g., NIRCam as primary) or for the MIRI MRS as the primary mode, can result in non-uniform master backgrounds and over-subtraction of flux in background-subtracted images. 

Both these strategies only work well if the background observations (or science observations of point-like sources) have large dithers. Therefore, we strongly recommend observers to always take MIRI images with {\em multiple} and {\em large} dithers.

\subsection{Long integrations - Deep observation Performance}
\label{sec:deep_obs}

The imager team investigated two issues to de-risk long integration deep imaging observing strategies. Firstly, Spitzer IRAC suffered from temporal signal drifts on detectors with the same technology as MIRI which could reduce the signal to noise of deep imaging programs. Signal drifts had been seen in MIRI detector ground testing, which we generally attributed to persistence, therefore we wanted to verify that no new drift component was seen in flight that may come from e.g. the telescope or spacecraft background. Also, MIRI imaging programs will have the maximum length of an integration limited from the brightness of a source or high background. For deep imaging program of e.g. high-$z$" targets at low background there is no notional limit to the length of an integration. However, there is no destructive reset in an integration as the detectors sample up the long ramps and the data might thus be affected by additional noise. For example, a residual noise component could be added by cosmic rays hits whose signal is only removed by a destructive reset and this could reduce the overall signal-to-noise ratio for long integrations\footnote{From pre-flight predictions we expected a hit rate of 55 per second for a MIRI array. The prediction is then that about 7\% of the MIRI pixels will be affected by cosmic ray hits in 100 seconds, and about 70\% in 1000 seconds. In flight we see the cosmic ray rate vary significantly, the cause of which is not well understood at the time of writing. } Therefore, in MIRI commissioning, we also investigated if there is any penalty for taking very long integrations ($>$100 groups) that could be used in deep imaging programs. 

This commissioning activity was based on six exposures in program PID~1027 using MIRI’s shortest wavelength and lowest background filter, F560W. Each exposure was 3000 seconds in length but used different integrations lengths of:

\begin{itemize}
\item 90 groups = 250 seconds (FASTR1) (Repeated twice) 
\item 180 groups = 500 seconds (FASTR1)
\item 360 groups = 1000 seconds (FASTR1)
\item 540 groups = 1500 seconds (FASTR1)
\item 42 groups = 1000 seconds (SLOWR1)
\end{itemize}

For each of the six exposures we used a multiple of integrations so that the total number of groups matched between each exposure i.e., two integrations for the 540 groups data and 12 integrations for the 90 group data sets. In the test we chose not to use dithers to limit the free parameters that could affect the test results. The target was a nominally empty background region that was chosen simply because it was close to a photometric calibration star also used in program 1027. Overall, this experiment provides 6 deep exposures (both in imaging and MRS) that only differ by the number of integration’s used and to the authors knowledge is the only set of deep MIRI imaging data of its kind - taking the same image of the sky but with different detector setups. 

\begin{figure}
\includegraphics[width=250pt]{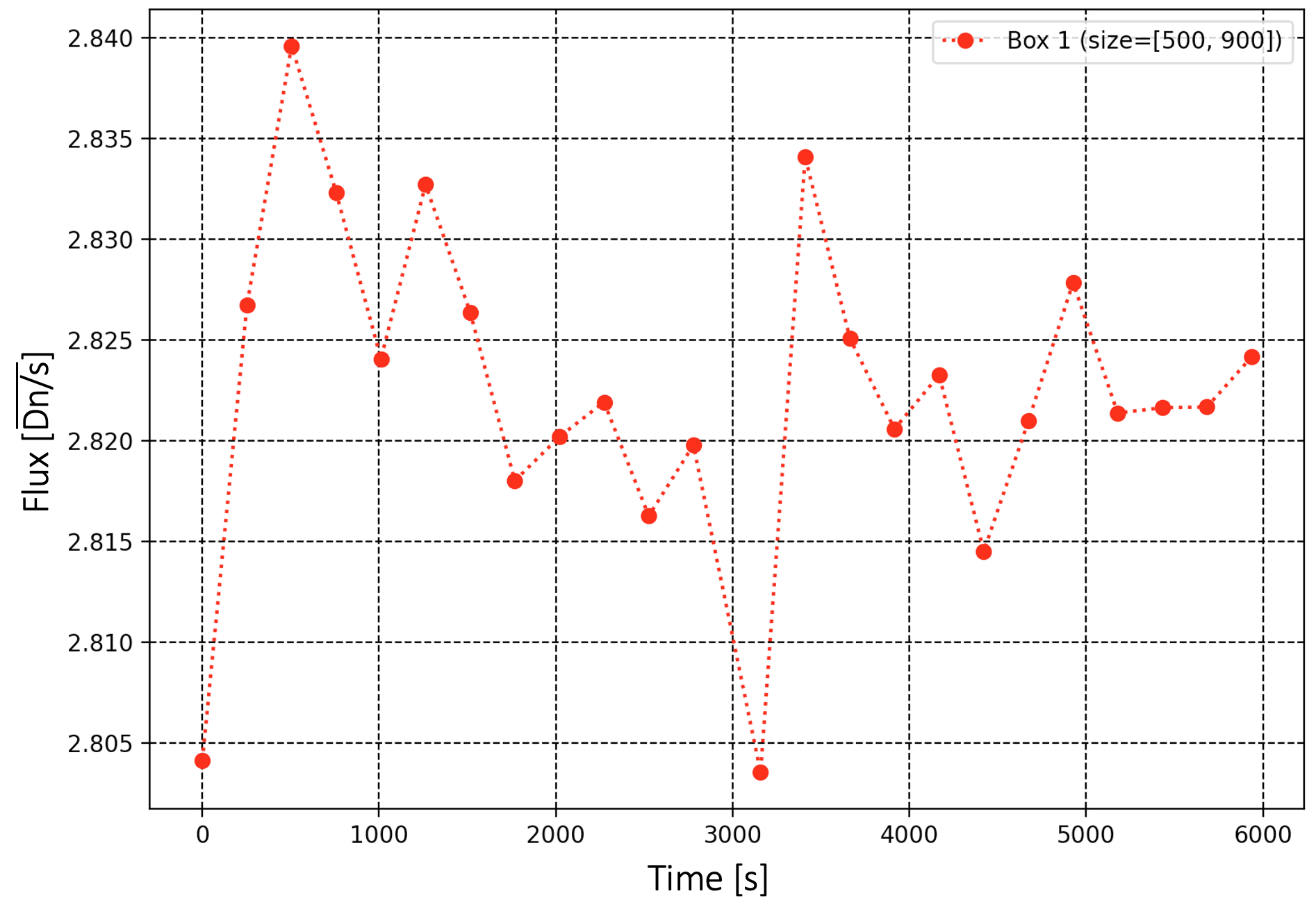}
\caption{Results of the drift investigation, showing the mean flux per integration for a 500$\times$900 pixels box centred on the imager field of view - from PID~1027 observation 2. All data is for the F560W filter and no dither or mechanism move was made during the 6000s observation. 
\label{fig:drift}}
\end{figure}

Figure~\ref{fig:drift} shows the results for the flight drift test – plotting the 24 consecutive integrations with 90 groups - a total observation time of 100 minutes. The plot shows the mean flux within a 500$\times$500 pixels box at the center of the imager for each integration. We do not see any evidence for a drift in the signal between integrations, where the difference in flux across the integrations is approximately 0.02 DN/s. We do see a first integration effect due to reset switch charge decay (RSCD - discussed in \cite{morrison2023}), where the first integration of the two exposures (each with 12 integrations) is lower in value than the subsequent integrations but overall there is no tendency to higher or lower signal over the 100 minute duration of the test. However, MIRI imager users might see drifts in their data which could be due to persistence. For example, when the instrument moves from long wavelength to short wavelength filters persistence may be seen in the short wavelength data as a decaying signal in the background (see Section~\ref{sec:artefacts-bkg}).

Next we can look at the maximum integration question using the whole data set of six exposures to investigate if there is any reduction in signal to noise or other cosmetic issue for data sets with long integrations ($>$100 groups). 

\begin{figure*}
\includegraphics[width=\linewidth]{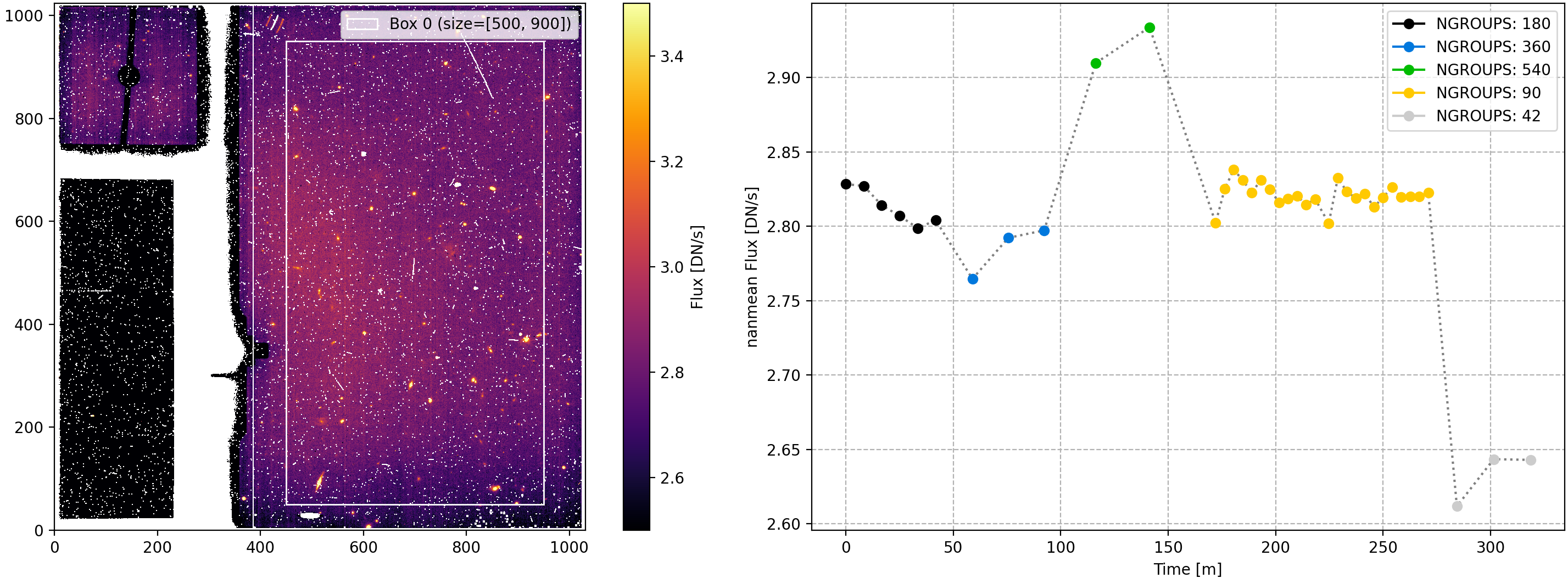}
\caption{Left: a stage 2 rate image showing the 500$\times$900 box area used to measure statistics. Right: the mean flux in the box for all integrations as a function of observing time. Note the 42 group data was taken in SLOWR1 mode and all other data in FASTR1.   
\label{fig:max_ramp_ints}}
\end{figure*}

As for the drift test above, Figure~\ref{fig:max_ramp_ints} shows a plot of the mean flux measured per integration, where the mean flux was measured within a 500$\times$900 pixels box centered on the imager field of view. Again, we expect the first integration of every exposure to be out of family with the subsequent exposures \citep{morrison2023} but it is clear there is an offset in the mean flux value between the data sets than can only be attributed to the difference in the number of integrations and readout mode used in the exposures - albeit small $<$ 0.2 MJy~sr$^{-1}$ and well within the flux calibration error. The biggest difference is between the SLOWR1 and FASTR1 data sets which is not surprising as the readout mode is very different in SLOWR1 - where groups are averaged on board the spacecraft. For the FASTR1 data there is no trend in the flux offset with integration length although the longest integration data (540 groups) has the highest offset compared to the other exposures. These small difference are likely due to a combination of persistence residuals, RSCD and detector settling effects. It is notable that the two 90 groups exposures show near identical flux results. This repeatability is an important strength of the MIRI instrument and its detectors where an observation tat is repeated exactly delivers the same result to a high level of accuracy. 

In Table~\ref{table:max_ramp} we show the results from data processed to the pipeline stage 2 (CRDS version = 11.16.22) where the resulting six images are the sum of all integrations for each exposure. Here we are looking to see if the noise, as measured by the standard deviation within the box, is higher for any of the data sets. The standard deviation varies only by 0.03 MJy~sr$^{-1}$ between the data sets with highest noise found for the longest ramps and the lowest noise found for the SLOWR1 and 180 group data\footnote{The SLOWR1 mode data performs well in this test but the fact that it is pre-averaged on board the telescope may affect the standard deviation measurement.}. 

\begin{table}
\centering
\begin{tabular}{ |c|c|c|c|  }
\hline
Read Pattern & Ngroups & Nints & Flux/stddev \\
\hline
FASTR1 & 180  & 6 & 2.81 $\pm$ 0.12\\
FASTR1 & 360  & 3 & 2.78 $\pm$ 0.12\\
FASTR1 & 540  & 2 & 2.92 $\pm$ 0.15\\
FASTR1 & 90  & 12 & 2.82 $\pm$ 0.13\\
FASTR1 & 90 & 12 & 2.82 $\pm$ 0.13\\
SLOWR1 & 42 & 3 & 2.63 $\pm$ 0.12 \\
\hline
\end{tabular}
\caption{Detector parameters and flux results for the long integration test discussed in Section~\ref{sec:deep_obs}. }
\label{table:max_ramp}
\end{table}

\begin{figure*}
\includegraphics[width=\linewidth]{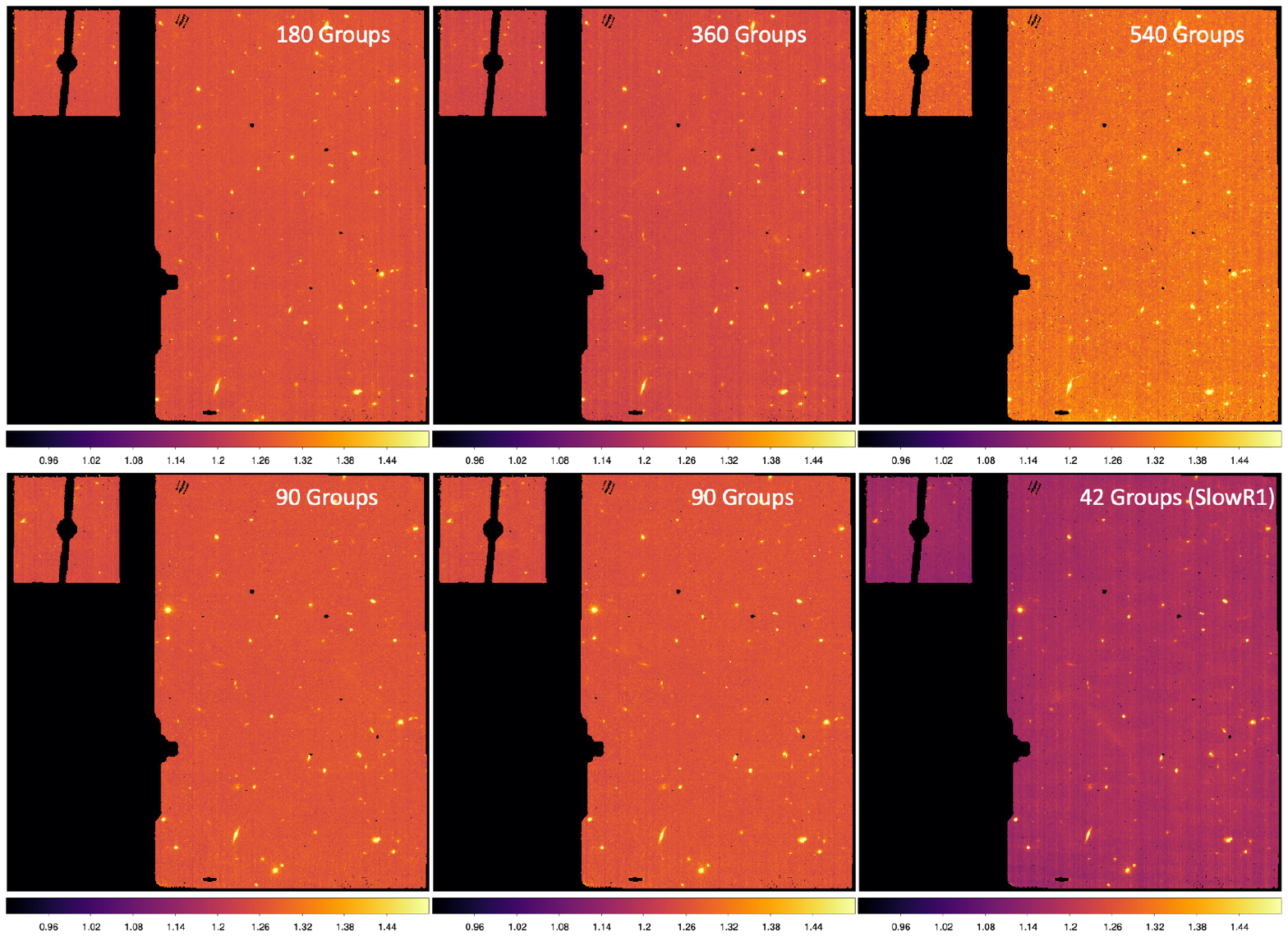}
\caption{Showing the stage 2 \texttt{cal} data for the six exposures obtained as part of the maximum integration and drift tests. The total depth of data in each image equals 3000 seconds. Flux units are MJy~sr$^{-1}$ and the colour scale is identical for all six images.  
\label{fig:max_ramp_images}}
\end{figure*}

Figure~\ref{fig:max_ramp_images} shows the resulting primary image of each of the six exposures i.e., the combination of all integrations with the total depth of 3000s per image. Here, by eye, we can confirm the results above that there is a flux offset between the data sets most notably for the 540 group and SLOWR1 data. 

From Figure~\ref{fig:max_ramp_images} we note that the two sets of 90 group data appear identical as expected. The 540 group data with the longest integrations does appear noisier or more granular as indicated in the statistics above. Because there is only one integration this could be due to the residual persistence from cosmic rays that is not removed by the destructive reset between integrations that takes place in the other exposures. Striping is also seen in the data which is associated with the reset anomaly \citep{morrison2023}, which is associated with the dark correction. The amplitude or contrast of the striping is normally very low ($<<$0.1 DN/s) and can vary depending on the processing of the different data sets but does not have a measurable impact on the signal or noise measured. Finally, the SLOWR1 data appears cosmetically worse than the other data this could be because the pre-averaging can make it harder to remove cosmic rays artefacts from the data. Also, some structure in the image could be because SLOWR1 flats from flight data have not been made yet, so the FASTR1 flat was used for this test.

Overall, the drift and maximum integration length commissioning tests show there is no important advantage or disadvantage from either short or long integrations from a signal-to-noise ratio point of view when considering integrations of 250 to 1500 seconds. From these data alone it is hard to recommend a “best” observing method for deep imaging programs. Good imaging is obtained for all data sets using at least 90 groups per integration.

\subsection{Short integrations - Bright Target Performance}\label{sec:bright_targets}

Pre-flight we had very limited MIRI imager data with bright targets using the flight system because ground test sources and backgrounds were optimized for the near-infrared instruments which are used for e.g., phasing the mirrors. Therefore, commissioning gave us the first opportunity to test how the imager would perform in science cases that require observations at the instruments bright limits.

As discussed in Section~\ref{sec:persistence}, as part of the anneal verification and testing program (PID~1023), we observed the bright $K$mag $=$ 2 star $\beta$ Doradus in four MIRI imager filters (F560W, F770W, F1000W and F1280W). The principle aim of the test was to analyse how the detectors recovered from the bright observation with and without the annealing function as well as verifying the anneal function worked. The test used four filters to investigate the extent of the cruciform artefact for this bright target and test the expectation that it would only be present at wavelengths less that 10 $\mathrm{\thinspace\mu m}$ as discussed in Section~\ref{subsec:cruciform} \citep[also see][]{gaspar2021}. The observations were made in one single integration of 90 groups which equals 250 seconds where the central core of the star saturates the detector in less than one group time ($<$2.7\,s). Therefore this data represents an extreme saturated data set for the MIRI imager. 

\begin{figure*}
\includegraphics[width=1.0\textwidth, center]{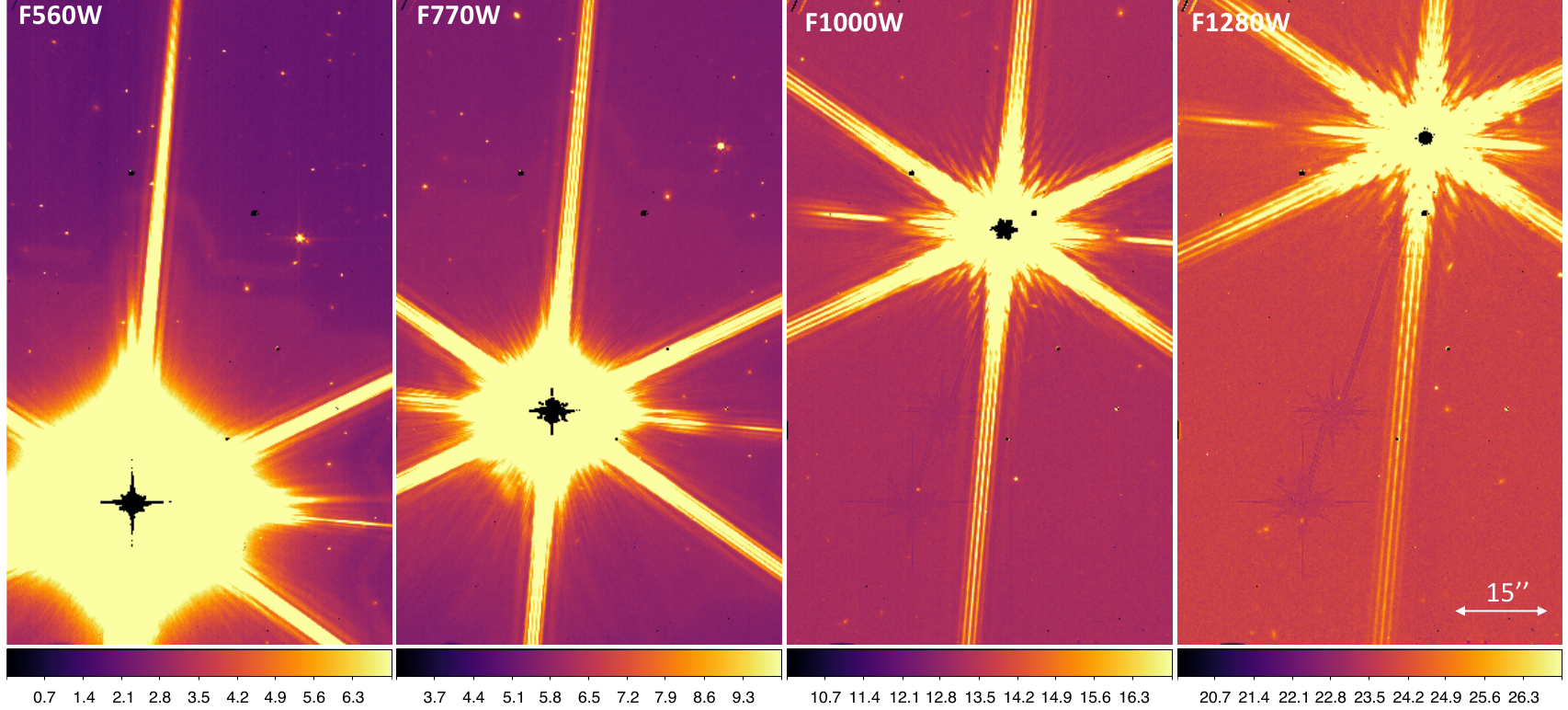}
\caption{Showing four images of the $K$mag $=$ 2 star $\beta$ Doradus taken as part of Program 1023 (observations 32--35) in four filters at 5.6, 7.7, 10 and 12.8 $\mu$m. The star was moved across the detector using offsets in the APT in order not to overlap the images. The bright star quickly (in less than 1 group $=$ 2.7\,s) saturates the detector at its centre (seen in black in these images). The images are presented in calibrated flux units (MJy sr$^{-1}$), processed to level 2b in the pipeline. At 5.6$\mathrm{\thinspace\mu m}$ an ``imprint" of the \emph{JWST} mirrors can be seen behind the source which is a persistence from slewing across the Flens as discussed in Section~\ref{sec:slew_persistence}.
\label{fig:beta_doradus}}
\end{figure*}

Figure~\ref{fig:beta_doradus} shows four \texttt{rate} images from the bright target test where the observations were made with offsets to allow the star to be imaged in different parts of the detectors, therefore any persistent signatures in the following exposures would not overlap. As discussed above in Section \ref{sec:persistence} the images clearly show persistence at the locations the star was observed. However what is also apparent from Figure~\ref{fig:beta_doradus} is that, although the star clearly saturates the detector in each of the images, there is no significant artefacts in the areas of the detector that were not illuminated by the saturating star as a consequence of the very bright star observation. In particular, the F560W image shows many galaxies in the background that are clear and well detected, although the number of background targets detected drops in the longer wavelength images due to the higher backgrounds. This result shows that the imager appears to function nominally in very high contrast imaging which will be important for science cases observing faint targets in the same field of view as a bright target. As also demonstrated in pre-flight ground testing, there is no penalty in terms of science performance for saturating on a source in the imager field of view that is not the principle science target. Therefore, the best strategy for imaging a faint source which contains bright targets in the same field of view may be to saturate the bright source(s) if the target of interest is of lower flux and not too close that the halo of the bright target overlaps with the science target. This would enable the observer to use longer ramps to observe the target of interest which are easier to calibrate as well as being more efficient\footnote{If in doubt of the best strategy contact the JWST help desk to get advice for your specific program.}.

\section{Normal operations post-commissioning}
\label{normalops}

Use of MIRI in the 1.5 years after commissioning has confirmed the promise of the instrument and its performance as verified in the commissioning tests. The overall imaging performance (and the other central instrument features) have generally returned data fully as expected and hoped. 

The sensitivity of the imager has proven to be moderately better than baselined in the pre-launch exposure time calculator (ETC), as a result of careful reductions performed with a super-background strategy \citep{2023Alvarez,2023Lyu}. This gain provides a bonus for the science programs and also margin against degradation.

One minor issue that still exists at the time of writing with regard to imager performance, as discussed in Section~\ref{sec:persistence_cosmic}, is the energetic particle hits on the detectors that produce extended artefacts made up of a small showers of particles freed within the detector material. This behaviour emphasises the importance of taking data in a way that provides a high level of redundancy in observations of any position within the region of interest, so these artefacts can be removed in processing without significant loss of signal to noise. 

In addition, On August 2023 the \textit{JWST} team announced that the MIRI Imager exhibits a reduced count rate in the long-wavelength filters. At the time of writing, the root cause of the issue is still under investigation, but the MIRI science operations are unchanged. To compensate for this issue, the JWST pipeline and flux calibration reference file has been updated to include this time-dependent throughput correction. The signal to noise change is about 20\% for the most strongly affected band (25.5 $\mu$m, F2550W) and this rate of change appears to be decreasing over the current mission lifetime. For more information refer to Gorden et al. in prep. and \emph{JWST} MIRI documentation.

The ``glowstick'' behaviour in the 4QPM coronagraphs described in \cite{wright2023PASP} is now mitigated with a suitable strategy to put the source on the centre of the coronagraph. In addition, reference files are available to remove the glowsticks themselves virtually perfectly, leaving at worst a small increase in noise at their positions. The resulting performance is illustrated well by \citet{2023Boccaletti}, who show strong detections of all four planets around HR 8799, including  planet \textit{e} at  0.4$''$ from star. At 11.4 $\mu$m, $\lambda$/D for the telescope is 0.4$''$, so the coronagraph is working well down to radii of $\lesssim$ $\lambda$/D, demonstrating the small inner working angle made possible by the 4QPM coronagraph design. An unanticipated result is how well PSF subtraction works for high contrast imaging \citet{2023Gaspar}. This is a result of the telescope figure being much more stable than had been anticipated. It is a valuable alternative to observing with the Lyot 23 $\mu$m coronagraph, since it can reveal areas close to the star that are behind the focal plane mask for the coronagraph.

\section{Summary} \label{sec:summary}
In summary, the pre-launch predictions for the MIRI imager, as verified during commissioning, are further confirmed and being exploited for a very broad range of mid-infrared science. The indications are that the excellent instrument performance will continue to serve the astronomical community for many years.

\begin{figure}
\centering\includegraphics[width=1\linewidth]{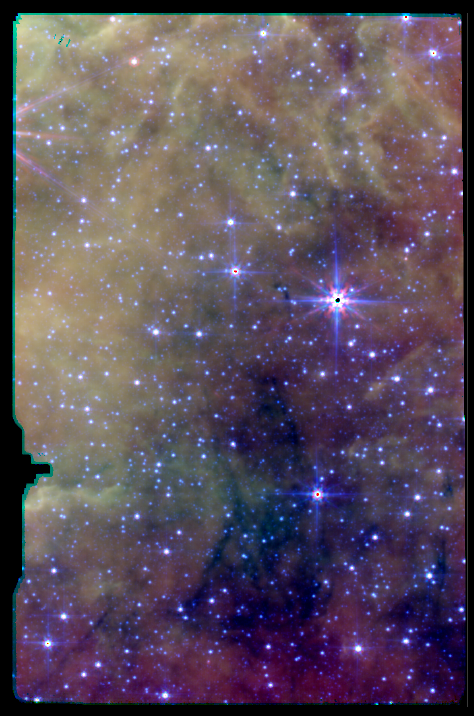}
\caption{MIRI commissioning image from near the galactic centre from program PID~1448. RGB image using F560W, F770W, F1280W.  \label{fig:galactic-centre}}
\end{figure}

From the MIRI commissioning program we were able to show that the MIRI imager meets and exceeds all the requirements that were set for the instrument mode pre-launch. Since the start of the \emph{JWST} science program in July 2022 the MIRI imager has met expectations and produced ground breaking data sets on topics across many disciplines of astrophysics. Highlights of the first imaging programs can be seen on the NASA and ESA websites from the early release data (ERS) and early release observations (ERO). Although this paper has outlined many of the challenges in producing high quality imaging products with MIRI these issues often refer to extreme cases of observation e.g., very high or very low background, bright targets or high contrast data sets. Figure~\ref{fig:galactic-centre} shows an RGB commissioning image of a region near the galactic centre. The image is noteworthy because no special processing was used in its production - it is simply the result of the default level 3 pipeline and calibration processing. Producing such high quality science product with minimal processing so early on in the missions stands to show the high quality of the observatory and instrument. 

\section{Conclusions} \label{sec:conclusions}

We have presented the in flight performance and calibration of the MIRI imager onboard \emph{JWST}. This work presents the results of the MIRI imager commissioning program and the performance of the instrument mode as best understood within the first year of science operations. 

As discussed in Section~\ref{sec:key} the key results show that MIRI's sensitivity was measured to be two orders of magnitude better than \textit{Spitzer} at 5.6~\mum\, and 1 order of magnitude better at 25.5~\mum\ and the relative photometric response of imaging was shown to be better than 5\% at 80\% encircled energy including the response between different subarrays and filters. We also demonstrated the performance and stability of the JWST background for MIRI showing the background level variation is within expectations at less than 5\% for hot and cold attitude temporal variation. The ability to correct distortion and flat field the imaging data was also demonstrated. 

Finally we have catalogued all image artefacts known at the time of writing and presented guides to the current best practices for imaging with JWST MIRI. The reader is again encouraged to refer to the JWST documentation for the most up to date performance and guidance for the imager instrument mode in the future: \href{https://jwst-docs.stsci.edu/jwst-mid-infrared-instrument#JWSTMidInfraredInstrument-Imager}{\textit{JWST} Documentation - MIRI imaging}. 

Further information on MIRI commissioning and performance can be found in \cite{rigby2022a, rigby2022b} and \cite{wright2023PASP} and more specifically in \cite{Argyriou2023a} for the Medium-Resolution Spectrometer,  \cite{morrison2023} for the Detector Effects and Data Reduction Algorithms,  \cite{boccaletti2022} for the coronagraphs performance, \cite{Libralato2023} for the point spread function and \cite{Argyriou2023b} for a discussion on the brighter-fatter effect for MIRI detectors. 

\begin{acknowledgements}
PG would like to thank the Sorbonne University, the Institut Universitaire de France, the Centre National d'Etudes Spatiales (CNES), the "Programme National de Cosmologie and Galaxies" (PNCG) and the "Physique Chimie du Milieu Interstellaire" (PCMI) programs of CNRS/INSU, with INC/INP co-funded by CEA and CNES,  for their financial supports. J.A-M. acknowledges support by grant PIB2021-127718NB-100 from the Spanish Ministry of Science and Innovation/State Agency of Research MCIN/AEI/10.13039/501100011033 and by “ERDF A way of making Europe.
\end{acknowledgements}

\newpage

\begin{appendix}

\section{PSF radial profiles and comparison with WebbPSF models in the corners of the image} 
\label{appendix_PSF_profiles}

Similarly to Figure~\ref{fig:PSF_F560W_F2500W}, in this appendix we show F560W PSF radial profiles for the other four positions in the corners of the imager field of view where standard stars were observed, from top left (position 2) to bottom right (position 5). We compare the super-resolved PSF flight data with WebbPSF simulations, with and without the addition of the cross-artefact. The radial profiles in the corners show a similar behaviour than the centre of the field of view, with the cross-artefact dominating  the flux profile at radii longer than the secondary Airy ring (see Sect.~\ref{subsec:cruciform} for details). The cumulative radial flux profiles (Encircled Energy, EE, normalised at a radius of 5~arcsec or 45 pixels) are shown to the right of the radial profiles for each position. The F560W flight data is more centrally concentrated than the WebbPSF simulations including the cross-artefact. This is due to the approximate representation (exponential profile) of the cruciform spatial distribution added on top of WebbPSF simulations at the time of commissioning. We note very small variations of the central parts of the profiles across the different positions over the FoV ($<$1\%), and overall agreement between WebbPSF profiles and flight data. However, WebbPSF models in the top left (position 2) and top right (position 3) produce PSF cores that are more asymmetric than flight data, because there are extrapolated from other positions in the FoV. Those two corners were indeed not covered by wavefront measurements during ground-based test campaigns.

\begin{figure*}
\centering
\includegraphics[width=0.5\linewidth]{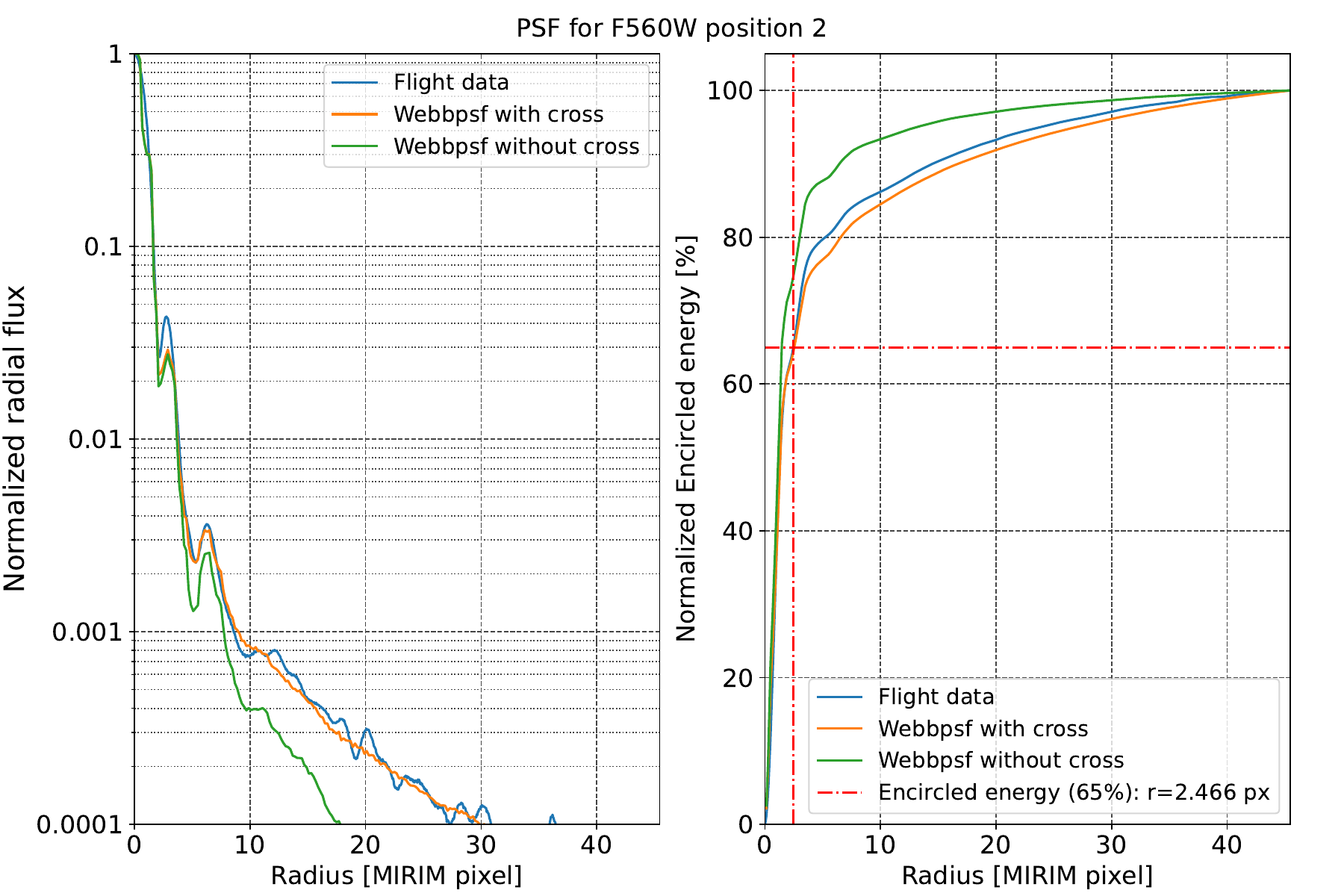}
\includegraphics[width=0.49\linewidth]{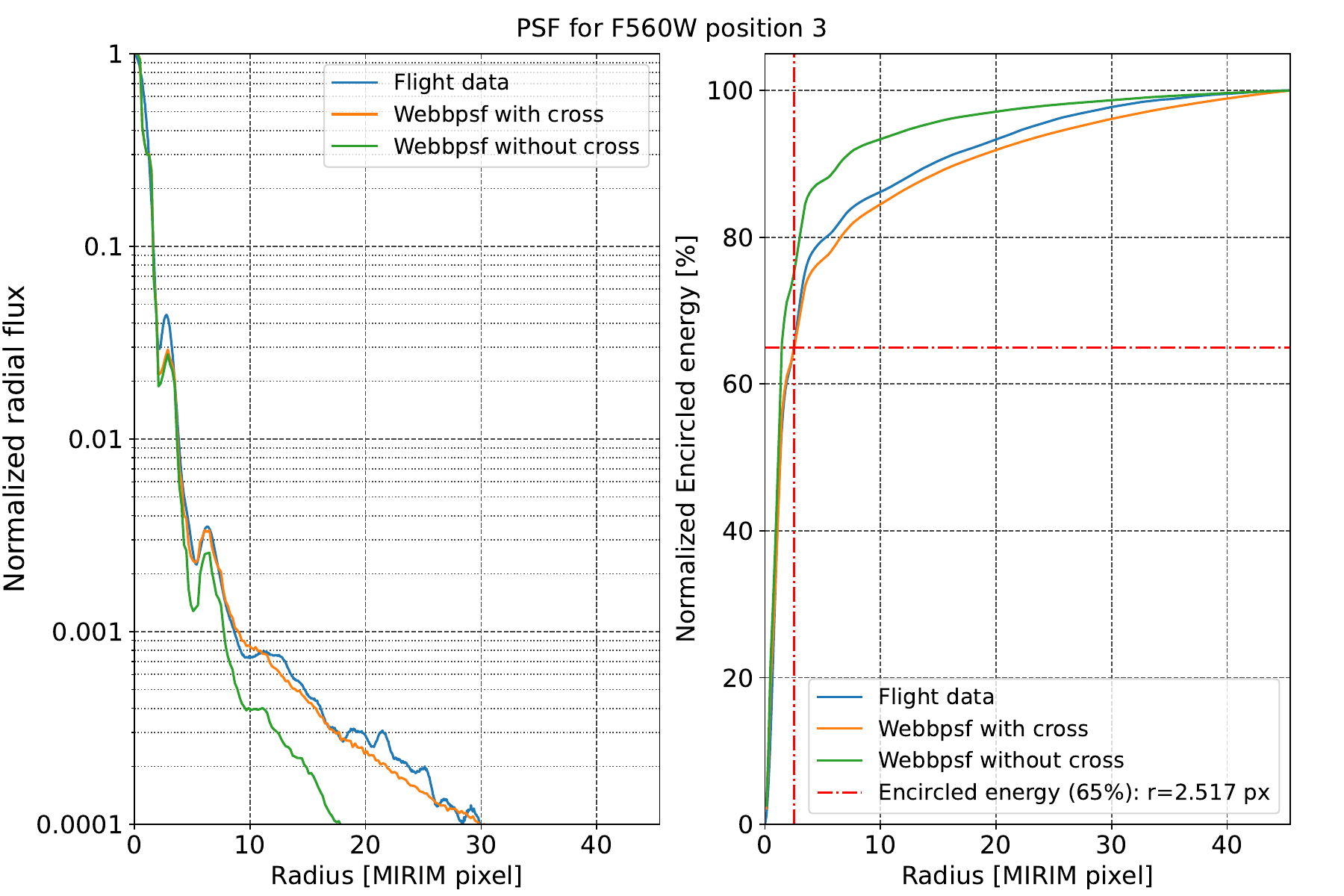}
\includegraphics[width=0.5\linewidth]{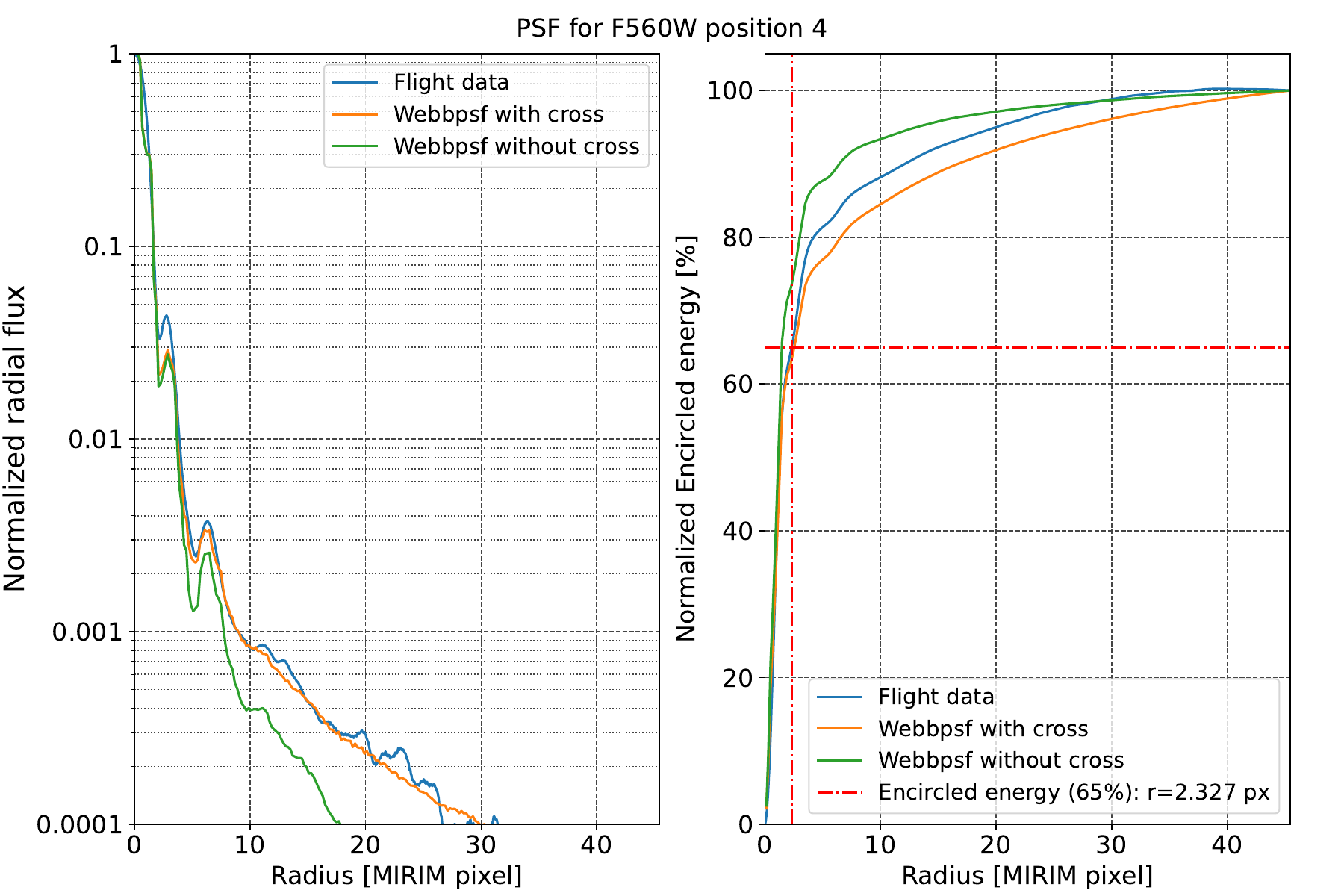}
\includegraphics[width=0.49\linewidth]{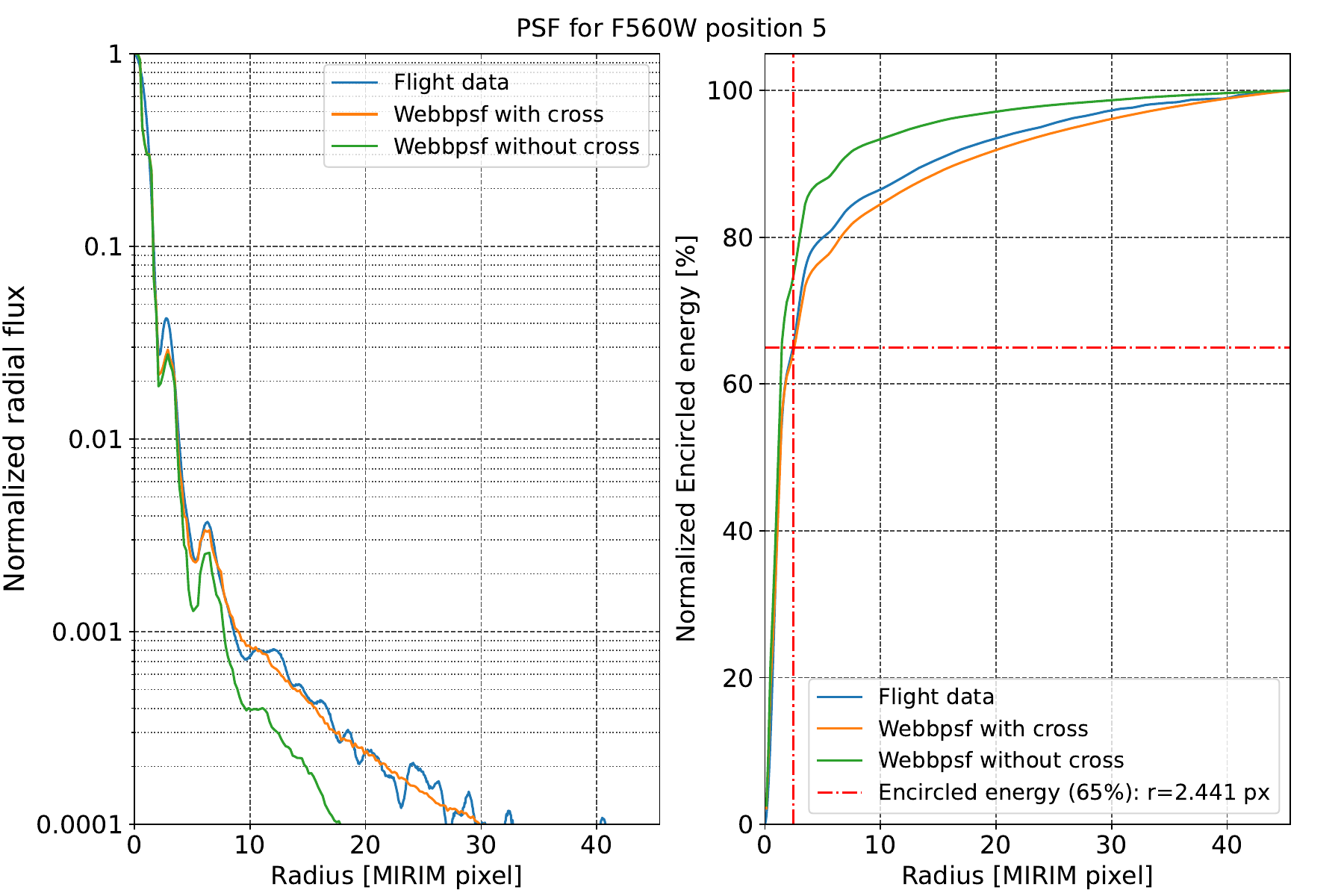}
\caption{Same as Figure~\ref{fig:PSF_F560W_F2500W}, for the 4 positions in the corners of the imager field of view. We compare the radial profiles of the flight F560W high-resolution PSFs (reconstructed from the microscanning data taken during commissioning) to WebbPSFs (with and without the cross-artefact, a.k.a the cruciform). }
\label{fig:PSF_F560W_all_positions}
\end{figure*}

\section{Microscanning dithers: pointing accuracy and estimated PSF positions on the detector.} 
\label{appendix_dithers}

\begin{figure*}
\centering
\includegraphics[width=0.61\linewidth, trim=-2cm -2cm 0 0]{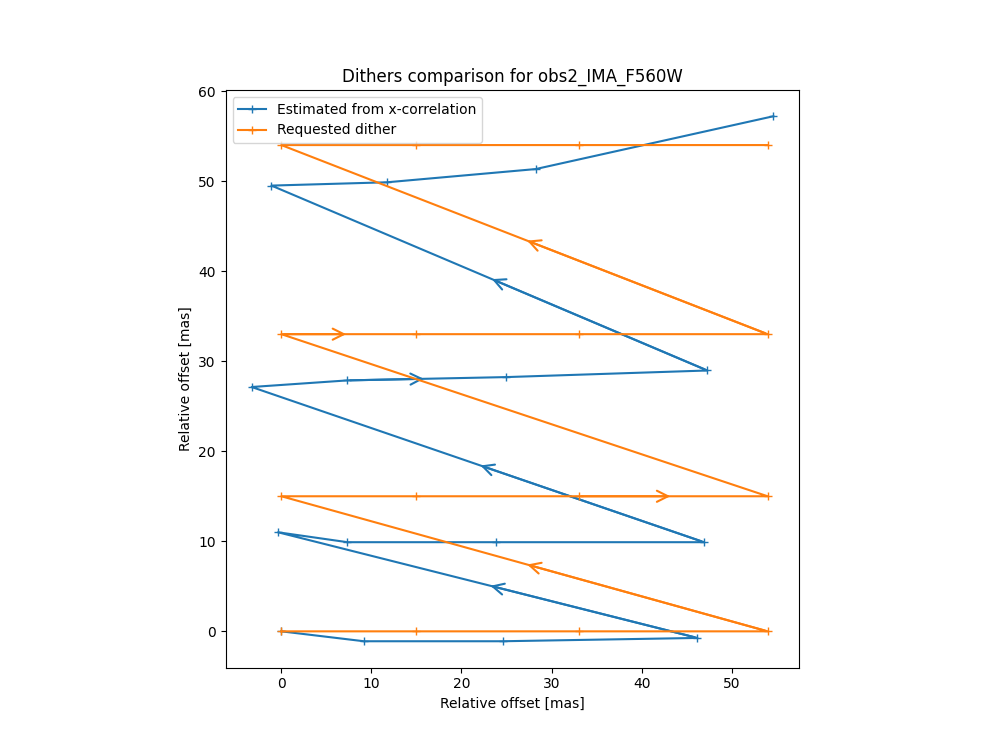}
\includegraphics[width=0.385\linewidth,  trim=0cm -2.5cm 0 0]{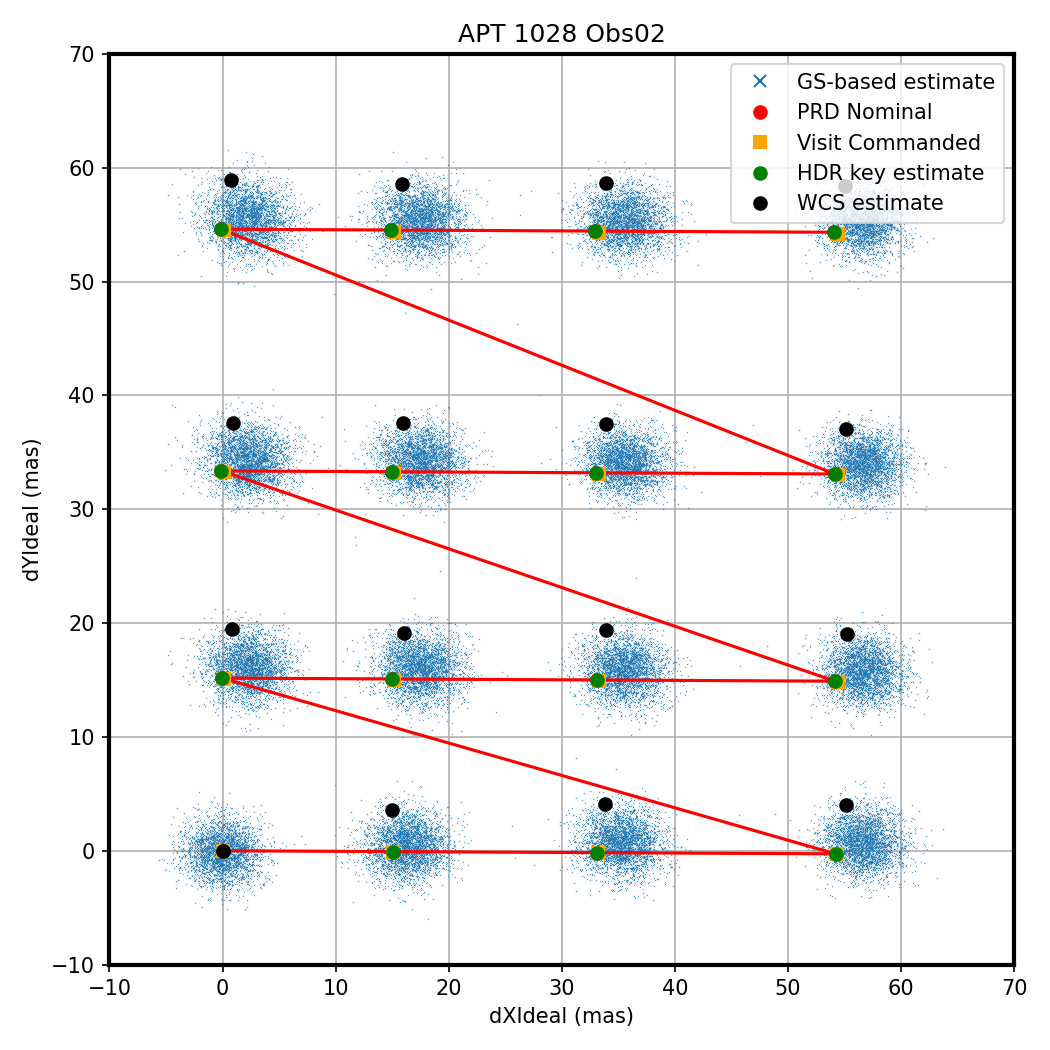}
\includegraphics[width=0.61\linewidth, trim=-2cm -2cm 0 0]{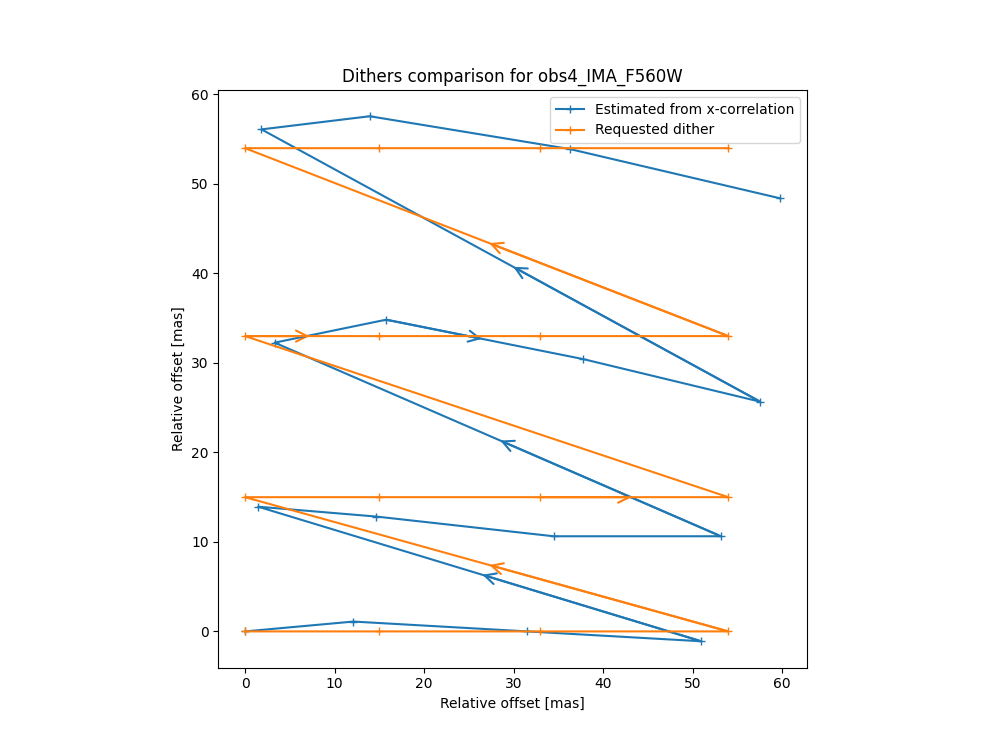}
\includegraphics[width=0.385\linewidth,  trim=0cm -2.5cm 0 0]{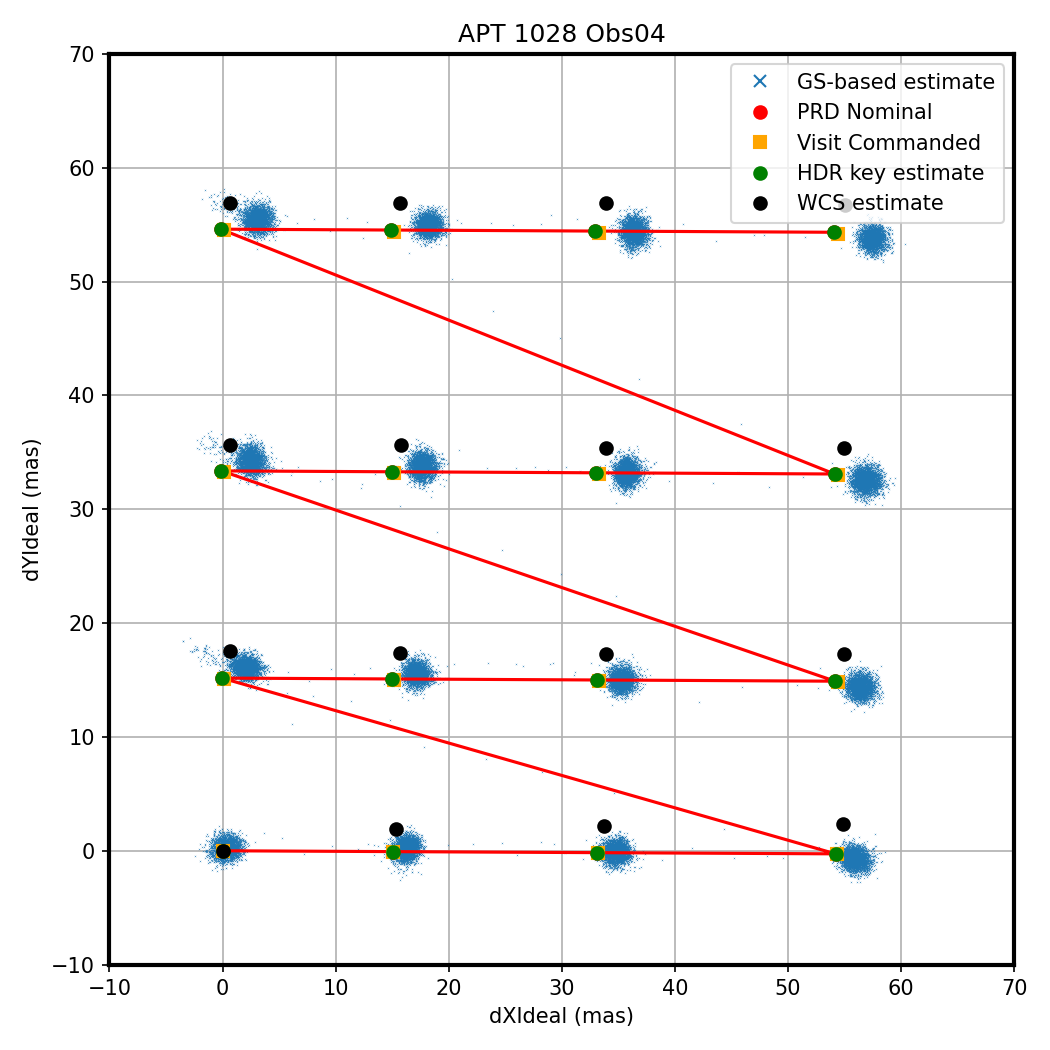}
\caption{Dither patterns used for super-resolution reconstruction of the PSF at 5.6~\mum. We used a $4\times 4$ pattern with sub-pixel shifts to sample the area of a pixel of the detector. We show the sub-pixel shifts (in mas, relative to the first pointing position) for Obs2 (top row) and Obs4 (bottom row). The left panel compares the nominal positions requested (blue crosses) with the actual shifts estimated from a two-pass Fourier transform cross-correlation.  The right panel compares the positions from the Fine Guiding Sensor data (blue jitter ball), the estimate from the WCS astrometric solution (black points), the file header coordinates (green points), and the commanded (resp. nominal) positions (orange, resp. red, points). }
\label{fig:microscan_dithers}
\end{figure*}

A $4\times 4 $~points sub-pixel microscanning dither pattern was used to sample the PSF at 5.6\mum, and reconstruct the super-resolved PSF shown in Fig.~\ref{fig:PSF_gallery}. We checked the accuracy of the dither positions by comparing the requested positions on the sky to the relative shifts estimated by a 2D FFT cross-correlation between individual images. The result is shown in the left panel of Fig.~\ref{fig:microscan_dithers} for two positions in the imager FoV (obs2: top left, and obs4: bottom right), as examples. The typical shift between the requested and estimated positions is less than 10~mas, which is compatible with the absolute pointing accuracy of the JWST. This is confirmed by an inspection of the Fine guidance System (FGS) data, shown on the right panel for both observations.

\end{appendix}

\bibliography{miri_imaging_flight}
\bibliographystyle{aa}

\end{document}